\documentclass[sigconf]{acmart}
\usepackage{amsfonts}

\usepackage{amssymb}
\usepackage{makecell}
\usepackage{pifont}
\usepackage{multirow}
\usepackage{graphicx}
\usepackage{booktabs}
\usepackage{siunitx,mathtools,amsmath,amssymb,bm}
\usepackage{booktabs}
\usepackage[capitalise,nameinlink,noabbrev]{cleveref}
\usepackage{subcaption}
\usepackage{microtype}
\usepackage{enumitem} % <--- added to allow itemize[...] optional args
\usepackage{soul,color,xcolor}

\DeclareRobustCommand{\wristp}{%
  \texorpdfstring{\textsc{WristP}\textsuperscript{2}}{WristP2}%
}
\newcommand{\handvqvae}{Hand--VQ--VAE}

\DeclareRobustCommand{\IoU}{\ensuremath{\operatorname{IoU}}}        % IoU symbol
\DeclareRobustCommand{\VolIoU}{\ensuremath{\operatorname{Vol.IoU}}} % Volumetric IoU
\newcommand{\cam}{wrist-mounted camera}

\AtBeginDocument{%
  }

\copyrightyear{2026}
\acmYear{2026}
\setcopyright{cc}
\setcctype{by}
\acmConference[CHI '26]{Proceedings of the 2026 CHI Conference on Human Factors in Computing Systems}{April 13--17, 2026}{Barcelona, Spain}
\acmBooktitle{Proceedings of the 2026 CHI Conference on Human Factors in Computing Systems (CHI '26), April 13--17, 2026, Barcelona, Spain}
\acmDOI{10.1145/3772318.3790626}
\acmISBN{979-8-4007-2278-3/2026/04}

\begin{document}

\title{\wristp{}: A Wrist-Worn System for Hand Pose and Pressure Estimation}
% \newcommand{\wristp}{WristP$^2$ }

\author{Ziheng Xi}
\authornote{These authors contributed equally to this work.}
\affiliation{%
  \institution{Department of Automation, BNRIst, Tsinghua University}
  \city{Beijing}
  \country{China}
}
\email{xizh21@mails.tsinghua.edu.cn}

\author{Zihang Ao}
\authornotemark[1]
\affiliation{%
  \institution{Department of Automation, Tsinghua University}
  \city{Beijing}
  \country{China}
}
\email{azh24@mails.tsinghua.edu.cn}

\author{Yitao Wang}
\affiliation{%
  \institution{Department of Automation, Tsinghua University}
  \city{Beijing}
  \country{China}
}
\email{wangyita23@mails.tsinghua.edu.cn}

\author{Mingze Gao}
\affiliation{%
  \institution{Tsinghua University}
  \city{Beijing}
  \country{China}
}
\email{gaomingze2022@gmail.com}

\author{Wanmei Zhang}
\affiliation{%
  \institution{Tsinghua University}
  \city{Beijing}
  \country{China}
}
\email{zwm23@mails.tsinghua.edu.cn}

\author{Jianjiang Feng}
\affiliation{%
  \institution{Department of Automation, Tsinghua University}
  \city{Beijing}
  \country{China}
}
\email{jfeng@tsinghua.edu.cn}

\author{Jie Zhou}
\affiliation{%
  \institution{Department of Automation, BNRIst, Tsinghua University}
  \city{Beijing}
  \country{China}
}
\email{jzhou@tsinghua.edu.cn}

\renewcommand{\shortauthors}{Ziheng Xi et al.}

\begin{abstract}
\label{sec:abstract}
Accurate 3D hand pose and pressure sensing is essential for immersive human-computer interaction, yet simultaneously achieving both in mobile scenarios remains a significant challenge. We present \wristp{}, a camera-based wrist-worn system that estimates 3D hand pose and per-vertex pressure from a single wide-FOV RGB frame in real time. A ViT (Vision Transformer) backbone with joint-aligned tokens predicts \handvqvae{} codebook indices for mesh recovery, while an extrinsics-conditioned branch jointly estimates per-vertex pressure. On a self-collected dataset of 133{,}000 frames (20 subjects; 48 on-plane and 28 mid-air gestures), \wristp{} attains MPJPE (Mean Per-Joint Position Error) of \SI{2.9}{\milli\meter}, Contact~\IoU{} of \num{0.712}, \VolIoU{} of \num{0.618}, and foreground pressure MAE of \SI{10.4}{g}. Across three user studies, \wristp{} delivers touchpad-level efficiency in mid-air pointing and robust multi-finger pressure control on an uninstrumented desktop. In a real-world large-display Whac-A-Mole task, 
\wristp{} also enables higher success ratio and lower arm fatigue than head-mounted camera-based baselines. These results position \wristp{} as an effective, mobile solution for versatile pose- and pressure-based interaction. Website: \url{https://zhenqis123.github.io/WristPP/}

% TODO: Add a summary sentence.

\end{abstract}



\begin{teaserfigure}
  \centering
  \includegraphics[width=\textwidth]{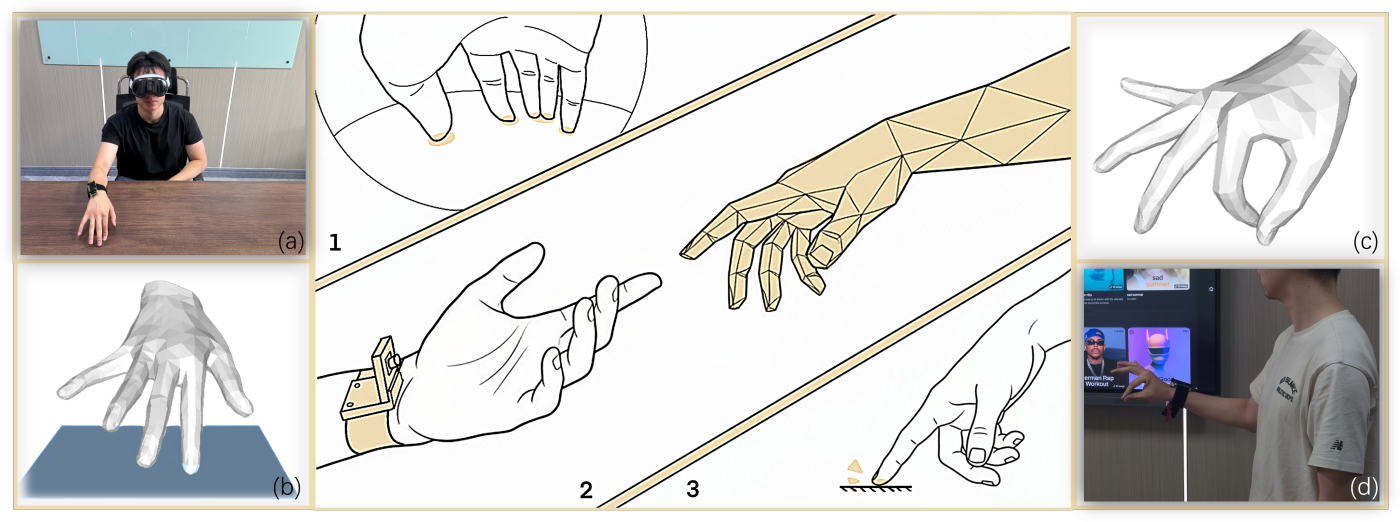}
  \caption{\wristp{} is a wrist-worn system with a wide-FOV fisheye camera that reconstructs 3D hand pose and per-vertex pressure in real time. Illustrated scenarios: (a) planar input on a virtual surface in XR; (b) reconstructed hand and planar contact corresponding to (a); (c) reconstructed hand corresponding to (d); (d) controlling slide presentations in a mobile scenario.}
\Description{Overview of the WristP2 system and example usage scenarios. 
The figure shows the wrist-worn fisheye camera, reconstructed 3D hand meshes with pressure visualization, planar virtual touchpad interaction in XR, and remote slide control on a large screen.}
  \label{fig:teaser}
\end{teaserfigure}

%% 
%% This command processes the author and affiliation and title.
\maketitle

\section{Introduction}
\label{sec:intro}
Accurate perception of hand pose and applied pressure is crucial for immersive, natural, and efficient interaction, especially in egocentric settings \cite{romero2022embodied,eyehand_movement,STMG}. In extended reality (XR), pose alone is often insufficient: touch contact and pressure provide additional control bandwidth and feedback \cite{ComforTable-User-Interfaces}. For example, when writing or drawing on a virtual canvas, variation in finger pressure and contact area can modulate stroke thickness \cite{pressurevision++}. In virtual piano playing, different finger pressures and poses can produce various sound effects such as soft, loud, vibrato, and staccato notes \cite{AirPiano}. Many laptops already feature pressure-sensitive touchpads that respond to different force levels with haptic feedback, underscoring the practical value of pressure-aware interaction.

At the same time, building a system that senses both rich 3D hand pose and fine-grained pressure in everyday settings remains challenging. Key difficulties include: (i) the large number of hand joints and the resulting articulatory complexity; (ii) out-of-view (OOV) and occlusion events during natural interaction; and (iii) the acquisition of high-resolution contact and pressure data that generalizes beyond instrumented surfaces.

Environment-mounted systems—such as depth cameras~\cite{V2v-posenet, mueller2019real}, multi-camera arrays~\cite{Interhand2_6m, khaleghi2022multiview, Voxelpose}, and optical motion capture (MoCap) systems~\cite{motion_capture, deepmocap}—provide high-accuracy pose, but typically focus on geometry alone and require fixed workspaces, careful calibration, and non-trivial deployment cost and space. Glove-based wearables equipped with flexible sensors~\cite{Actionsense,IMU_Based_glove} can directly measure pose and pressure, yet often reduce comfort, interfere with tactile perception, and restrict finger motion. Lightweight on-body solutions based on inertial measurement units (IMUs)~\cite{IMU_measurements,IMU_Based,TapType}, electromyography (EMG)~\cite{sEMG_and_IMU,meg2pose,neuromotor_emg}, and acoustic sensing~\cite{echoWrist,VersaTouch} bring sensing closer to the hand, but are often limited to recognizing predefined gesture sets without continuous 3D articulation~\cite{IMUSensor,TapType,echoWrist,meg2pose}, or cannot estimate pressure~\cite{WearableSystemsReview}. 

Head-mounted displays (HMDs) provide egocentric visual access to the hands and underpin commercial hand tracking~\cite{HMD-Poser,Mocap_Everyone_Everywhere}. However, hands generally need to operate within the headset’s tracking volume to remain reliably trackable, which can be challenging near the edges of the field of view and during prolonged mid-air pointing~\cite{eyehand_movement,HOOV}. From the HMD perspective, it is also difficult to distinguish whether the hand is hovering just above a surface or in actual contact~\cite{EgoTouch}. In addition, HMDs tend to have higher cost and power consumption, which can limit everyday adoption.

Recent prototypes and products based on wrist-mounted cameras~\cite{TapXR1,IMOO,FineType,FingerTrak,Back-Hand-Pose,WristCam} suggest an alternative: wrist-anchored, eyes-free sensing that moves with the user. But these devices provide only limited hand pose estimation and, to our knowledge, do not estimate hand pressure. Achieving high-precision 3D pose and pressure estimation with a portable, low-cost device, especially under complex finger motions in mobile scenarios, remains an open challenge.

%TODO: 优化交互平面种类的表述
We present \wristp{}, a wrist-mounted system that reconstructs hand pose and per-vertex contact and pressure in real time from a single wide-FOV (Field-Of-View) fisheye RGB frame. The system employs a deep neural network to recover an accurate 3D hand mesh and surface-pressure distribution, enabling precise, multidimensional input. With these outputs, \wristp{} turns rigid planar or quasi-planar surfaces into touchpad-like, pressure-sensitive input areas and also supports fine-grained mid-air gestures, broadening the space of mobile and XR interaction.

We designed 48 plane-based and 28 mid-air gestures, and collected a dataset of ~133,000 frames from 20 participants. The data collection system consists of an infrared FZMotion Motion Capture System, a Kinect camera, and a Sensel Morph \cite{SenselMorph} pressure-sensitive pad, enabling the acquisition of sub-millimeter marker positions and pressure maps. High-fidelity 3D hand meshes and per-vertex pressure annotations were then obtained through a multi-objective optimization process. Building on this dataset, we propose a novel framework that tokenizes the ground-truth 3D hand mesh into a discrete latent space with \handvqvae{}, then leverages a ViT (Vision Transformer) \cite{vaswani2017attention} based neural network to predict the corresponding token indices in the codebook to get hand mesh prediction. In parallel, the per-vertex surface pressure distribution is inferred through a cross-attention mechanism. The results show that the mean per-joint position error (MPJPE) was 2.9~mm, the mean joint angle error (MJAE) was 3.2°, the intersection over union (IoU) for pressure was 71.2\%, the volume intersection over union (Vol.IoU) for surface pressure was 61.8\%, the mean absolute error (MAE) on single vertex with positive value was 10.4~g. To assess the practical applicability of \wristp{} for human-computer interaction (HCI), we conducted three user studies spanning both planar and mid-air interaction. Across the three studies, \wristp{} delivered input efficiency comparable to standard input devices and robust performance on demanding tasks requiring fine-grained pressure and displacement control. In an additional application study, we compared \wristp{} with head-mounted camera-based input methods in a real-world large-screen interaction task; participants using \wristp{} completed the task more effectively and reported lower arm fatigue. In user-experience questionnaires, participants also highlighted the naturalness and accuracy of interaction with \wristp{}.

In summary, our main contributions include:
\begin{itemize}
    \item We introduce \wristp{}, an eyes-free, low-cost, wide-FOV wrist-worn device that enables high-precision hand pose and pressure estimation.
    \item We collect a large‑scale dataset of wrist‑mounted fisheye images with aligned hand meshes and surface pressure, and design an extrinsics-aware \handvqvae{} based architecture for joint pose and pressure prediction.
    \item We extensively evaluate \wristp{} through offline experiments and four studies. \wristp{} achieves low-millimeter pose and high pressure accuracy, and delivers touchpad-level pointing efficiency and robust multi-finger pressure control (Studies~1–3), and, in a large-display Whac-A-Mole task (Study~4), yields higher task success and lower arm fatigue than head-mounted baselines.
\end{itemize}

\section{Related Work}

\begin{table*}[htbp]
  \centering
  \caption{Hand pose and pressure estimation techniques that have been proposed or hold potential applications in HCI, as well as the performance metrics achieved on their own datasets. Here, “$\bigcirc$” indicates that the technique supports partial hand mesh estimation for pose or pressure, limited to keypoints, wrist points, individual fingers, or other localized hand positions. “--” signifies that the corresponding metric is not reported in the paper, or the technique required to achieve this metric is not implemented in the paper.}

  \resizebox{\textwidth}{!}{  % Scale the table to fit the text width
  \renewcommand{\arraystretch}{1.3}
    \begin{tabular}{c|ccccccccccc}
    \toprule
    \textbf{Category} & \textbf{Study}  & \textbf{Year} & \textbf{Sensor}& \textbf{Eyes-free} &\textbf{\makecell[c]{ Hand Pose \\ Estimation}}& \textbf{\makecell[c]{Contact \\ Detection}} & \textbf{\makecell[c]{Hand Pressure \\ Estimation}} & \textbf{\makecell[c]{Pose \\ MPJPE(mm)$\downarrow$}}&\textbf{\makecell[c]{Pose \\MJAE($^\circ$)$\downarrow$}}& \textbf{\makecell[c]{Contact\\ Accuracy(\%)$\uparrow$}}& \textbf{\makecell[c]{Pressure \\ MAE(kPa/\%/g)$\downarrow$}} \\
    \hline
    \multirow{3}[0]{*}{Head-worn}
          & EgoTouch \cite{EgoTouch}    & 2024  & HMD   & \textcolor{red}{$\times$}        & \textcolor{red}{$\times$}  & \textcolor{green!65!black!65!black}{\checkmark}   &\textcolor{orange}{$\bigcirc$} & --&--&95.6&5.6-16.4\% \\
          & PressureVision++ \cite{pressurevision++}   & 2024  & HMD & \textcolor{red}{$\times$} & \textcolor{red}{$\times$}  & \textcolor{green!65!black}{\checkmark}   &\textcolor{green!65!black}{\checkmark}  & --&--& 89.3 & 44~kPa\\
          & EgoPressure \cite{Egopressure} & 2024  & HMD   & \textcolor{red}{$\times$}      & \textcolor{green!65!black}{\checkmark} & \textcolor{green!65!black}{\checkmark}    & \textcolor{green!65!black}{\checkmark}    & 5.68&--&92&71 kPa \\
    \hline
    \multirow{2}[0]{*}{Arm-worn}
          & PiMForce \cite{PiMForce}   & 2024   & EMG+RGB camera & \textcolor{red}{$\times$} & \textcolor{red}{$\times$}  & \textcolor{red}{$\times$}   & \textcolor{orange}{$\bigcirc$}& --&--&--&6.98-10.1\% \\
          & EITPose \cite{kyu2024eitpose}   & 2024   & EIT sensor & \textcolor{green!65!black}{\checkmark} & \textcolor{green!65!black}{\checkmark}  & \textcolor{red}{$\times$}   & \textcolor{red}{$\times$}& 11.06-18.91&--&--&-- \\
    \hline
    \multirow{2}[0]{*}{Hand-worn}
          & VibeMesh \cite{mao2025visuo}  & 2024   & microphone+ speaker+RGB-D camera   &  \textcolor{red}{$\times$}  &  \textcolor{green!65!black}{\checkmark} & \textcolor{green!65!black}{\checkmark} & \textcolor{red}{$\times$} & --&--&--&-- \\
          & Ring-a-Pose \cite{yu2024ring}  & 2024   & microphone + speaker   &  \textcolor{green!65!black}{\checkmark}  &  \textcolor{orange}{$\bigcirc$} & \textcolor{red}{$\times$} & \textcolor{red}{$\times$} & 10.30-14.10&--& --&-- \\
    \hline
    \multirow{9}[0]{*}{Wrist-worn}
        & Digits \cite{Digits}  & 2012  & IR camera   & \textcolor{green!65!black}{\checkmark}  & \textcolor{orange}{$\bigcirc$}  & \textcolor{red}{$\times$}   & \textcolor{red}{$\times$}     & --&5.34-7.51&--&-- \\
          & WristCam \cite{WristCam}   & 2019  & RGB camera & \textcolor{green!65!black}{\checkmark}  & \textcolor{orange}{$\bigcirc$}  & \textcolor{red}{$\times$}   & \textcolor{red}{$\times$}     & --&--&--&-- \\
          & FingerTrak \cite{FingerTrak} & 2020  & IR camera & \textcolor{green!65!black}{\checkmark}  & \textcolor{green!65!black}{\checkmark}  & \textcolor{red}{$\times$}   & \textcolor{red}{$\times$} & 12.6&6.46&--&--\\
          & Back-Hand-Pose \cite{Back-Hand-Pose}  & 2020  & IR camera & \textcolor{green!65!black}{\checkmark}  & \textcolor{orange}{$\bigcirc$}   & \textcolor{red}{$\times$}   & \textcolor{red}{$\times$}   & --&8.81-9.77&--&-- \\
          & RotoWrist \cite{RotoWrist}   & 2021  & IR light modules & \textcolor{green!65!black}{\checkmark}  & \textcolor{orange}{$\bigcirc$}  & \textcolor{red}{$\times$}   & \textcolor{red}{$\times$}     & --&5.9-7.8&--&-- \\
          & emg2pose \cite{meg2pose}  & 2024  & EMG   & \textcolor{green!65!black}{\checkmark} & \textcolor{green!65!black}{\checkmark}  & \textcolor{red}{$\times$}    & \textcolor{red}{$\times$}    & 15.8-27.2&12.2-15.8&--&-- \\
          & EchoWrist \cite{echoWrist}  & 2024  & microphone + speaker   & \textcolor{green!65!black}{\checkmark}  & \textcolor{orange}{$\bigcirc$}  & \textcolor{red}{$\times$}   & \textcolor{red}{$\times$}     & 4.81&3.79&--&-- \\
           & FineType \cite{FineType}    & 2025  & IMU+IR camera   & \textcolor{green!65!black}{\checkmark}  & \textcolor{orange}{$\bigcirc$}   & \textcolor{green!65!black}{\checkmark}  & \textcolor{red}{$\times$}   & --&--&--&-- \\
          \cline{2-12}
    & \textbf{Ours} & -  & RGB camera & \textcolor{green!65!black}{\checkmark} & \textcolor{green!65!black}{\checkmark}  & \textcolor{green!65!black}{\checkmark}  & \textcolor{green!65!black}{\checkmark}   & 2.9 & 3.2 & 97 & 10.4g \\
    \bottomrule
    \end{tabular}
}
\label{tab:related}
\end{table*} 
Accurate hand pose and pressure estimation underpins immersive human-computer interaction. Beyond 
algorithmic design, a system's capability is strongly influenced by its sensing placement, which determines 
(i) what cues are observable, (ii) whether hands remain continuously trackable through natural 
postures, and (iii) how ergonomic and deployable the interaction is. We therefore organize related 
work by placement—environment-mounted, head-mounted (egocentric), and wrist-proximal—and, 
within each, discuss pose and pressure jointly as two facets of hand understanding. 
Representative methods and 
performance appear in Table \ref{tab:related}.

\subsection{Environment-Mounted Sensing}
Environment-mounted systems observe from fixed external viewpoints or instrument the world itself. 
For pose, classical geometric pipelines combine calibrated multi‑view/depth/IR with triangulation, 
kinematic fitting, and optimization 
\cite{xu2013efficient,supancic2015depth,oikonomidis2011efficient,tagliasacchi2015robust,motion_capture}, which enabled 
high‑accuracy capture and the creation of canonical benchmarks—NYU \cite{Tompson2014ICCV_NYU}, ICVL \cite{Tang2014ICVL}, 
STB \cite{Zhang2017STB}, BigHand2.2M \cite{BigHand2_2M}, etc \cite{FreiHAND_2019_ICCV,Interhand2_6m,GigaHands_2025_CVPR}. Learning-based third-person vision then regresses joints and meshes directly from RGB or depth using heatmaps~\cite{Ge_2016_CVPR,Moon2018V2V}, volumetric encoding~\cite{Zimmermann2017ICCV}, occlusion-aware modeling for hand–object interaction~\cite{choi2017robust,choi2017learning}, and end-to-end MANO-based reconstruction with CNNs and transformers~\cite{Baek2019Mesh,Kulon2020Hand,Handy_2023_CVPR,Hamer_2024_CVPR,WiLoR_2025_CVPR,Mueller_2017_ICCV}.
For pressure, two strategies dominate: (a) instrumented surfaces such as force/pressure mats~\cite{tekscan,kasai2018development,valero2016interfacial}, fingertip sensors~\cite{BUSCHER2015244,PressureProfile}, or tactile-sensing gloves~\cite{sundaram2019learning} directly measure pressure distributions but alter surfaces or require worn hardware; (b) external visual inference, where pressure is estimated from color changes~\cite{chen2020estimating,Photoplethysmograph,finger_posture}, soft-tissue deformation~\cite{hwang2017inferring}, and cast shadows~\cite{hu2002visual,Visual_cues}. Other approaches infer force through object-motion cues during grasping~\cite{Hand-Object_Contact,Li_2019_CVPR,Ehsani_2020_CVPR}.
Dataset-driven methods further pair RGB with high-resolution pressure maps from Sensel Morph~\cite{SenselMorph} to learn appearance-to-pressure mappings~\cite{pressureVision,pressurevision++}.

However, these systems inherently confine interaction to instrumented workspaces. The reliance on fixed camera frustums and complex calibration creates a high deployment burden and severely restricts mobility. Furthermore, external viewpoints are susceptible to frequent occlusions and out-of-view motions, which break both pose tracking and visual pressure inference pipelines. Finally, methods that infer force via object motion are fundamentally inapplicable to interactions with static surfaces (e.g., tables, walls). Collectively, these constraints render environment-mounted sensing unsuitable for ubiquitous, everyday scenarios.

\subsection{Head-Mounted Sensing}

Head-mounted cameras co-locate sensing with head, yielding long-horizon hand visibility and stable reference frames in mixed reality; 
this configuration underpins commercial hand tracking (e.g., Meta Quest) \cite{meta_quest_2022} and research systems that unify 
multi-view egocentric imagery with end-to-end 3D hand reconstruction \cite{han2022umetrack}. 

Beyond pose, egocentric viewpoints expose contact appearance on visible skin, enabling pressure inference from purely visual cues. 
In particular, estimating fingertip pressure from local tissue deformation and shading demonstrates appearance-to-pressure mappings 
in the wild \cite{EgoTouch}, while jointly estimating hand mesh and pressure shows bidirectional benefits between geometry and contact 
intensity \cite{Egopressure}. 

Recent work broadens egocentric pressure and touch beyond fingertips to broader contact regions and tasks. Real-time index-to-palm touch with 
single head-mounted RGB illustrates monocular pressure-sensitive interaction on the palm \cite{he2025palmpad}; uncertainty-aware 
egocentric pipelines support rapid touch and text input in MR by explicitly modeling confidence \cite{streli2024touchinsight}; egocentric 
sequences enable thumb-tip force estimation during everyday manipulation \cite{jeong2025thumb}; and hand-bound pads demonstrate
 pressure-like input using hand-tracking signals alone, without instrumented surfaces \cite{dupre2025investigating}. 

While egocentric cameras offer rich cues of hands, they inherently restrict observability to the HMD's limited FOV. This creates a usability dilemma: ensuring reliable tracking requires users to maintain hands within the specific volume, forcing elevated postures that induce "gorilla-arm" fatigue~\cite{Keyboard-Text-Input,Consumed—endurance,Modeling-Cumulative}. Conversely, resting hands naturally outside this frustum causes immediate out-of-view (OOV) failures, breaking both pose and pressure estimation~\cite{HOOV,reimer2023evaluation, gemici2025before}. 
Consequently, interaction is fundamentally bound by the user's head orientation.

Commercial HMDs partly mitigate these constraints by deploying multiple sensors to enlarge the tracking volume. However, even distinct multi-camera setups struggle to observe hands in natural resting positions (e.g., alongside the body) due to torso occlusion. This limitation is exacerbated by the parallel trend toward lightweight, glasses-style devices \cite{Meta2020ProjectAria,Meta2024Orion,Rokid2024ARLite}, where strict hardware budgets often restrict sensing to a single, narrow-FOV camera. In both regimes, hands frequently exit the visible frustum; specifically addressing this "hands-down" blind spot is critical for ubiquitous interaction.

\subsection{Wrist-Proximal Sensing}

Wrist-Proximal On-body sensing moves perception with the user, improving continuity through natural postures and reducing reliance on external infrastructure. Non-optical wearables (EMG \cite{meg2pose,neuromotor_emg,PiMForce}, 
acoustics \cite{echoWrist,yu2024ring,Interferi}, capacitive \cite{CapBand}, IR laser/tomography\cite{Digits, SensIR}, electrical
impedance \cite{Electrical_Impedance,Electrical_Impedance_1}, IMU~\cite{shen_mousering_2024}) bring sensors near the hand.
These modalities offer promising continuous tracking (e.g., EMG) or precise gesture classification (e.g., acoustic), but
often trade off against noisy or ambiguous signals, environmental sensitivity, customized hardware, and coverage biased
toward discrete poses rather than full 3D articulation.

Wrist-mounted optical systems leverage commodity cameras to get fine details of the fingers. RotoWrist estimates wrist orientation with surrounding IR modules~\cite{RotoWrist}; WristCam observes the back-of-hand or environment for wrist trajectories~\cite{WristCam}; FineType augments an under-wrist IR camera with IMU for tap detection~\cite{FineType}; multi-view wrist-mounted thermal imaging (FingerTrak) reconstructs full hand pose from hand silhouettes captured by four miniature thermal cameras on a wristband~\cite{FingerTrak}. However, because distal fingertips and contacts are not directly observed in these setups, full-hand joint angles and the subtle appearance cues necessary for pressure inference (skin deformation, contact shadows, millimeter-scale fingertip displacements) remain under-captured.

\wristp{} mounts a fisheye RGB camera (180° FOV) on the palmar side of the wrist, providing a stable, body-anchored
vantage that simultaneously exposes distal articulation and contact appearance. This enables continuous pose and pressure
estimation in comfortable, hands-down postures, improves distal joint precision, and supports bare-hand
interaction on uninstrumented surfaces without additional wearables. Notably, the wrist perspective directly observes
fingertip skin deformation and contact shading—signals that are difficult to sustain from head-mounted viewpoints—thereby
benefiting appearance-based pressure inference.

Taken together, existing placements present a trade-off: environment-mounted systems offer accuracy but limited mobility; head-mounted systems support mobility but suffer from FOV restrictions and "gorilla-arm" fatigue; and prior wrist-mounted systems lack the vantage point to capture contact mechanics. \wristp{} addresses these limitations by mounting a fisheye RGB camera on the palmar side of the wrist. This stable, body-anchored vantage directly exposes distal articulation and fingertip skin deformation. This configuration enables continuous pose and pressure estimation in comfortable, hands-down postures, effectively unbinding interaction from the head's field of view and the environment's infrastructure.

\section{Hardware Implementation}
\label{sec:System_Design}
%整个系统的硬件组成
% --- Figure 1: hardware ---
\begin{figure*}[htbp]
    \centering
    \includegraphics[width=1\textwidth]{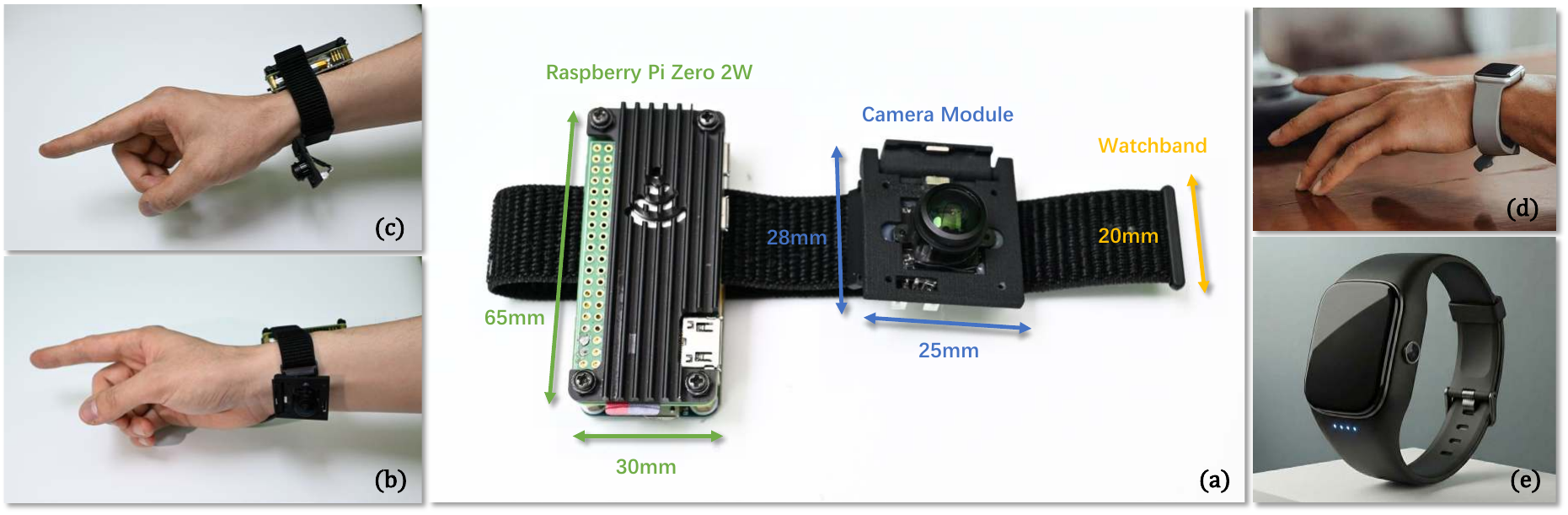}
    \caption{
      \textbf{Hardware design of \wristp{}}.
    (a) System modules: right—180° FOV ultra-wide RGB camera module (28 $\times$ 25 mm), left—Raspberry Pi Zero 2 W with a battery-backed micro-UPS;
    (b) Stowed configuration worn on the wrist, with the overall thickness reduced to $\approx$ 10 mm for comfortable long-term wear;
    (c) In-use configuration with the camera deployed via a 90° magnetic fold-out hinge to provide a wrist-centric view from below the hand (height $\approx$ 28 mm);
    (d, e) Future integrated concept toward a smartwatch-style form factor.
}
\Description{
Overview of the WRISTP hardware design. The system includes an ultra-wide
fisheye RGB camera and a Raspberry Pi Zero 2 W module. The camera deploys
on a fold-out hinge for use, providing a wrist-mounted view of the hand,
and folds back into a slim stowed configuration for comfortable wear.
A conceptual smartwatch-style integration is also illustrated.
}
    \label{fig:hardware_prototype}
\end{figure*}

Our prototype system is primarily composed of a 180° FOV RGB camera module, a Raspberry Pi Zero 2 W, a nylon strap, and a custom 3D-printed enclosure, with a total hardware cost of approximately USD 50. While the electronic components and the strap are commercially available off-the-shelf, the enclosure was custom-designed to ensure wearing comfort and ergonomics.

As shown in Fig.~\ref{fig:hardware_prototype}, we designed a 3D-printed camera housing with a magnetized rotational hinge, enabling the camera to be quickly switched between a 90° working position and a 0° stowed position, minimizing the wearing burden when the device is not in use. The camera module is sized at 28~$\times$~25~mm. The Raspberry Pi Zero 2 W and a 2000~mAh Li-Po battery are integrated into a compact 60~$\times$~30~$\times$~15~mm unit, enabling wireless image transmission and stable power supply within a small form factor.

For the data collection and user-study procedures described in Sections~\ref{sec:dataset} and~\ref{sec:user-study}, a wired connection was employed to ensure real-time performance and transmission stability. In the real-world scenario evaluation presented in Section~\ref{sec:application}, we further validated a wireless version: images of 512~$\times$~512 resolution at 30~fps were transmitted over a 2.4~GHz WLAN. The system operated continuously for approximately three hours under a maximum power consumption of about 1.5~W, demonstrating its practicality in everyday interaction scenarios.

Looking ahead, as illustrated in Fig.~\ref{fig:hardware_prototype}(d, e), we plan to integrate the camera and wireless communication module into a compact smartwatch form factor. By employing a high-integration SoC, a custom flexible PCB, and an optimized mechanical design, we envision two distinct hardware configurations.

The first design places the camera on the watch bezel or side. In daily use, the watch face remains dorsal (face-up); for interaction with VR headsets or other devices, users simply rotate the watch face toward the palmar side to activate \wristp{}. This rotatable paradigm shares similarities with commercial devices like Omate TrueSmart and imoo \cite{Omate,IMOO}, which integrate side-mounted cameras.

The second design embeds the camera directly at the bottom of the watch chassis, enabling immediate interaction without mechanical rotation, a form factor successfully adopted by TapXR \cite{TapXR1}. Both designs demonstrate the feasibility and strong commercial potential of wrist-worn optical sensing.

\section{\wristp{} Dataset}
\label{sec:dataset}
\subsection{Data Capture Setup}
\begin{figure*}[htbp]
    \centering
    \includegraphics[width=\textwidth]{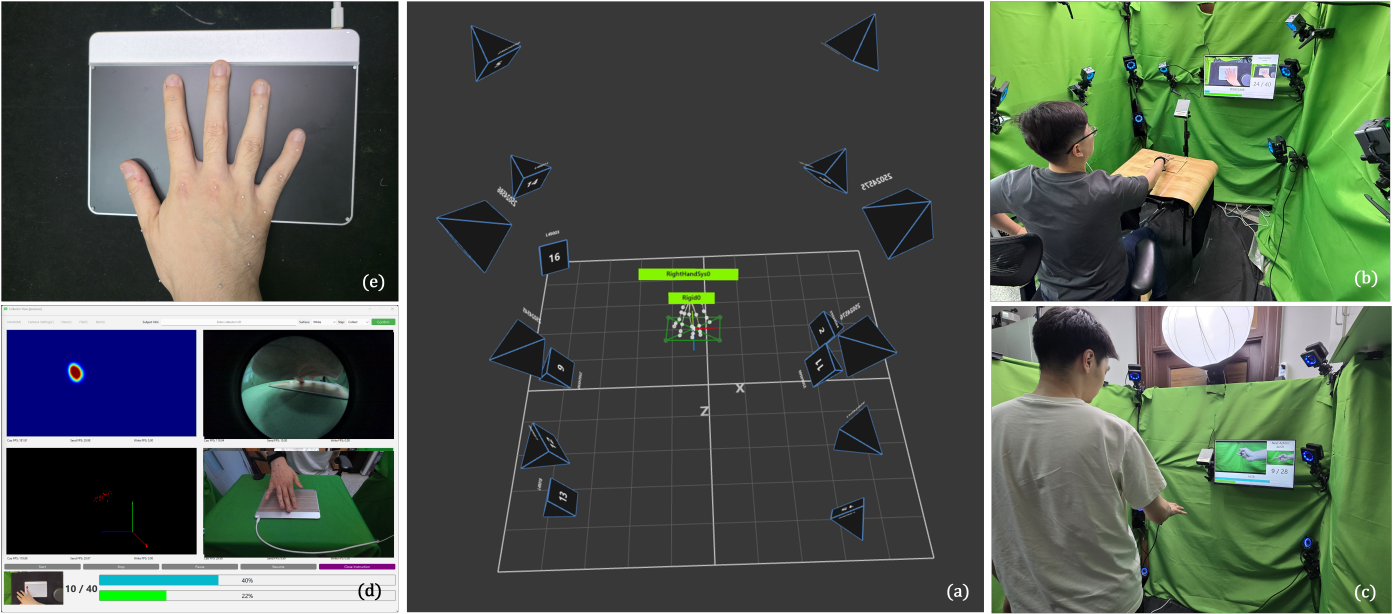}
    \caption{\textbf{Collection Scene.} (a) Visualization of the collection environment; (b) Planar interaction data collection; (c) Mid-air hand pose data collection; (d) The GUI of data collection software; (e) The 21 markers attached at the positions of the hand joint points and the 4 markers attached at the four corners of the Sensel Morph.}
    \Description{Overview of the data collection setup, showing the motion capture system, planar interaction setup, mid-air interaction setup, the GUI, and the marker layout on the hand and the pressure pad.}
    \label{fig:Collection_Scene}
\end{figure*}
As shown in Fig.~\ref{fig:Collection_Scene}, to obtain high-fidelity annotations of hand pose and pressure, we designed a data collection system. The system integrates a sub-millimeter precision FZMotion MoCap system, a high-resolution pressure-sensitive touchpad (Sensel Morph) \cite{SenselMorph}, a wrist-mounted camera, an Azure Kinect camera and a display terminal to guide user interactions. The Sensel Morph is placed horizontally on a tabletop to capture pressure distributions during hand-plane interactions. A wrist-mounted camera captures wrist-view images of hand, while a Kinect RGB camera mounted on a tripod captures third-person color images of global hand movements. 

For a subset of participants, we additionally deployed a head-mounted RGB camera (as in Section~\ref{sec:STUDY 4}) to capture synchronized egocentric views. While this dataset enables future multi-view fusion studies, here we use it solely to benchmark our wrist-based method against standard vision-based baselines. To ensure a fair comparison, the head-mounted camera featured a 150°~FOV, selected to maximize the field of view without introducing significant distortion. Details are provided in Appendix~\ref{sec:appendix_eval}.

We developed a data collection software with a user-friendly GUI that provides gesture videos and textual instructions, displays task timing and progress, and supports real-time monitoring, allowing the experimenter to promptly detect and address anomalies.
\subsection{Data Capture Environment}
To improve robustness and generalization, we randomized key environmental factors known to affect vision-based perception. 

\paragraph{Lighting Diversity.}
We used fixed illumination fixtures that provided controllable lighting conditions.
For each session, one of three illuminance levels---high, medium, or low---was randomly selected, 
where each level corresponded to a predefined range of illuminance 
(e.g., high: $I \in [500, 2000]$\,lux; 
      medium: $I \in [100, 500]$\,lux; 
      low: $I \in [1, 100]$\,lux).
This design ensured controlled yet stochastic variability across lighting conditions.

\paragraph{Background Control.}
Green screens were installed on three vertical sides and the base of the capture frame to stabilize the background and facilitate post-processing (e.g., hand segmentation and random background replacement).

\paragraph{Surface Texture Diversity.}
To model common desktop appearances, we prepared six textures: pure white, pure black, light wood grain, dark wood grain, light marble, and dark marble. For each participant, every recording session was conducted with a different texture, ensuring that all textures were covered across sessions. The assignment of textures to sessions was randomized and counterbalanced across participants to reduce ordering effects.

\begin{figure}[t]
  \centering

  \begin{subfigure}[b]{0.68\columnwidth}
    \centering
    \includegraphics[width=\linewidth]{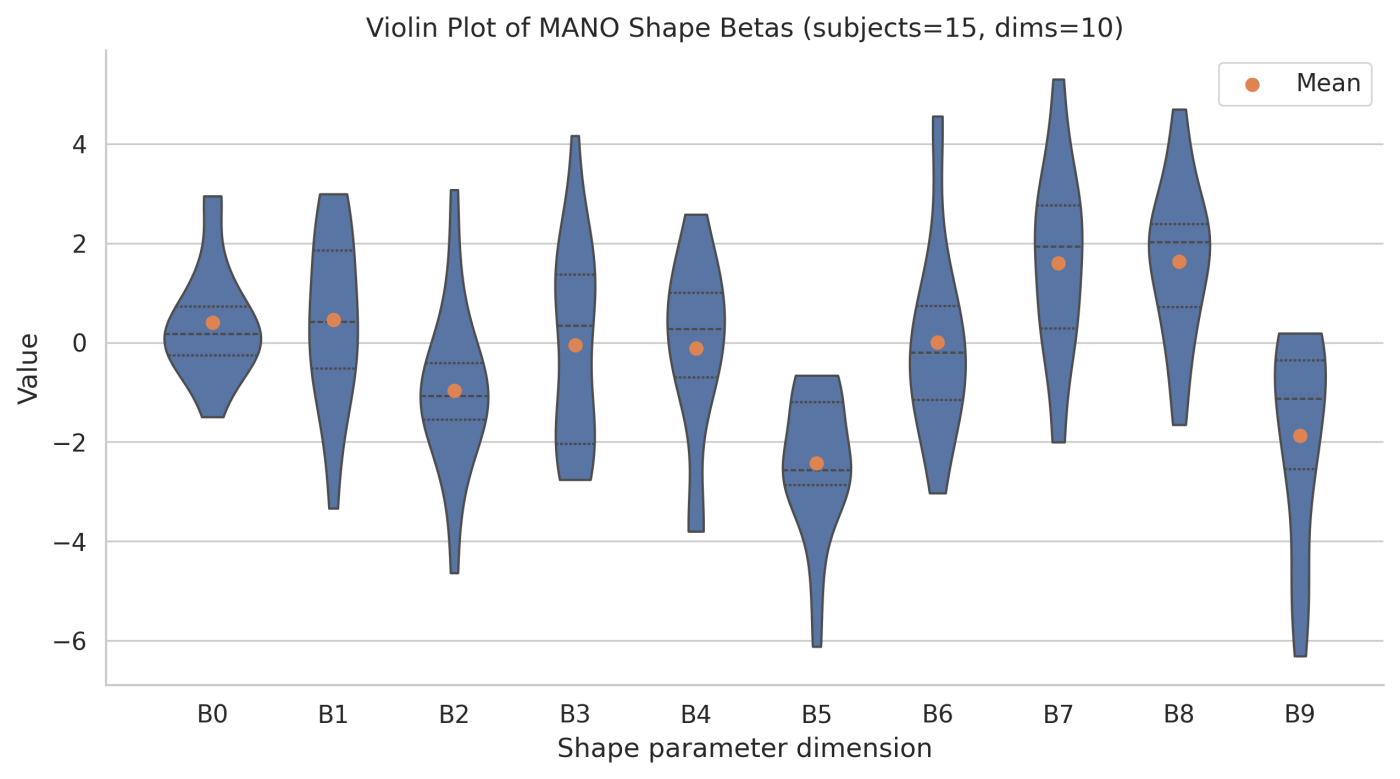}
    \subcaption{}\label{fig:Participants_distribution}
  \end{subfigure}\hfill
  \begin{subfigure}[b]{0.30\columnwidth}
    \centering
    \includegraphics[width=\linewidth]{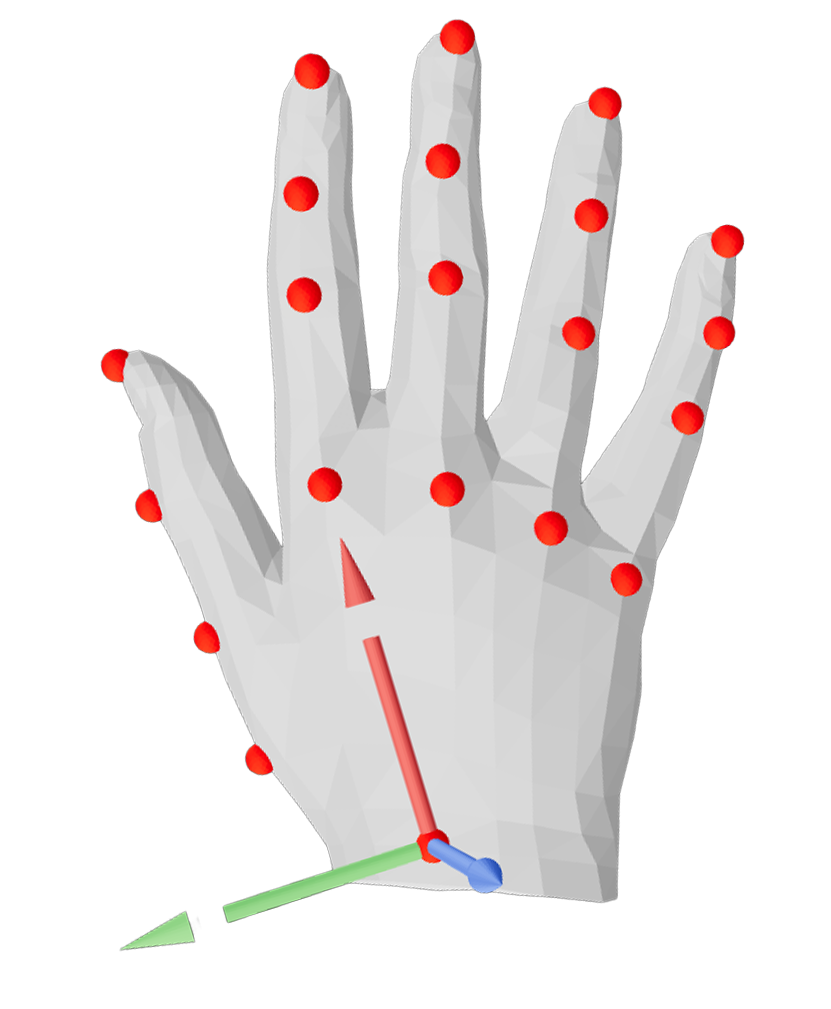}
    \subcaption{}\label{fig:local_frame}
  \end{subfigure}

  \caption{
    (a) Distribution of shape parameters $\beta$ across participants.
    (b) Canonical hand-local coordinate system defined from anatomical markers.
  }
  \Description{(a) A violin plot showing the distribution of MANO hand-shape parameters $\beta$ across participants. (b) A schematic of the hand-local coordinate system and marker definitions based on anatomical markers.}
  \label{fig:data_capture_overview}
\end{figure}

\subsection{Participants}
We recruited 20 participants from our institution (15 male, 5 female; ages 19--28 years, $M = 24$).
Among them, a subset of 5 participants additionally captured egocentric data using the head-mounted camera.
The lengths of the middle finger ranged from 6.5 to 9.3~cm ($M = 7.1$, $SD = 0.48$), reflecting a diverse set of hand shapes within this demographic (see Fig.~\ref{fig:Participants_distribution} for the distribution of MANO~$\beta$ values).
We further categorized participants' skin tones using the Fitzpatrick skin type scale (I--VI), which provides a standardized measure of skin pigmentation.
The participants spanned Types~II--IV, corresponding to light to medium skin tones typical among East Asian populations.
Our current dataset does not cover the full spectrum of skin tones and demographics; extending to broader populations is an important direction for future work.
Each participant who successfully completed data collection received a compensation of~\$15.

\subsubsection{Preparatory Work}
\label{sec:Preparatory Work}
During preparation, participants were briefed on the tasks, provided informed consent, and viewed a demonstration video. We then placed 21 infrared-reflective markers on the dorsal joints of the right hand and affixed four markers to the corners of the Sensel Morph touchpad to delineate the pressure plane. Participants sat on adjustable stools and wore a wrist-mounted camera. Before data collection, they performed a calibration gesture by spreading their fingers and rotating their hand 3--4 times before the Kinect camera.
\subsubsection{Planar Interaction Data Collection}
The planar interaction data were collected in three sessions, with a different surface texture randomly selected from aforementioned six overlays for each session. These overlays represent common desktop surface finishes. Participants followed the sequential prompts on the screen to perform 48 actions, each lasting approximately 15 seconds. The actions covered common touchpad interactions—single- and multi-finger sliding, pressing, and pinching—as well as more complex behaviors such as symbol drawing, fingertip rotation, and rolling presses using different parts of the hand. Each session concluded with (i) a non-contact hover trial (fingers near but not touching the surface) to obtain data for non-contact conditions and (ii) a brief free-interaction segment without prompts. To simulate real-world use and introduce natural variation in fit and camera pose, participants removed and re-donned the wristband between consecutive sessions.
A complete list and detailed definitions of these actions are provided in the Appendix~\ref{sec:extended_dataset}.

\subsubsection{Mid-air Hand Pose Data Collection}
Mid-air data were collected in two sessions. Across the sessions, participants performed a set of 28 gestures, each for approximately 15 seconds, including 10 American Sign Language (ASL) letters and common daily interaction gestures such as fist, OK sign, finger heart, grasping, pinching, pointing, rock gesture, thumbs-up, and claw pose. During these actions, participants were instructed to actively vary their wrist pitch and yaw angles to increase pose diversity. Each session ended with a short free-form movement segment to further enrich the dataset. As with the planar protocol, participants removed and re-donned the wristband between the two mid-air sessions to emulate realistic donning/doffing variability. Detailed descriptions of all gestures are provided in Appendix~\ref{sec:extended_dataset}.

\subsubsection{Data Synchronization}
Data collection from all sensors is triggered by a server, ensuring unified timebase. After correcting hardware delays, we aligned streams from high- to low-frame-rate devices, keeping the residual synchronization error within 5\,ms. Additionally, participants are instructed to move smoothly to minimize synchronization errors.

An experimenter continuously monitored the recording interface and provided guidance. The total duration for each participant was approximately one hour, with breaks available on demand.

\subsection{Annotation Method}
\begin{figure*}[htbp]
    \centering
    \includegraphics[width=1\textwidth]{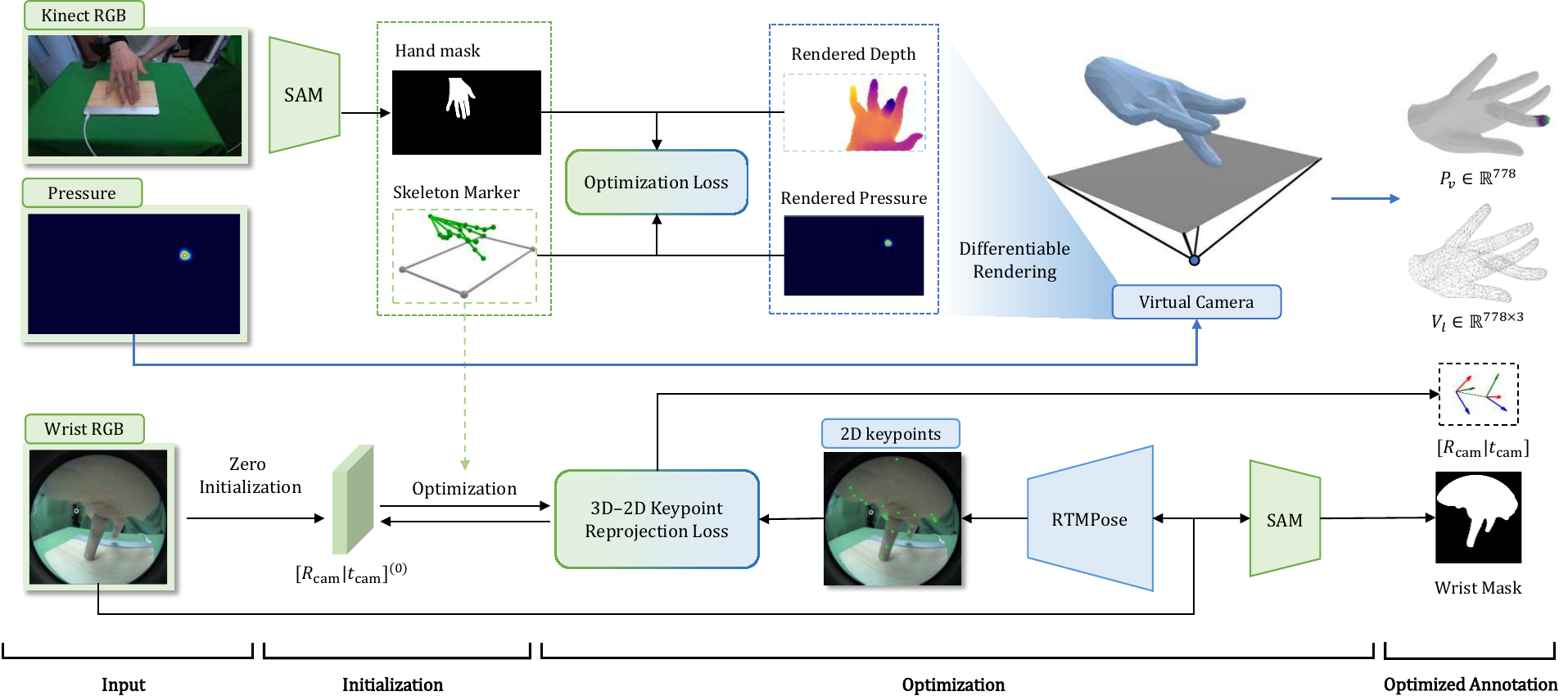}
    \caption{\textbf{Annotation pipeline overview.} 
The pipeline consists of two parallel stages to generate ground-truth data. 
\textbf{Top (Hand \& Pressure Optimization):} We fuse multi-modal inputs—including third-person Kinect RGB images, motion-capture markers, and tactile pressure maps—to jointly optimize the hand mesh geometry $V_l$ and the per-vertex pressure field $P_v$. A differentiable rendering module minimizes the discrepancy between simulated and observed pressure/depth maps.
\textbf{Bottom (Wrist Camera Calibration):} Using the reconstructed 3D hand as a reference, we estimate the wrist camera extrinsics $[R_{cam}|t_{cam}]$ relative to the hand-local frame. This is achieved by minimizing the 3D-2D reprojection loss  between the projected 3D joints and 2D keypoints detected by RTMPose  from the raw wrist-camera inputs.}
\Description{An overview diagram illustrating the annotation pipeline.
The figure shows the synchronized data sources used for annotation,
including the wrist-mounted fisheye camera, the motion-capture markers,
the depth camera, and the pressure-sensing pad. It visualizes how these
modalities are combined to reconstruct high-fidelity 3D hand meshes and
per-vertex pressure annotations through multi-stage optimization.}
    \label{fig:Annotation_Method}
\end{figure*}

To obtain high-fidelity hand meshes and per-vertex pressure from raw measurements, we develop a marker-based, multi-objective optimization pipeline built on the MANO model~\cite{romero2022embodied} (overview in Fig.~\ref{fig:Annotation_Method}). 
Given (i) 21 infrared-reflective markers stuck on the hand in the world frame, (ii) synchronized third-person RGB images, and (iii) calibrated pressure frames from a Sensel Morph touchpad, the pipeline fuses these observations to jointly estimate pose, shape, extrinsics, and a per-vertex pressure field.
All quantities are formulated in a canonical hand-local coordinate system that we define algorithmically from anatomical landmarks (Sec.~\ref{subsec:hand-local-frame}); accordingly, we estimate the rigid transformation between the world frame and the hand-local frame.
\subsubsection{Hand-Local Canonical Coordinate System}
\label{subsec:hand-local-frame}
To provide a consistent reference frame across participants and sessions, we introduce a hand-local canonical coordinate system, which serves as the basis for both annotation optimization and network prediction. This local frame is defined algorithmically from anatomical markers on the hand mesh. Performing optimization and prediction in this canonical hand-local frame removes global translation and rotation degrees of freedom, improving numerical stability and convergence. An illustration of this coordinate system is provided in Fig.~\ref{fig:local_frame}, while its detailed definition is described in the supplementary material.

\paragraph{Initialization.}
% The calibrated MANO shape $\beta^{*}$ is fixed throughout optimization.
We follow the approach of~\cite{karunratanakul2023harp} to obtain the shape parameter $\beta$, which is fixed during subsequent optimization.
The initial MANO pose $\theta_{\mathrm{init}}$ is predicted by a graph neural network pretrained on our self-collected large-scale dataset ~\ref{sec:handvqvae}.
This network takes as input the 3D coordinates of 21 fixed mesh vertices in the canonical hand-local frame,
on the hand mesh,
and outputs the MANO pose parameters.
The same dataset is also used to train our \handvqvae{} to provide a strong prior knowledge of hand geometry;
details of the dataset construction and statistics are provided in supplementary.

For extrinsics, the initial rigid transformation is obtained from markers (denoted by $[R^{(0)}_{L}{}^{W}\mid T^{(0)}_{L}{}^{W}]$ or its inverse), but due to calibration noise it is not treated as ground truth. 
Therefore, in optimization we refine extrinsics \emph{per frame} via incremental updates:
\[
R_{L}^{W}=R^{(0)}_{L}{}^{W}\,\Delta R,\qquad
T_{L}^{W}=T^{(0)}_{L}{}^{W}+\Delta T,
\]
where $(\Delta R,\Delta T)$ are the learnable deltas.

\subsubsection{Annotation Optimization}
The hand mesh in the hand-local coordinate system is written as $\mathcal{V}_l=\mathrm{MANO}(\theta,\beta^{*})$, and the world-space vertices are
\[
\mathcal{V}_w \;=\; R_{L}^{W}\,\mathcal{V}_l + T_{L}^{W}
\;=\; \big(R^{(0)}_{L}{}^{W}\Delta R\big)\,\mathcal{V}_l \;+\; \big(T^{(0)}_{L}{}^{W}+\Delta T\big).
\]
We optimize
\[
\Theta=\{\theta,\;\Delta R,\;\Delta T,\;\mathcal{P}_v\}
\quad(\text{$\beta$ fixed to $\beta^{*}$})
\]
by minimizing
\begin{equation}
\mathcal{L}_{\text{opti}}(\Theta)=
\mathcal{L}_{\text{markers}}(\Theta)+
\mathcal{L}_{\text{mask}}(\Theta)+
\mathcal{L}_{\text{render}}(\Theta)+
\mathcal{L}_{\text{anat}}(\Theta).
\end{equation}

\paragraph{Orthographic camera and differentiable rendering.}
We use a virtual orthographic camera placed beneath the center of the Sensel Morph plane and looking upward along the plane normal. The camera resolution matches the sensor grid ($H\times W$).
Two render passes are computed with a differentiable renderer: a \emph{pressure} pass and a \emph{depth} pass.
For the pressure pass, the per-vertex pressure $\mathcal{P}_v$ is assigned as the surface texture of $\mathcal{V}_w$, yielding a rendered pressure map $\tilde{\mathcal{R}}_p$.
After a vertical flip to align the sensor convention, we penalize the pixel-wise MSE to the ground-truth pressure map $P_{\mathrm{gt}}$.
For the depth pass, we render a depth map $\tilde{\mathcal{R}}_d$ and convert it to the \emph{relative-to-plane depth} by subtracting a calibrated offset $\delta=0.1$:
$d_{\mathrm{rel}}=\tilde{\mathcal{R}}_d-\delta$.
We then produce a soft contact probability
$\hat C=\sigma\!\big((\varepsilon-d_{\mathrm{rel}})/\tau\big)$
and apply a BCE-with-logits loss against the binary contact mask
$M_{\mathrm{contact}}=\mathbf{1}_{\{P_{\mathrm{gt}}>\gamma\}}$
within the projected hand region $M_{\mathrm{hand}}$.

\paragraph{Rendering losses.}
We therefore write
\begin{equation}
\begin{aligned}
\mathcal{L}_{\text{render}}(\Theta)
&=
\omega_{\text{press}}\,
\mathcal{L}_{\text{press}}\!\big(\mathrm{Flip}(\tilde{\mathcal{R}}_p),\,P_{\mathrm{gt}}\big)
+\omega_{\text{hand}}\,\mathcal{L}_{\text{hand}}.
\end{aligned}
\end{equation}

\noindent with:
\begin{subequations}\label{eq:render-losses}
\begin{align}
\mathcal{L}_{\mathrm{press}}
&= \frac{1}{HW}\,\|\tilde{\mathcal{R}}_p - P_{\mathrm{gt}}\|_F^2, \label{eq:render-losses-a}\\
\mathcal{L}_{\mathrm{hand}}
&= \frac{1}{\|M_{\mathrm{hand}}\|_{1}}
\Big\|\, M_{\mathrm{hand}} \odot
\ell\!\Big(\frac{\varepsilon - (\tilde{\mathcal{R}}_d - \delta)}{\tau},\, M_{\mathrm{contact}}\Big)
\,\Big\|_{1}. \label{eq:render-losses-b}
\end{align}
\end{subequations}

Here, $\mathrm{Flip}(\cdot)$ denotes a vertical flip to match the sensor's raster order,
$\sigma(\cdot)$ is the logistic function and $\ell(\cdot,\cdot)$ denotes the element-wise BCE-with-logits.
$M_{\mathrm{hand}}$ is the projected hand mask, and
$M_{\mathrm{contact}}=\mathbf{1}_{\{P_{\mathrm{gt}}>\gamma\}}$ is the binary contact mask.

\paragraph{Anatomical loss.}
The anatomical loss $\mathcal{L}_{\text{anat}}$ regularizes the pose parameters to ensure 
biologically plausible hand configurations ~\cite{yang2021cpf}.
This discourages physically implausible finger bending, twisting, spreading, 
and improves the stability of the optimization.

\paragraph{Intuition.}
This design serves two purposes:
(i) it constrains the predicted \emph{2D} pressure projection to match the ground-truth sensor map, and
(ii) it encourages the \emph{contact foreground} to lie close to the plane while pushing the \emph{background} away from it,
while the per-frame extrinsic deltas $(\Delta R,\Delta T)$ compensate for calibration bias in the marker-derived initialization.

\subsubsection{Wrist-mounted Camera Extrinsic Annotation}
To annotate the extrinsics of the wrist-mounted camera, we adopt a 2D--3D joint alignment approach. 
Specifically, we randomly sampled 3000 images from the wrist-mounted camera stream and manually annotated 
the 2D positions of 21 hand joints in each image. A 2D joint detection model was then trained within the 
MMPose~\cite{mm-Pose} framework by finetuning the RTMPose model~\cite{https://doi.org/10.48550/arxiv.2303.07399}, achieving an MPJPE of 6.2 pixels on 
the held-out test set at $512 \times 512$ resolution. This model was used to produce 2D joint annotations for the 
entire dataset.

For extrinsic calibration, we employ the ocamcalib polynomial fisheye camera model~\cite{scaramuzza2006toolbox} 
to project the ground-truth 3D joints onto the image plane using the pre-calibrated intrinsics. 
The camera extrinsics $[R_{\text{cam}} \mid t_{\text{cam}}]$ are then estimated by solving the following 
least-squares optimization problem:
\begin{equation}
\min_{R_{\text{cam}}\in \mathrm{SO}(3),\,t_{\text{cam}}\in \mathbb{R}^3}
\;\sum_{i=1}^{N}\Big\|
\pi_{\text{ocam}}\!\big(R_{\text{cam}}X_i+t_{\text{cam}}\big)-u_i
\Big\|_2^{2},
\end{equation}
where $X_i$ are the 3D joints in the hand-local frame, $u_i$ are the detected 2D joint coordinates, 
and $\pi_{\text{ocam}}(\cdot)$ denotes the ocamcalib projection function. 
It is important to note that these extrinsics are defined with respect to the hand-local coordinate 
system introduced in Sec.~\ref{subsec:hand-local-frame}, ensuring consistency with our annotation pipeline.

\section{\wristp{} Algorithm}
\label{sec:method}

\subsection{Model Architecture}
\begin{figure*}[htbp]
    \centering
    \includegraphics[width=\textwidth]{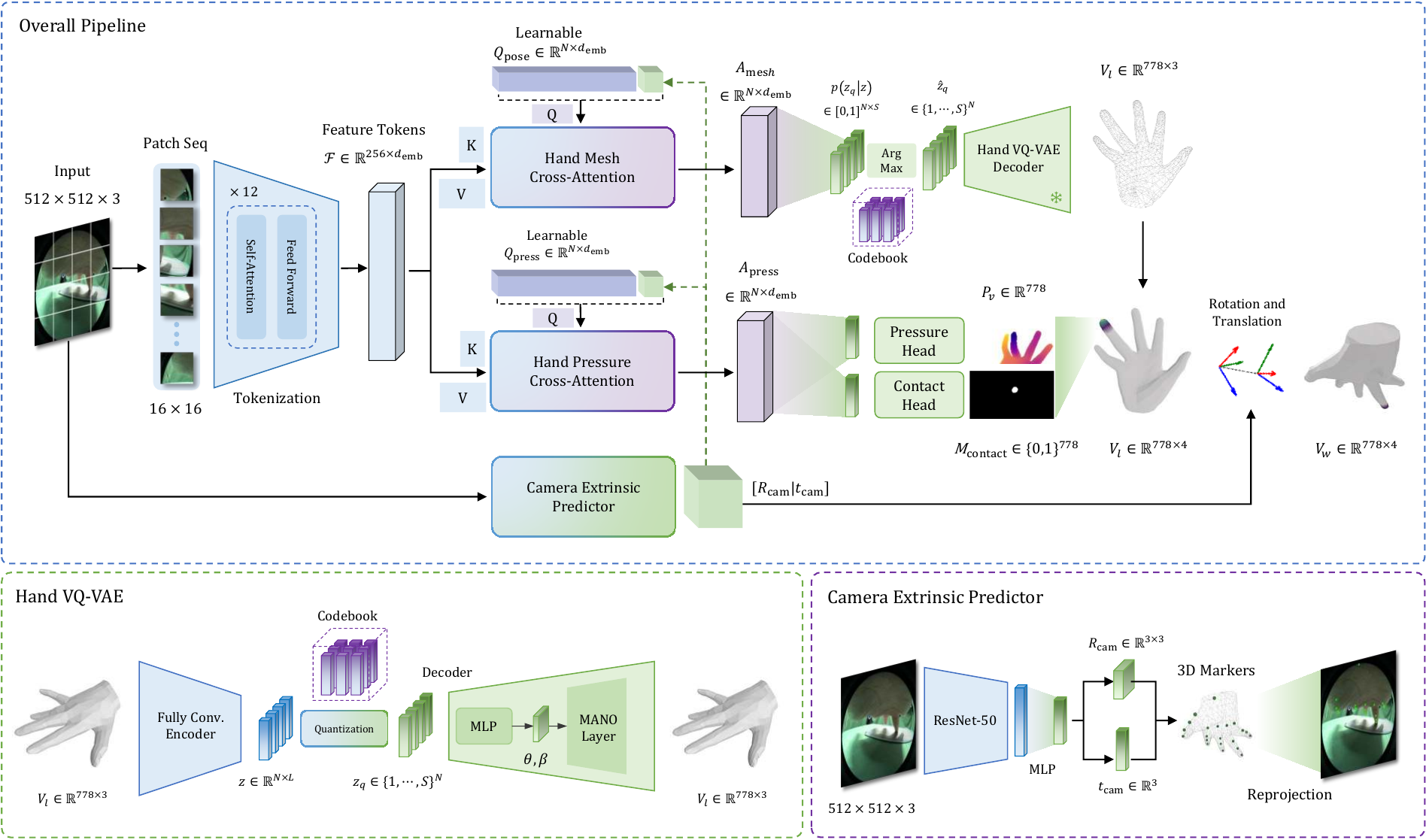}
    \caption{\textbf{\wristp{} pipeline.} A ViT-based backbone produces image features that are queried by two sets of learnable tokens (pose- and pressure-tokens, both of length 21). The wrist-camera extrinsics are embedded and \emph{concatenated channel-wise to both token sets before cross-attention}, yielding extrinsics-aware token features. The pooled token features are fed into task-specific heads: a codebook indices classification head (decoded by \handvqvae{} decoder), and a \emph{pressure branch with two heads} for contact classification and pressure regression.}
\Description{A diagram of the WristP2 model pipeline. 
The figure shows a wrist-view image processed by a ViT backbone, followed 
by two groups of learnable tokens for pose and pressure. Camera extrinsics 
are embedded and concatenated with the tokens before cross-attention. 
The resulting token features are passed to a mesh codebook classification 
head for pose reconstruction and to two pressure heads for contact prediction 
and pressure regression.}
    \label{fig:Model_pipeline}
\end{figure*}

\noindent\textbf{Overview.}
\wristp{} jointly predicts (i) the canonical hand mesh, (ii) per-vertex contact and pressure, and (iii) the wrist-mounted camera extrinsics.
A shared ViT-based backbone produces patch embeddings. 
From these features, two \emph{joint-aligned} token sets (21 tokens each, aligned to anatomical joints) query the backbone features tokens via cross-attention for pose and pressure regression.
In parallel, a lightweight regression branch predicts the wrist-camera extrinsics in the hand-local frame from raw RGB image; the predicted extrinsics are embedded and concatenated to both token sets before cross-attention, ensuring that token interactions are conditioned on viewpoint geometry.

For the pose pathway, each cross-attended \emph{pose token} independently solves a classification problem over the \handvqvae{}.
The resulting discrete indices are used to retrieve the corresponding codebook embeddings, which are fed into the \handvqvae{} decoder to obtain MANO parameters $(\theta,\beta)$; the MANO layer then maps $(\theta,\beta)$ to mesh vertices, reconstructing the canonical hand mesh.
In contrast, the 21 pressure tokens are globally pooled into a single embedding that is mapped to two $778$-dimensional outputs: a contact classification head (per-vertex contact logits) and a pressure regression head (per-vertex pressure magnitudes).

\subsubsection{Canonical Hand Mesh Tokenization}
\label{subsec:canonical-tokenization}
Vector-Quantized VAE (VQ-VAE)\cite{VQ_VAE} is a discrete representation learning technique originally developed for image generation.
Recent studies\cite{fiche2024vq, fiche2025mega, dwivedi2024tokenhmr} show its strong performance in pose estimation and mesh reconstruction, especially under limited data size.
Following them, we adopt a variant of Mesh-VQ-VAE\cite{zhou2020fully} whose encoder is a Fully convolutional mesh autoencoder, while the decoder is replaced by a MLP followed by a MANO layer. We denotes it \handvqvae{}.
This design tokenizes the canonical hand mesh into latent representation space, thus making the mesh reconstruction problem into a simple classification problem. In \handvqvae, the codebook comprises 512 embedding vectors in $\mathbb{R}^9$. For each input frame, the encoder produces 21 latent vectors, which are quantized via nearest-neighbor lookup in the 512-way codebook, yielding 21 code indices.

\subsubsection{Pressure Branch}
\label{subsec:pressure}
The 21 extrinsics-aware \emph{pressure tokens} are globally pooled into a fixed-length vector $\bar t_{\text{press}}$:
\[
\bar t_{\text{press}}=\mathrm{Pool}(T_{\text{press}})\in\mathbb{R}^{d_{\text{emb}}}.
\]
From this embedding, two parallel heads produce per-vertex predictions:
(i) a \textbf{contact classification head}, outputting logits $m\in\mathbb{R}^{778}$ (after sigmoid to obtain contact probabilities), and 
(ii) a \textbf{pressure regression head}, outputting $p\in\mathbb{R}^{778}$ as continuous pressure values. We simply use a pressure threshold to generate gt contact mask.

This decoupled design is particularly beneficial under the foreground-sparsity regime, where only a small subset of hand vertices are in contact while the majority are background. 
By letting the contact head explicitly address the highly imbalanced classification problem (with focal-style supervision), and conditioning pressure regression on the detected foreground, the model suppresses background leakage and stabilizes training, leading to more reliable pressure estimation in sparse-contact scenarios.

\subsubsection{Pose Branch}
\label{subsec:pose}
The 21 extrinsics-aware \emph{pose tokens} are aligned to anatomical joints. 
Each token independently solves a classification problem over the \handvqvae{} codebook:
\[
\mathbf{P}(z_q\!\mid z)\in[0,1]^{21\times S},
\]
where $S$ is the codebook size.
The predicted discrete indices are used to retrieve the corresponding codebook embeddings, which are then passed through the \handvqvae{} decoder to obtain MANO parameters $(\theta,\beta)$.
Finally, a differentiable MANO layer maps these parameters to the mesh vertices
\[
V_H = \mathrm{MANO}(\theta,\beta)\in\mathbb{R}^{778\times 3},
\]
reconstructing the canonical hand mesh.  
Thanks to the asymmetric encoder–decoder design of \handvqvae{}, the shape parameter $\beta$ can be fixed during inference. This reduces inter-frame drift and suppresses spurious variations, which is crucial for identity consistency in interactive use.

\subsubsection{Camera Extrinsics Branch}
\label{subsec:extrinsics}
A lightweight predictor regresses the \cam{} extrinsics relative to the hand-local frame $\{H\}$.
Instead of directly predicting absolute camera extrinsics, we let the network estimate a \emph{delta} with respect to the dataset-averaged extrinsics, which empirically stabilizes optimization. 
Concretely, a ResNet-50 backbone (initialized from ImageNet pretraining) extracts image features that are fed into a small MLP head.
The branch outputs a 6D rotation representation $\Delta \mathbf{r}_{\text{6D}}\in\mathbb{R}^{6}$ and a translation offset $\Delta t_{\text{cam}}\in\mathbb{R}^3$.
These deltas are composed with the dataset mean extrinsics $(\bar R_{\text{cam}},\bar t_{\text{cam}})$ to yield
\[
\hat R_{\text{cam}} = \exp([\Delta\omega]_\times)\,\bar R_{\text{cam}}, 
\qquad 
\hat t_{\text{cam}} = \bar t_{\text{cam}} + \Delta t_{\text{cam}},
\]
where $\Delta\omega$ is the axis–angle vector converted from $\Delta \mathbf{r}_{\text{6D}}$ by orthonormalization~\cite{zhou2019continuity}.
Finally, the embedding $e_{\text{cam}}$ is replicated and concatenated to both pose and pressure token sets before cross-attention:
\[
\tilde Q_{\text{pose}}=[Q_{\text{pose}};E_{\text{cam}}],\quad
\tilde Q_{\text{press}}=[Q_{\text{press}};E_{\text{cam}}],
\]
making both branches explicitly conditioned on viewpoint geometry.

\subsection{Model Training}
\subsubsection{\handvqvae}
\label{sec:handvqvae}
We pretrain \handvqvae{} on public datasets (Gigahand \cite{GigaHands_2025_CVPR}, HO3D \cite{Ho3D}, Interhand \cite{Interhand2_6m}, HanCo \cite{HanCo}, DexYCB \cite{DexYCB_2021_CVPR}, FreiHand \cite{FreiHAND_2019_ICCV}) after transforming meshes to the hand-local frame, and then finetune on our dataset, achieving 2.1\,mm per-vertex reconstruction error.

\subsubsection{Data Augmentation}
We apply random background replacement tailored to wrist-mounted fisheye imagery. 
First, we segment the hand with SAM2 to obtain a binary mask. 
Then, panoramas from an indoor dataset~\cite{xiao2012recognizing} are warped to the fisheye image plane using the pre-calibrated intrinsics $\mathbf{K}_{\text{fish}}$~\cite{scaramuzza2006toolbox} and a pose sampled from $\mathrm{SE}(3)$. 
During training, we replace the background with probability $0.5$. 
Representative visualizations of this background replacement are shown in Fig.~\ref{fig:bg-aug-viz}; algorithmic details are provided in the supplementary material. 
\begin{figure}[t]
    \centering
    \includegraphics[width=0.85\linewidth]{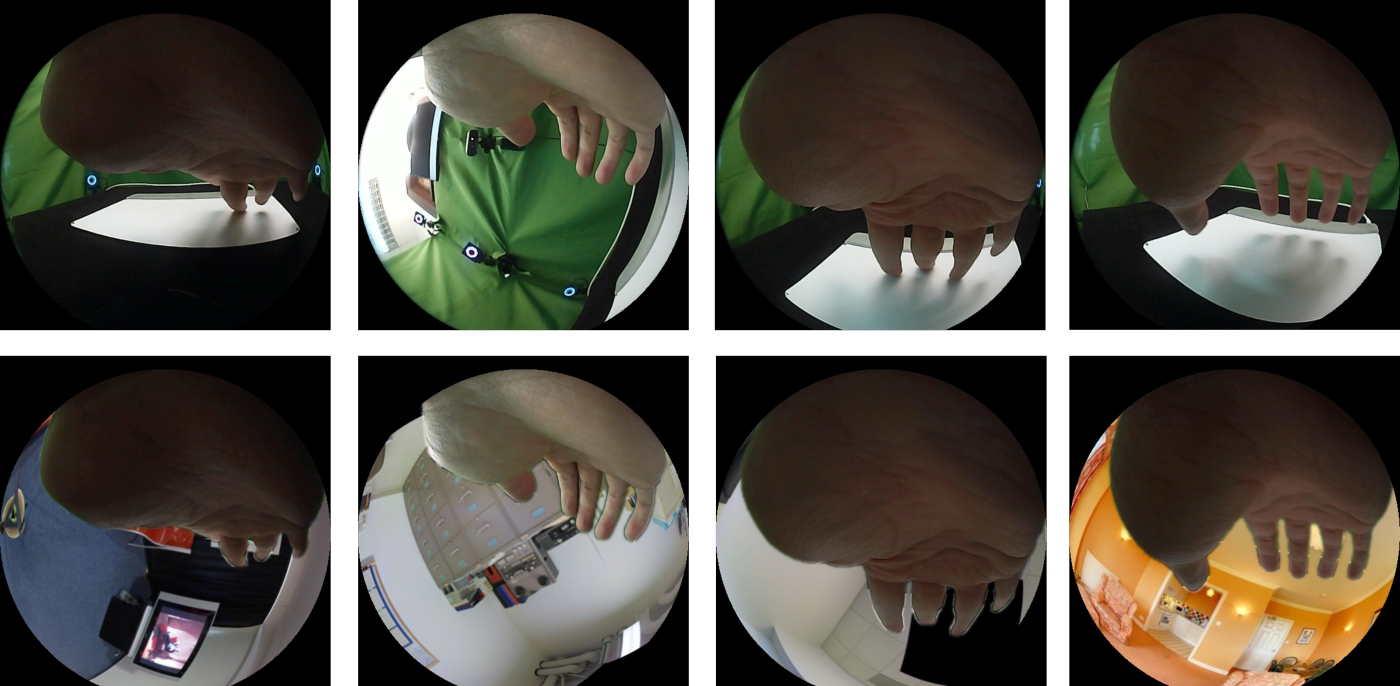}
    \caption{Example of the wrist-mounted camera image with random replaced background}
\Description{An example wrist-mounted fisheye camera image showing the hand 
from the palmar-side wrist viewpoint, with the original background replaced 
by a randomized scene as part of background augmentation.}
    \label{fig:bg-aug-viz}
\end{figure}

For both the replaced-background samples and pose-only samples, we manually set pressure labels to zero.
Standard photometric augmentations are also applied: illumination jitter, random gamma, and additive noise.

\subsubsection{Loss Function}
Let $\hat R_{\text{cam}},\hat t_{\text{cam}}$ be predicted extrinsics; $R_{\text{cam}}^*,t_{\text{cam}}^*$ ground truth.
Let $J\in\mathbb{R}^{K\times 3}$ denote 3D joints in $\{H\}$ and $\pi(\cdot)$ denote projection under fisheye intrinsics.

\paragraph{(1) Extrinsics loss with residualization.}
We predict a residual $\Delta\xi=(\Delta\omega,\Delta t)$ and compose it with a running estimate $(R_0,t_0)$:
\[
\hat R_{\text{cam}}=\exp([\Delta\omega]_\times)\,R_0,\qquad
\hat t_{\text{cam}}=t_0+\Delta t.
\]
The loss combines 2D keypoint reprojection and direct pose terms:
\[
\begin{aligned}
\mathcal{L}_{\text{ext}}
&=
\lambda_{\text{kp}}\frac{1}{K}\sum_{k=1}^K
\big\|\pi(\hat R_{\text{cam}}J_k+\hat t_{\text{cam}})-u_k^*\big\|_1
\\
&\quad+\lambda_R\, d_{\mathrm{geo}}(\hat R_{\text{cam}},R_{\text{cam}}^*)
+\lambda_t\,\|\hat t_{\text{cam}}-t_{\text{cam}}^*\|_1 .
\end{aligned}
\]

where $d_{\mathrm{geo}}(R_1,R_2)=\|\log(R_2^\top R_1)\|_2$ is the geodesic distance on $\mathrm{SO}(3)$.
Residualization (predicting $\Delta\xi$) empirically stabilizes training.

\paragraph{(2) VQ-VAE index loss.}
For $N{=}21$ tokens and codebook size $S{=}512$, we use token-wise cross-entropy:
\[
\mathcal{L}_{\text{vq}}=\frac{1}{N}\sum_{n=1}^{N}\mathrm{CE}\big(\hat{\mathbf{p}}_n,\; q_n^*\big),
\quad \hat{\mathbf{p}}_n\in\mathbb{R}^{S}.
\]

\paragraph{(3) Pressure and contact losses.}
Let contact logits be $m\in\mathbb{R}^{778}$, pressure prediction $p\in\mathbb{R}^{778}$,
and ground-truth pressure $\mathbf{p}^*\in\mathbb{R}^{778}$ (with zeros on replaced/pure-pose samples).
Define positive-contact mask
$\mathbf{c}^*=\mathbf{1}(\mathbf{p}^* \cdot h_{\max} > \tau)$ with scale $h_{\max}$ and threshold $\tau$.
We use focal loss for contact:
\[
\mathcal{L}_{\text{contact}}=
\mathrm{FocalBCE}\big(m,\mathbf{c}^*;\alpha,\gamma\big).
\]
For pressure regression we apply a soft gate $g=\sigma(m)$ and enforce non-negativity of predictions using \emph{softplus}:
\[
\tilde p=(g^{\gamma_{\text{gate}}})\cdot \mathrm{softplus}(p).
\]
Foreground/background are split by $\mathbf{c}^*$:
\[
\mathcal{L}_{\text{press}}
=\frac{1}{|\mathrm{FG}|}\!\!\sum_{v\in \mathrm{FG}}\!\!\big(\tilde p_v-\mathbf{p}^*_v\big)^2
+\lambda_{\text{bg}}\frac{1}{|\mathrm{BG}|}\!\!\sum_{v\in \mathrm{BG}}\!\!|\tilde p_v|.
\]
The overall pressure loss is
\[
\mathcal{L}_{\text{prs}}=\mathcal{L}_{\text{contact}}+\mathcal{L}_{\text{press}}.
\]

\paragraph{(4) Total loss.}
\[
\mathcal{L}=\lambda_{\text{ext}}\mathcal{L}_{\text{ext}}
+\lambda_{\text{vq}}\mathcal{L}_{\text{vq}}
+\lambda_{\text{prs}}\mathcal{L}_{\text{prs}}.
\]

\subsubsection{Training Setting}
The experiment was conducted on six RTX 4090 GPUs, with a total training time of approximately six hours.
We used the AdamW optimizer with a learning rate of $1\times 10^{-5}$ and a global batch size of 32.
The learning rate followed a OneCycle schedule.

\section{Offline Evaluation}
\label{sec:offline-eval}

\begin{figure}[t]
  \centering
  \includegraphics[width=0.65\linewidth]{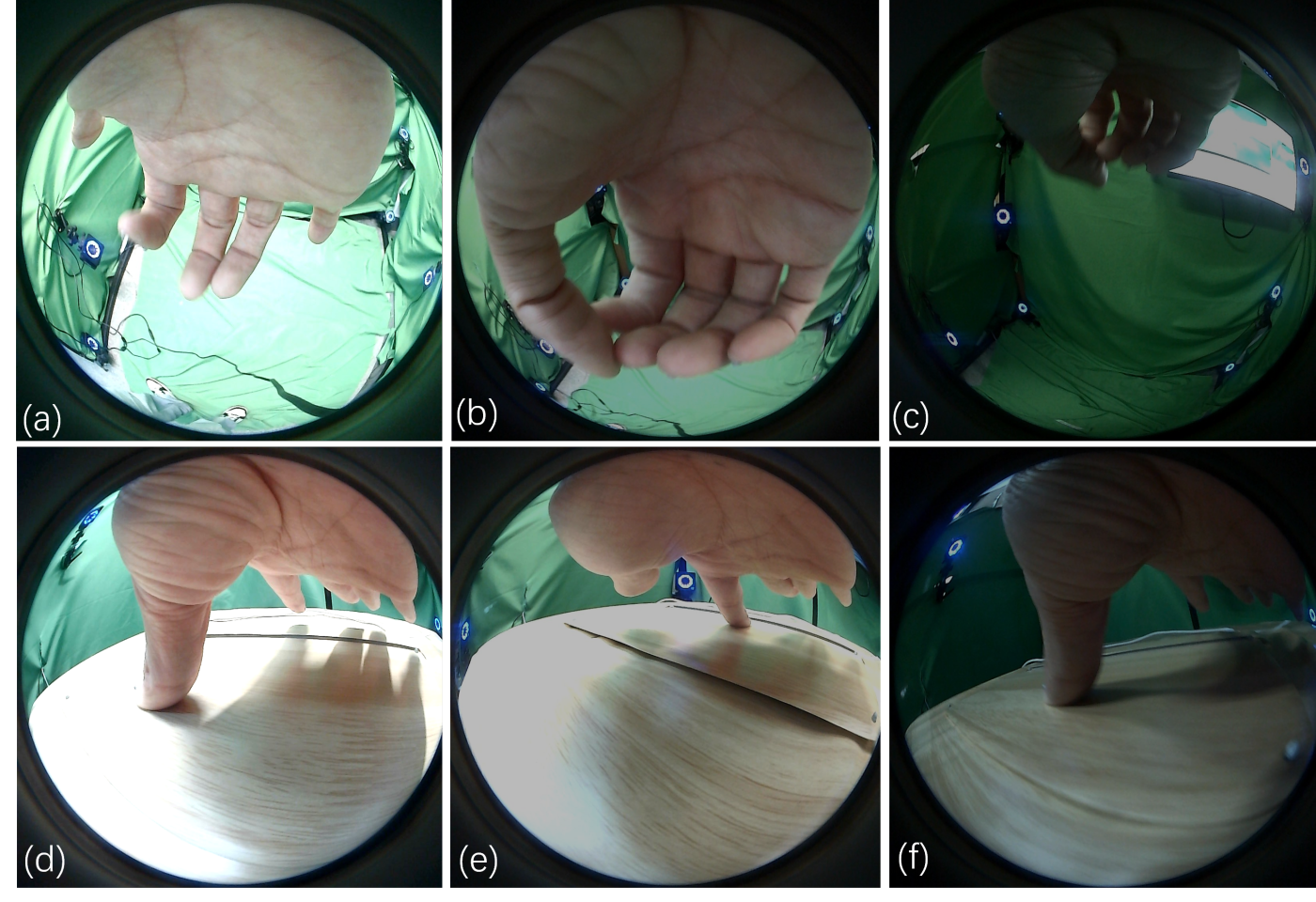}
  \caption{\textbf{Example wrist-camera views under different illumination conditions.}
  Top row: mid-air hand poses; bottom row: planar interaction on the surface.
  From left to right: high, medium, and low illumination.}
\Description{Wrist-mounted camera images illustrating three illumination 
conditions—high, medium, and low—shown for both mid-air hand poses and 
planar on-surface interaction.}
  \label{fig:illum_examples}
\end{figure}

We evaluate \wristp{} on three core tasks: (i) 3D hand pose reconstruction, (ii) per-vertex pressure estimation, and (iii) wrist-camera extrinsics estimation. 
Beyond reporting overall performance, we explicitly analyze the system's robustness to two critical factors: environmental illumination and wrist orientation. 
First, we report results across three illumination levels $I$: High (H, $2000 \geq I \geq 500$\,lux), Medium (M, $100 \leq I < 500$\,lux), and Low (L, $I < 100$\,lux), as visualized in Figure~\ref{fig:illum_examples}. Second, we investigate the impact of wrist orientation on reconstruction accuracy to validate the effectiveness of our camera design in handling extreme viewpoints. In addition, we compare \wristp{} against several existing methods for 3D hand pose reconstruction and pressure estimation under all these different settings. 
Due to space constraints, the full experimental setup and comparison details is reported in Appendix~\ref{sec:appendix_eval}.

\subsection{Pose Reconstruction}

We first evaluate the accuracy of 3D hand pose reconstruction in the hand-local frame. We use standard metrics: MPJPE (Mean Per-Joint Position Error), PA-MPJPE (Procrustes-Aligned Mean Per-Joint Position Error), PVE (Per-Vertex Error), PA-PVE, and MJAE (Mean Joint Angle Error). 
Table~\ref{tab:self-eval} summarizes the results.

\begin{table}[t]
  \centering
  \caption{\textbf{Pose accuracy across illumination conditions.} 
  Position errors in mm; MJAE in degrees ($^\circ$). Lower is better.}
  \label{tab:self-eval}
  \renewcommand{\arraystretch}{1.2}
  \setlength{\tabcolsep}{4pt} % 默认通常是 6pt，这里缩减到 4pt
  \begin{tabular}{lccccc}
    \toprule
    Illumination & MPJPE & PA-MPJPE & PVE & PA-PVE & MJAE [$^\circ$] \\
    \midrule
    High (H)    & 2.7 & 2.3 & 2.8 & 2.5 & 3.0 \\
    Medium (M)  & 2.8 & 2.4 & 2.9 & 2.6 & 3.1 \\
    Low (L)     & 3.1 & 2.7 & 3.2 & 2.9 & 3.3 \\
    \midrule
    All         & 2.9 & 2.4 & 3.0 & 2.6 & 3.2 \\
    \bottomrule
  \end{tabular}
\end{table}

% MPJPE of 2.9\,mm and MJAE of 3.11$^\circ$. should 3.11 be rounded to two decimal places?
\wristp{} achieves high accuracy with an overall MPJPE of 2.9\,mm and MJAE of 3.2$^\circ$. Notably, the system demonstrates strong robustness to illumination changes. As lighting conditions degrade from High to Low, the error increase is marginal (e.g., MPJPE increases by 0.4\,mm).

\subsection{Pressure Estimation}

Next, we evaluate per-vertex contact and pressure estimation. Metrics include Contact IoU (Contact Intersection over Union), Vol.~IoU (Volumetric Intersection over Union), Contact Accuracy (threshold $>$10\,g), and MAE (Mean Absolute Error) for foreground ($MAE_{\mathrm{FG}}$) and all vertices ($MAE_{\mathrm{ALL}}$).
Table~\ref{tab:press-eval} and Table~\ref{tab:press-eval-illum} detail the performance across different surfaces and lighting conditions.

\begin{table}[t]
  \centering
  \caption{\textbf{Contact and pressure accuracy across surfaces.} IoU $\in[0,1]$ and Contact Accuracy in \% (higher is better). MAE in gram-force (g; lower is better).}
  \label{tab:press-eval}
  \small
  \setlength{\tabcolsep}{4pt} % 默认6pt，缩小列间距
  \renewcommand{\arraystretch}{1.15}
  \begin{tabular}{lccccc}
    \toprule
    Surface & Cont.\ IoU & Vol.\ IoU & Cont.\ Acc.\ (\%) & MAE$_{\mathrm{FG}}$ & MAE$_{\mathrm{ALL}}$ \\
    \midrule
    Light Wood    & 0.723 & 0.624 & 97.5 & 9.7  & 0.107 \\
    Dark Wood     & 0.716 & 0.618 & 94.8 & 10.2 & 0.108 \\
    Light Marble  & 0.679 & 0.606 & 96.2 & 11.8 & 0.122 \\
    Dark Marble   & 0.673 & 0.591 & 95.5 & 12.4 & 0.127 \\
    Black         & 0.703 & 0.610 & 95.4 & 10.8 & 0.118 \\
    White         & 0.730 & 0.631 & 98.1 & 9.2  & 0.104 \\
    \midrule
    All           & 0.712 & 0.618 & 97.1 & 10.4 & 0.111 \\
    \bottomrule
  \end{tabular}
\end{table}

\begin{table}[t]
  \centering
  \caption{\textbf{Contact and pressure accuracy under different illumination conditions.} IoU $\in[0,1]$ and Contact Accuracy in \% (higher is better). MAE in gram-force (g; lower is better).}
  \label{tab:press-eval-illum}
  \small
  \setlength{\tabcolsep}{4pt}
  \renewcommand{\arraystretch}{1.15}
  \begin{tabular}{lccccc}
    \toprule
    Illumination & Cont.\ IoU & Vol.\ IoU & Cont.\ Acc.\ (\%) & MAE$_{\mathrm{FG}}$ & MAE$_{\mathrm{ALL}}$ \\
    \midrule
    High (H)    & 0.730 & 0.635 & 98.9 & 9.7  & 0.102 \\
    Medium (M)  & 0.712 & 0.614 & 97.4 & 10.4 & 0.109 \\
    Low (L)     & 0.690 & 0.600 & 96.1 & 11.1 & 0.120 \\
    \midrule
    All         & 0.712 & 0.618 & 97.1 & 10.4 & 0.111 \\
    \bottomrule
  \end{tabular}
\end{table}

\begin{figure*}[t]
    \centering
    \includegraphics[width=0.88\textwidth]{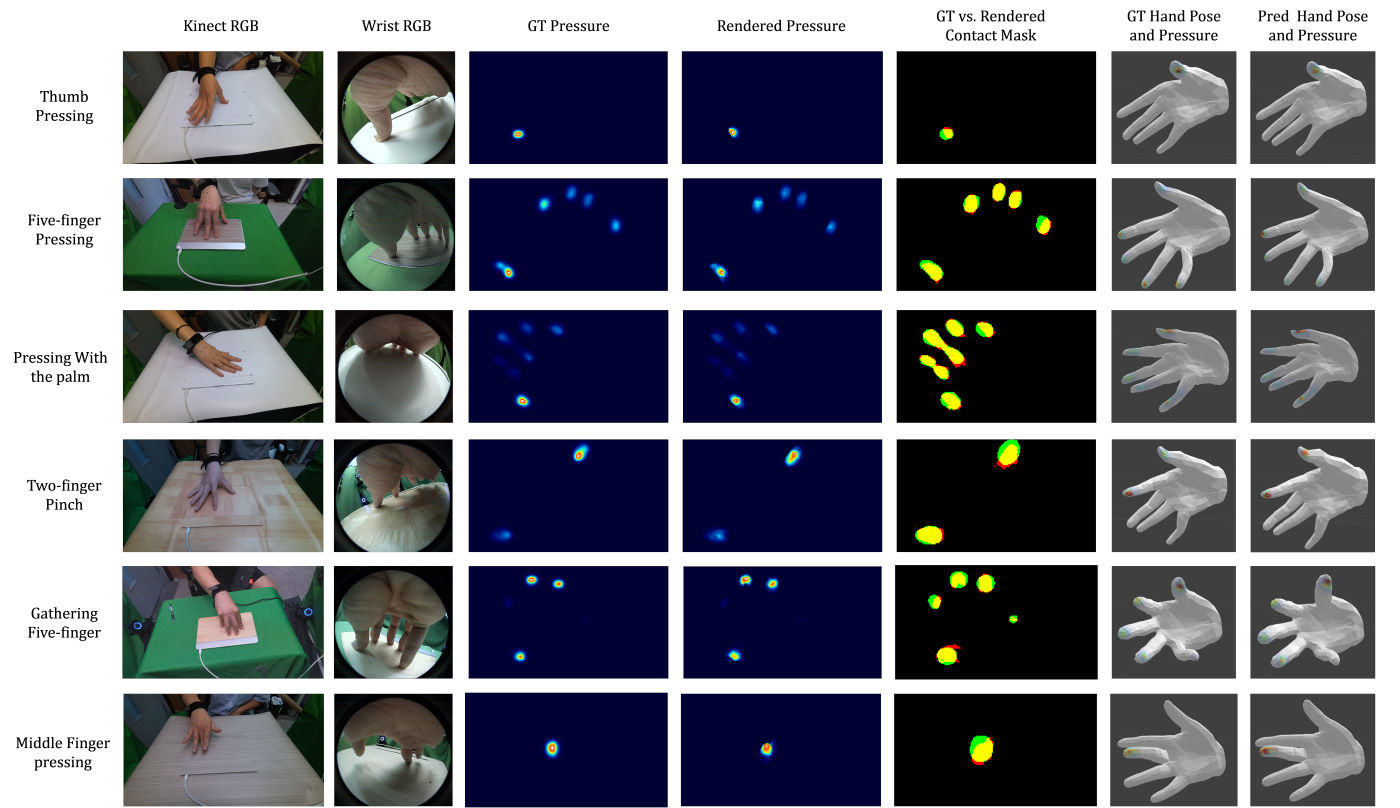}
    \caption{\textbf{Qualitative results of planar interaction.} The fifth column compares Ground Truth (Red) and Prediction (Green), with Yellow indicating overlap.}
  \Description{Qualitative examples of planar interaction showing wrist-camera 
inputs, reconstructed hand meshes, and surface pressure maps. The final column 
visualizes ground-truth pressure in red, predicted pressure in green, and their 
overlap in yellow.}
    \label{fig:press_results}
\end{figure*}

The system maintains high reliability across different surface textures, as shown in Table~\ref{tab:press-eval}. Clean and bright textures yield optimal results; for instance, the White surface achieves the lowest $MAE_{\mathrm{FG}}$ of 9.2\,g. While complex textures like Dark Marble present a challenge, increasing $MAE_{\mathrm{FG}}$ to 12.4\,g. Notably, the Contact Accuracy remains consistently above 90\% across all surfaces. This suggests that our network successfully learns to decouple geometric deformation cues from textural noise, ensuring reliable contact detection even on visually challenging surfaces. 

Similarly, the system demonstrates resilience to illumination changes (Table~\ref{tab:press-eval-illum}). Performance is stable between High and Medium settings, with $MAE_{\mathrm{ALL}}$ increasing marginally. Although Low-light conditions inevitably reduce the Signal-to-Noise Ratio (SNR), leading to a drop in Contact IoU to 0.690, the performance remains at a usable level, indicating that our proximal wrist-camera setup captures sufficient close-range visual features to support continuous interaction, preventing the complete failure often seen in environment-based systems under dim lighting.

\subsection{Wrist-Camera Extrinsics Estimation}
\label{sec:geometric_analysis}

Finally, we analyze the camera extrinsics prediction capability of \wristp{}. Table~\ref{tab:extrinsics-final} reports the results. The system achieves precise extrinsics estimation with a rotation error of 2.3$^\circ$ and translation error of 8.9\,mm.

\begin{table}[t]
  \centering
  \caption{\textbf{Wrist-camera extrinsics accuracy across illumination conditions.} Rotation in degrees ($^\circ$), translation in mm, reprojection in px. All metrics lower is better.}
  \label{tab:extrinsics-final}
  \small
  \setlength{\tabcolsep}{4pt}
  \renewcommand{\arraystretch}{1.15}
  \begin{tabular}{lccc}
    \toprule
    Illum. & Rot.\ Err.\ ($^\circ$) & Trans.\ Err.\ (mm) & 2D Reproj.\ (px) \\
    \midrule
    High (H)    & 2.1 & 8.5 & 13.1 \\
    Medium (M)  & 2.3 & 9.0 & 13.8 \\
    Low (L)     & 2.4 & 9.3 & 14.4 \\
    \midrule
    All         & 2.3 & 8.9 & 13.7 \\
    \bottomrule
  \end{tabular}
\end{table}

\begin{figure}[t]
  \centering
  \includegraphics[width=0.82\linewidth]{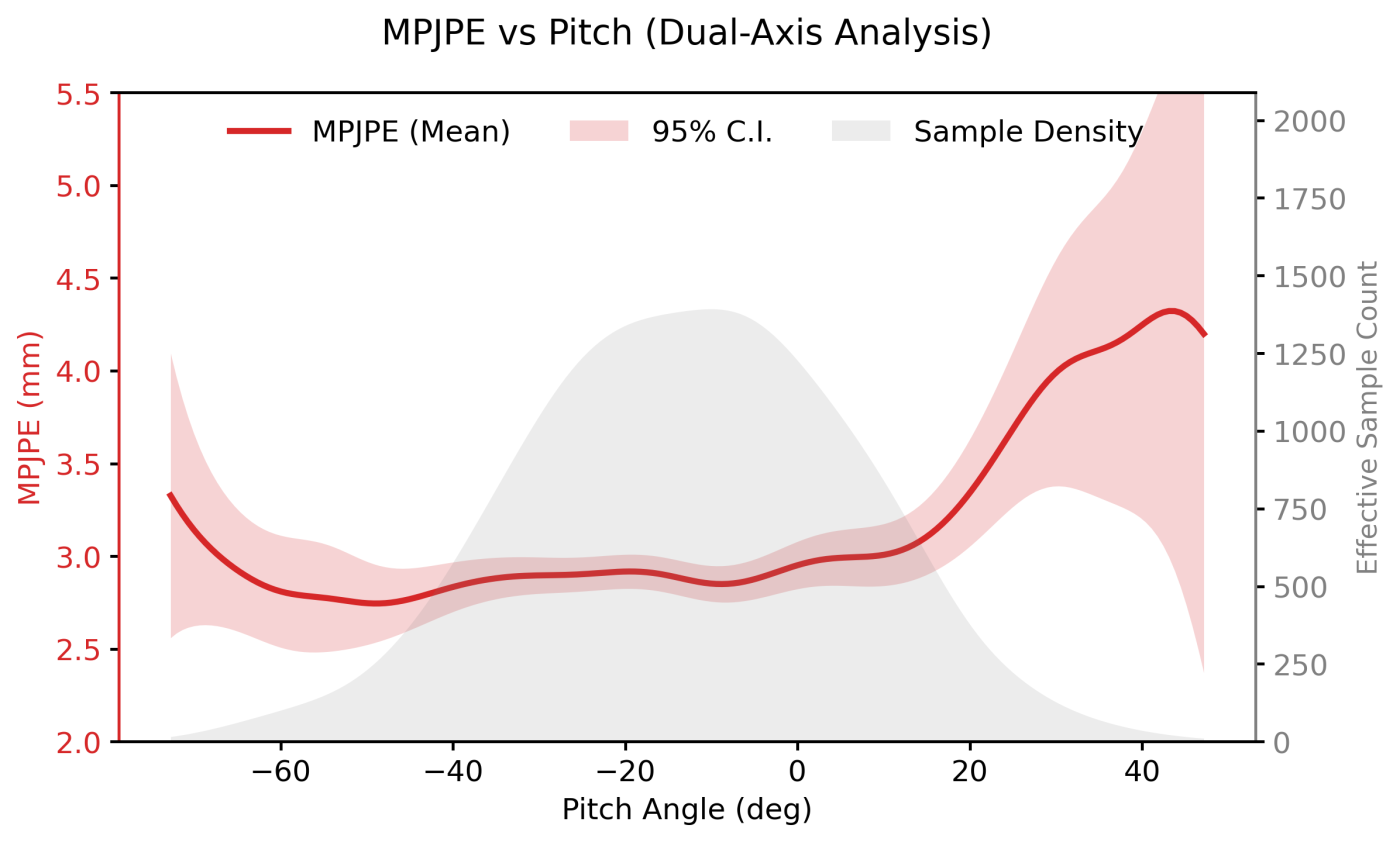} 
  \caption{\textbf{Error analysis w.r.t wrist pitch angle.} The red curve shows the MPJPE (mean with 95\% confidence interval). While error is minimized in the canonical range, the system maintains stable performance (sub-4.5\,mm error) even at extreme angles, demonstrating the robustness of the fisheye design.}
  \Description{A line plot showing mean per-joint position error versus wrist pitch angle. 
  Error is lowest in the canonical range and increases moderately at extreme negative 
  and positive angles, remaining below 4.5 mm overall.}
  \label{fig:pitch_curve}
\end{figure}

\noindent\textbf{Impact of Wrist Orientation.} 
The pitch angle determines the proportion of the hand in the camera frame. A larger proportion provides more visual details but may introduce near-field distortion, while a smaller proportion reduces the effective resolution. 
We analyze the MPJPE w.r.t wrist pitch angle in Figure~\ref{fig:pitch_curve}.

\begin{figure}[t]
  \centering
  \includegraphics[width=0.85\linewidth]{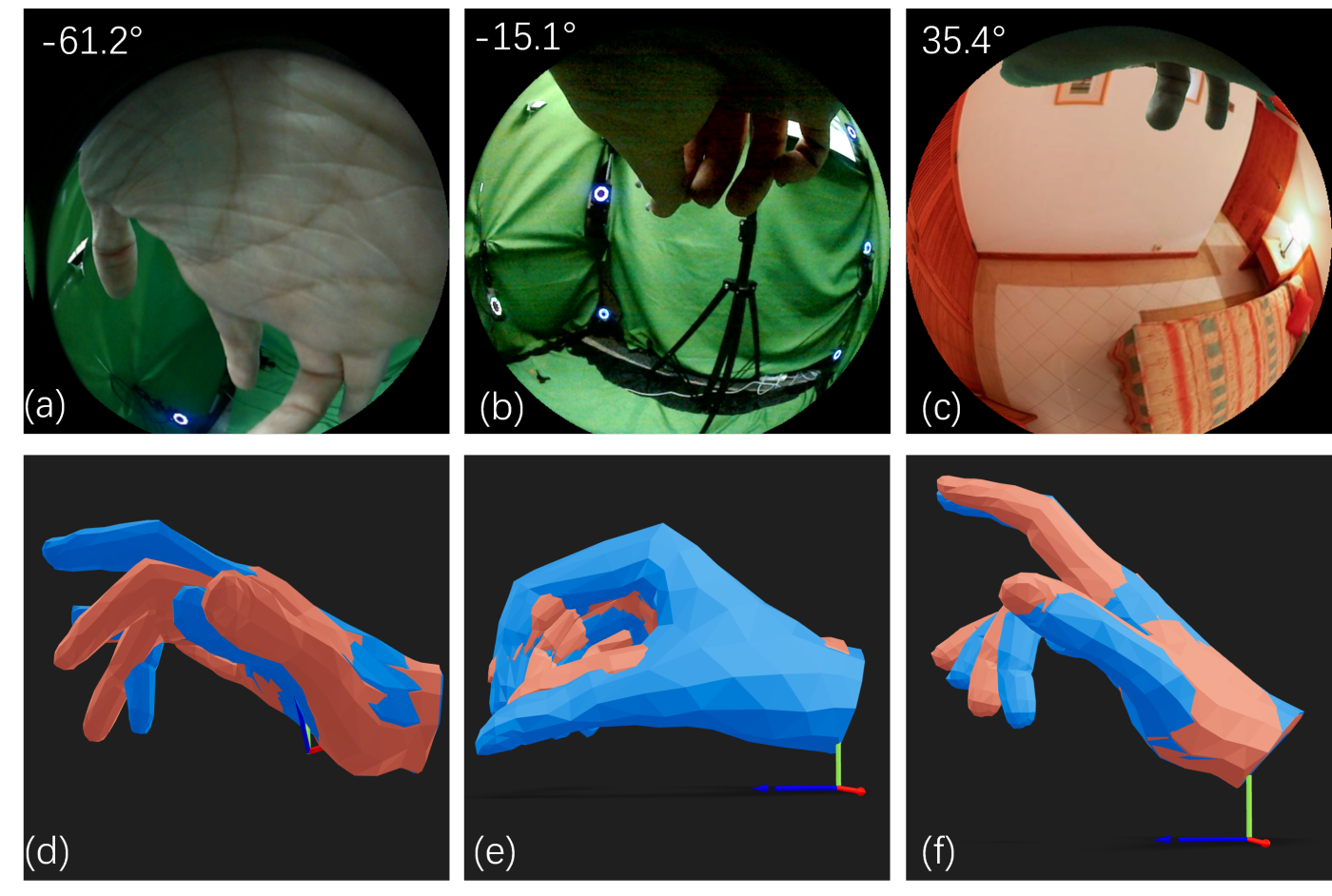} 
  \caption{\textbf{Qualitative visualization across pitch angles.} 
  The top row displays raw fisheye inputs, while the bottom row shows the corresponding 3D mesh reconstructions in the camera coordinate system. 
  \textbf{Legend:} The \textcolor{blue}{Blue} mesh represents Ground Truth (GT), and the \textcolor{red}{Red} mesh represents the Prediction. The coordinate triad at the origin marks the camera center, where the \textbf{blue axis indicates the Z-axis} (camera viewing direction).
  \textbf{(a, d) Extreme Negative Pitch ($-61.2^\circ$):} The hand occupies a significant portion of the frame with strong fisheye distortion.
  \textbf{(b, e) Canonical Range ($-15.1^\circ$):} The hand is centered with optimal scale and visibility.
  \textbf{(c, f) Extreme Positive Pitch ($35.4^\circ$):} The hand recedes to the periphery, yet the system maintains robust tracking.}
  \Description{Qualitative examples of wrist pitch angles. The top row shows fisheye camera images, and the bottom row shows overlaid 3D meshes in the camera coordinate system. The Blue mesh is Ground Truth, and the Red mesh is Prediction. A coordinate axis shows the blue Z-axis pointing in the camera's viewing direction. Columns show extreme negative pitch (hand large/distorted), canonical range (hand centered), and extreme positive pitch (hand small/distant).}
  \label{fig:qualitative_pitch}
\end{figure}

As illustrated in Figure~\ref{fig:pitch_curve}, although the error distribution correlates with wrist orientation, the system maintains high accuracy within the most densely populated angular ranges. Figure~\ref{fig:qualitative_pitch} presents representative examples across these ranges to illustrate the system's robustness:
\begin{itemize}
    \item In the canonical range ($-30^\circ$ to $0^\circ$), the hand is centered with an optimal proportion. For the sample shown in Fig.~\ref{fig:qualitative_pitch}b ($-15.1^\circ$), the system yields a low MPJPE of $\approx$2.9\,mm.
    \item At extreme negative angles (e.g., $-60^\circ$), the hand occupies a large proportion of the frame. Although perspective distortion increases, the specific case in Fig.~\ref{fig:qualitative_pitch}a ($-61.2^\circ$) shows an error of $\approx$3.3\,mm, verifying that the fisheye lens effectively captures sufficient detail even under severe distortion.
    \item At positive extremes ($>30^\circ$), the hand recedes to the periphery, reducing its proportion in the frame. While this generally challenges the algorithm, the example in Fig.~\ref{fig:qualitative_pitch}c ($35.4^\circ$) demonstrates functional accuracy with an MPJPE of $\approx$4.3\,mm, confirming that the wide Field-of-View prevents tracking loss.
\end{itemize}
We refer to Appendix~\ref{app:qua} for additional visualization results of pose prediction under varying wrist pitch angles.

In conclusion, \wristp{} delivers relatively consistent performance across varying illumination, surfaces, and wrist orientations. Our pitch-sweep analysis indicates that \wristp{} maintain high accuracy around the canonical wrist angles. In practice, this suggests that a palm-facing fisheye camera mounted near the wrist can robustly accommodate natural hand postures, while interactions that require fully pronated or supinated forearms remain challenging. These findings provide guidance on where wrist-mounted vision is best suited and where additional sensing (e.g., head-mounted or non-visual modalities) may be needed.

\section{USER STUDY}
\label{sec:user-study}
To systematically evaluate the performance of \wristp{} for both on-plane and mid-air interaction, we conducted three user studies: (1) mid-air cursor control; (2) multi-finger pressure input on a desktop surface; and (3) a composite virtual pressure-sensitive touchpad task integrating both fingertip displacement and pressure cues.

\begin{figure}[t]
  \centering
  \includegraphics[width=0.45\textwidth]{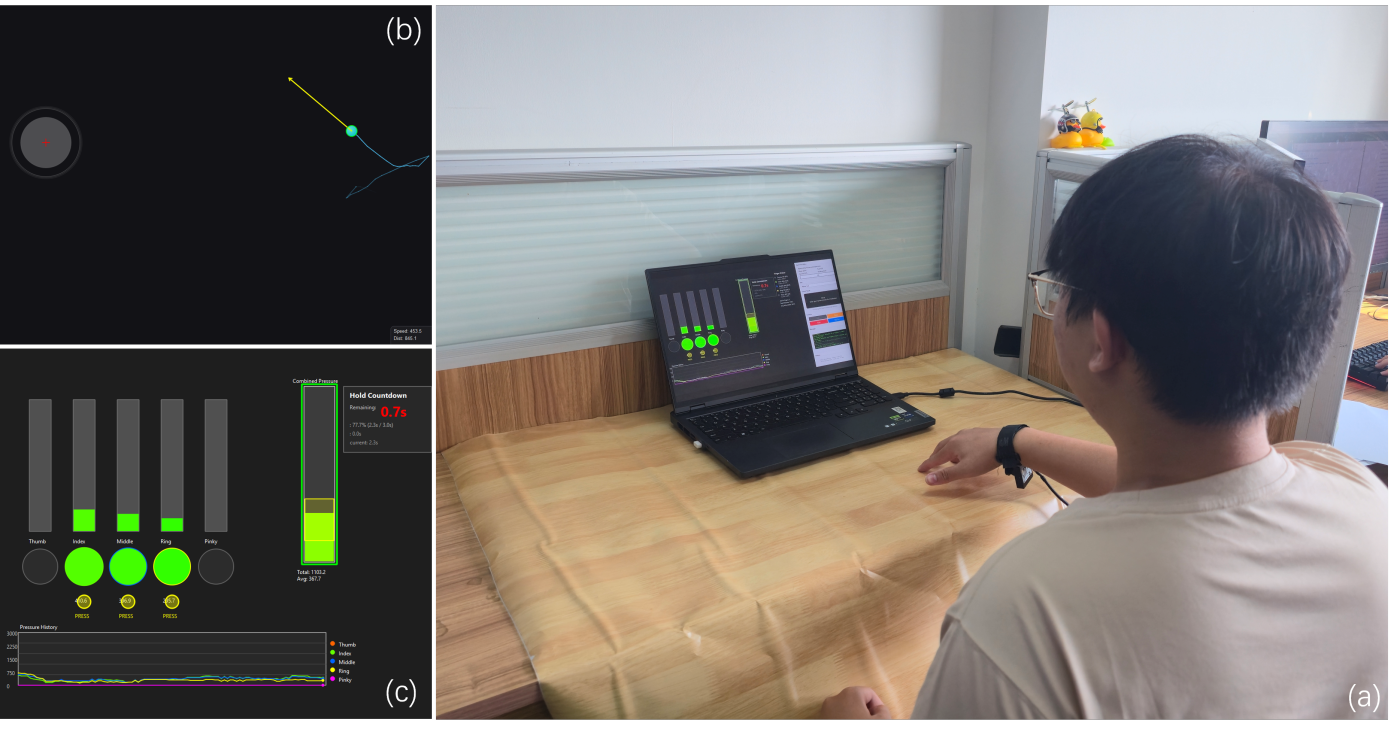}
  \caption{\textbf{Experimental setup for the first three user studies.}
  All studies were conducted in the same environment (a), under varying light conditions and surface textures.
  User Studies 1 and 3 employed the GUI shown in (b), while User Study 2 used the GUI in (c).}
\Description{Photographs showing the environment and graphical interfaces used in the first 
three user studies. The left image depicts the desk setup under different lighting and 
surface conditions. The middle image shows the GUI used for mid-air cursor control and the 
drag-and-press task. The right image shows the GUI with five pressure bars used for the 
multi-finger pressure control study.}
  \label{fig:us_env}
\end{figure}

To ensure stable data collection and consistent evaluation, all three studies used a wired data transmission setup. Dedicated GUIs were developed for each task. We further randomized lighting conditions and surface textures following the same protocol as in the dataset collection phase (Section~\ref{sec:dataset}). The experimental setups for these studies are illustrated in Fig.~\ref{fig:us_env}.

\subsection{Participants}
A total of 18 right-handed participants (3 females, aged 20–25, $M = 24.0$) from our institution took part in the studies consecutively after a 5-minute tutorial and a brief familiarization period. All participants were familiar with conventional mice and touchpads, but none had prior experience with the \wristp{} system.

\subsection{Apparatus}
All Python-based applications for User Studies~1-3 were executed on a Lenovo Legion Y9000P laptop
equipped with an Intel Core i9-13900HX processor (13th Gen, 2.2\,GHz),
16\,GB of RAM (5600\,MT/s), and an NVIDIA GeForce RTX 4060 Laptop GPU (8\,GB VRAM).
The system was running Windows 11 Home (64-bit, version 24H2).
In the Virtual Air Mouse study (Study~1), we additionally included two conventional input devices
as baselines: the laptop's built-in touchpad and a Lenovo LEGION M500 RGB gaming mouse.

\subsection{USER STUDY 1: Mid-Air Cursor Control}
\label{sec:STUDY_1}
\subsubsection{Design and Procedure}
The first study evaluated the system's capability to estimate hand postures to accomplish precise cursor movements. In this study, the participants' index finger served as a virtual air mouse to control the cursor in a Fitts' Law experiment. The index-finger coordinates (x, y) in the hand-local coordinate system were linearly mapped to the computer mouse coordinates on the screen using scaling factors of 23{,}000.0 (x direction) and 11{,}000.0 (y direction). We computed the Euclidean distance between the thumb and the index finger. When the distance was smaller than the contact threshold (5\,mm), the virtual air mouse was recognized as a press, thereby enabling click and long-press functions similar to those of conventional input devices. 

In each trial, a green button and a gray circular target were generated at random locations on the screen. The green button had a fixed diameter of 10\,mm, whereas the diameter of the gray target varied from 30\,mm to 90\,mm. The center-to-center distances ranged from 40\,mm to 400\,mm, and initial positions were constrained to preclude spatial overlap. Participants initiated each trial by long-pressing the green button. Timing began when they started moving the button and stopped once it had been dragged into the gray circular target and held there for 250\,ms. Each participant completed 50 trials for each input device (virtual air mouse, touchpad, and mouse). After every block of 10 trials, a 2-minute rest interval was provided, during which participants were required to remove the wristband.
\subsubsection{Measures and Analysis}
For each trial, the diameter of the gray circular target, the initial center-to-center distance between the green button and the gray target, and the trial completion time were recorded.
\subsubsection{Results}
To assess the significance of differences in task performance among the three input modalities, we conducted Friedman tests (as the data did not meet normality assumptions). The Friedman test results indicated a significant difference in the median completion times across the three input methods ($\chi^2=41.05, p<0.001$). The mouse (Median $= 861.3$\,ms, IQR $= 530.0$\,ms) demonstrated superior selection speed, with both the touchpad (Median $= 1574.2$\,ms, IQR $= 799.0$\,ms) and the virtual air mouse (Median $= 2111.1$\,ms, IQR $= 1199.9$\,ms) showing significantly lower performance relative to the mouse. The difference in selection speed between the virtual air mouse and the touchpad was approximately equivalent to the selection speed gap observed between the touchpad and the mouse. These findings indicated that the virtual air mouse constitutes a potentially viable input device.

\begin{figure}[t]
  \centering

  % 左图: violin_completion_time
  \begin{subfigure}[t]{0.42\textwidth}
    \centering
    \includegraphics[width=\textwidth]{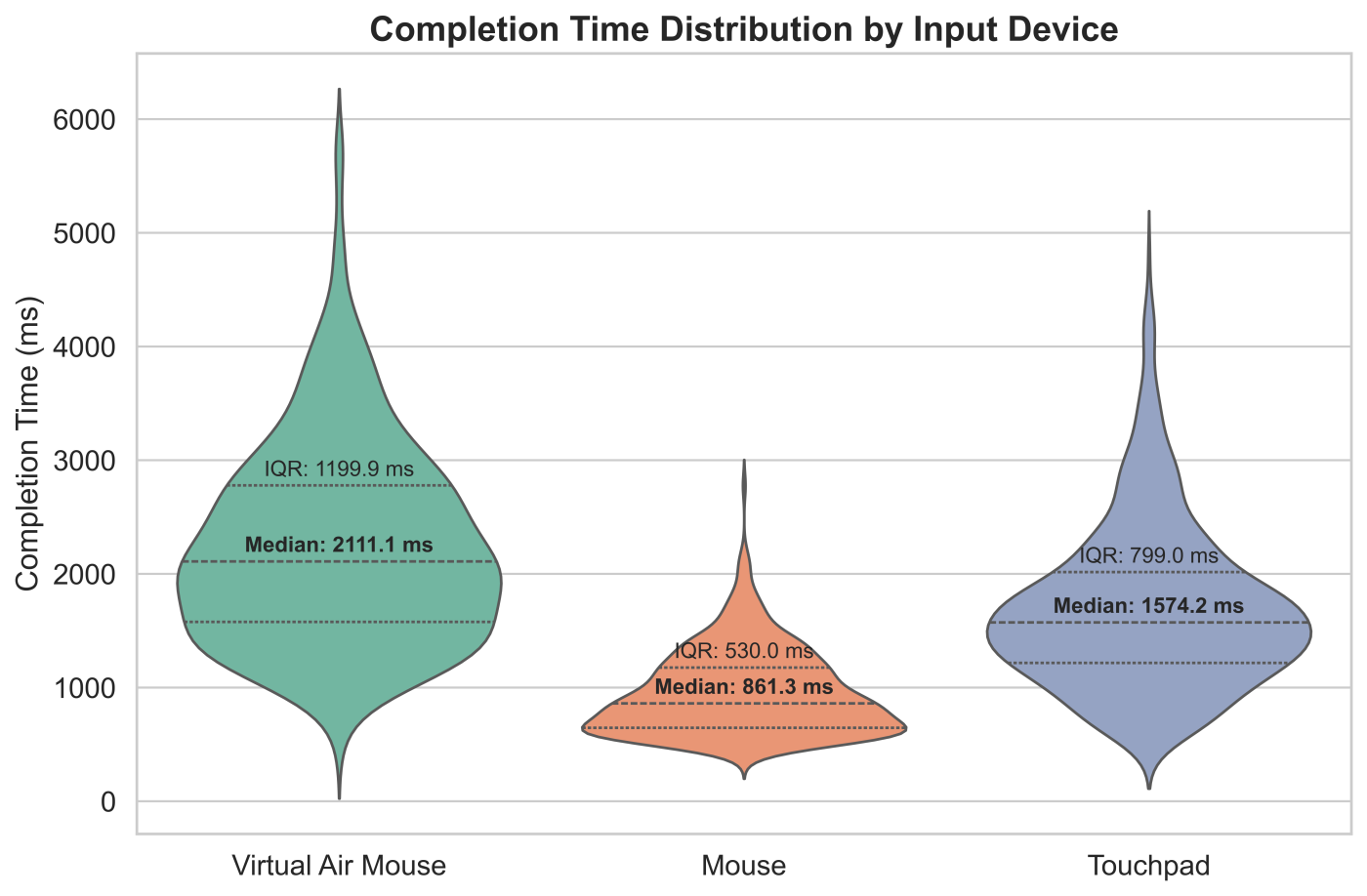}
    \caption{Completion time for different input devices.}
    \label{fig:violin_completion_time}
  \end{subfigure}
  \hfill
  % 右图: MCT
  \begin{subfigure}[t]{0.42\textwidth}
    \centering
    \includegraphics[width=\textwidth]{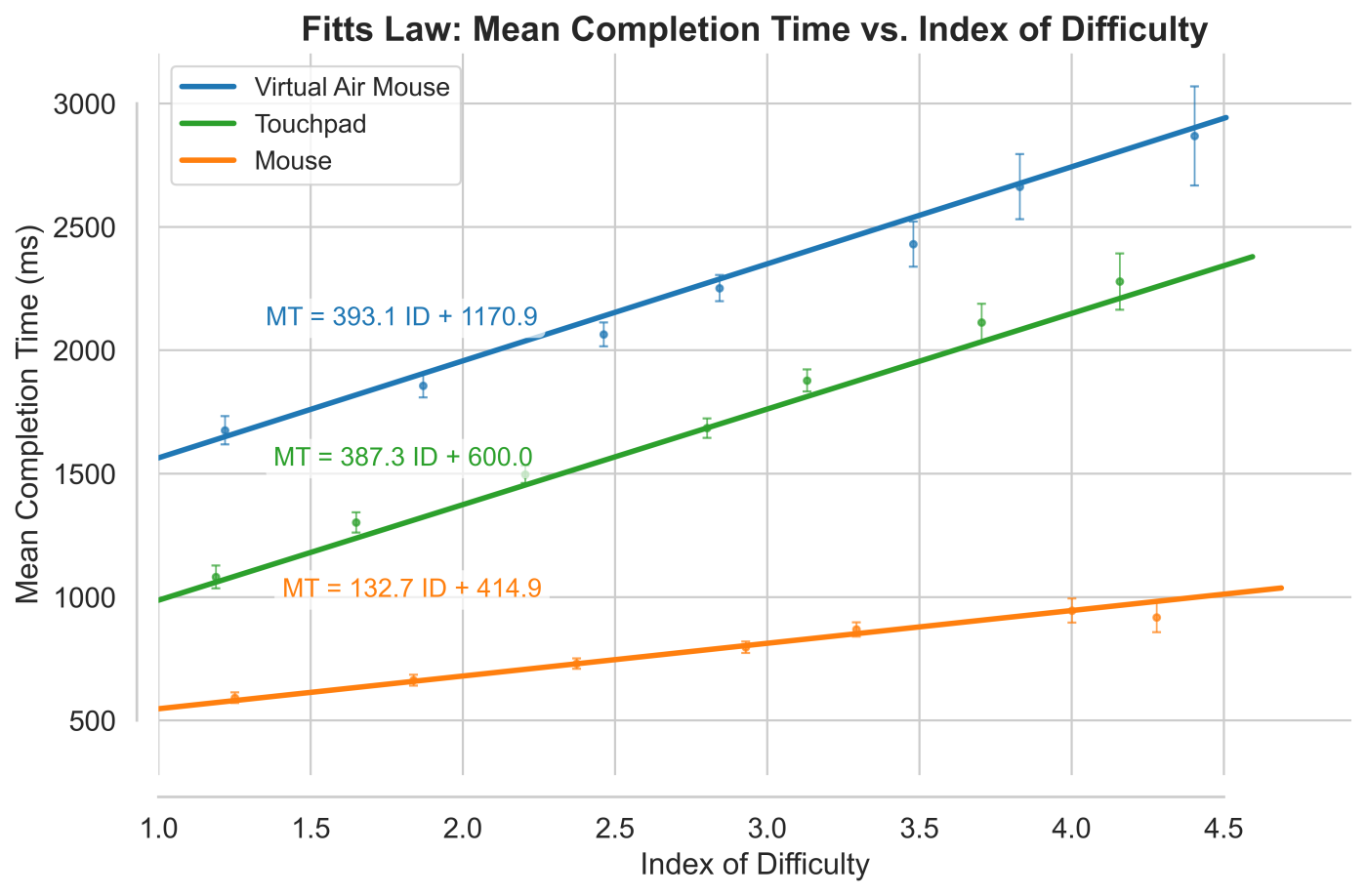}
    \caption{Fitted line of MT versus ID.}
    \label{fig:fitts_law_fit}
  \end{subfigure}

  \caption{\textbf{Performance comparison among input devices.}
  (a) Performance was quantified using the median and interquartile range (IQR). 
  (b) For each device, a regression was fitted using the aggregated data across all trials.}
\Description{
The figure contains two subplots comparing pointing performance across three input devices:
a mouse, a laptop touchpad, and the WristP\textsuperscript{2} virtual air mouse.
The left subplot is a violin plot showing the distribution of completion times for the three devices, 
where the mouse is fastest, followed by the touchpad, and then the virtual air mouse.
The right subplot shows Fitts’ Law regression lines of movement time versus index of difficulty 
for the three devices, with the mouse exhibiting the lowest slope and highest throughput.
}
  \label{fig:sr_mct_split_air}
\end{figure}

A regression line was fitted to describe the relationship between mean completion time (MT) and index of difficulty (ID = $\log_2\frac{2D}{W}$), where the slope served as an index of throughput and the intercept represented the constant time overhead (Fig.~\ref{fig:fitts_law_fit}). The mouse exhibited the highest throughput (k $= 132.7$, TP $= 7.5$\,bit/s), whereas the touchpad (k $= 387.3$, TP $= 2.6$\,bit/s) and the virtual air mouse (k $= 393.1$, TP $= 2.5$\,bit/s) demonstrated comparable throughput. This indicates that the virtual air mouse and the touchpad exhibit comparable cursor movement speed and similar input performance. The intercept for the virtual air mouse was the highest, reflecting extra constant time overhead caused by hand movement variability in three-dimensional space.

The Fitts' Law study suggested that the virtual air mouse and the touchpad exhibit comparable performance on specific metrics, highlighting the high precision of the hand mesh predictions. These results further suggest that \wristp{} may serve as a portable, lightweight, and precise input modality, capable of supporting interaction in a wide range of contexts.

% \subsubsection{Results}
% A Friedman test revealed significant differences in completion time across the three input methods ($\chi^2=41.05, p<0.001$). The mouse was fastest (Median $= 861.3$\,ms), followed by the touchpad (Median $= 1574.2$\,ms) and the virtual air mouse (Median $= 2111.1$\,ms); the gap between the virtual air mouse and the touchpad was comparable to that between the touchpad and the mouse.

% Fitts' Law regressions (Fig.~\ref{fig:fitts_law_fit}) yielded throughputs of 7.5\,bit/s for the mouse, 2.6\,bit/s for the touchpad, and 2.5\,bit/s for the virtual air mouse, indicating touchpad-level pointing efficiency for WristP2 with a higher intercept due to additional overhead from 3D hand motion and click confirmation.

\subsection{USER STUDY 2: Multi-Finger Pressure Control}
\label{sec:STUDY 2}
\subsubsection{Design and Procedure}
The second study evaluated the system's capability for multi-finger pressure control.
During each trial, participants were instructed to press the desk using a specific set of fingers. The system estimated fingertip pressure by aggregating vertex values within predefined fingertip regions on the hand mesh. These pressure estimates were mapped to the GUI, which provided five individual progress bars—one per finger—for fine-grained feedback, along with a combined bar indicating the average pressure of the fingers required in the current trial (Fig.~\ref{fig:us_env}). This combined bar served as a reference for coordinating multi-finger input.

At the start of each trial, one of 13 finger combinations was randomly selected, covering single-, two-, three-, four-, and five-finger configurations. These combinations were chosen because they reflect natural and frequently used finger postures during planar hand interactions; their detailed specification is reported in the supplementary. A target pressure range was displayed on the combined bar, with its center uniformly sampled from $[500,1500]\,\mathrm{g}$ and width from $[500,1500]\,\mathrm{g}$. Participants were required to apply force with the specified fingers so that the average pressure remained within the target range for $3\,\mathrm{s}$. A trial was considered a failure if the target range was not reached and stabilized within $10\,\mathrm{s}$. Each participant completed 100 trials.

\subsubsection{Measures and Analysis}
\begin{figure}[t]
  \centering

  % 左图: SR
  \begin{subfigure}[t]{0.42\textwidth}
    \centering
    \includegraphics[width=\textwidth]{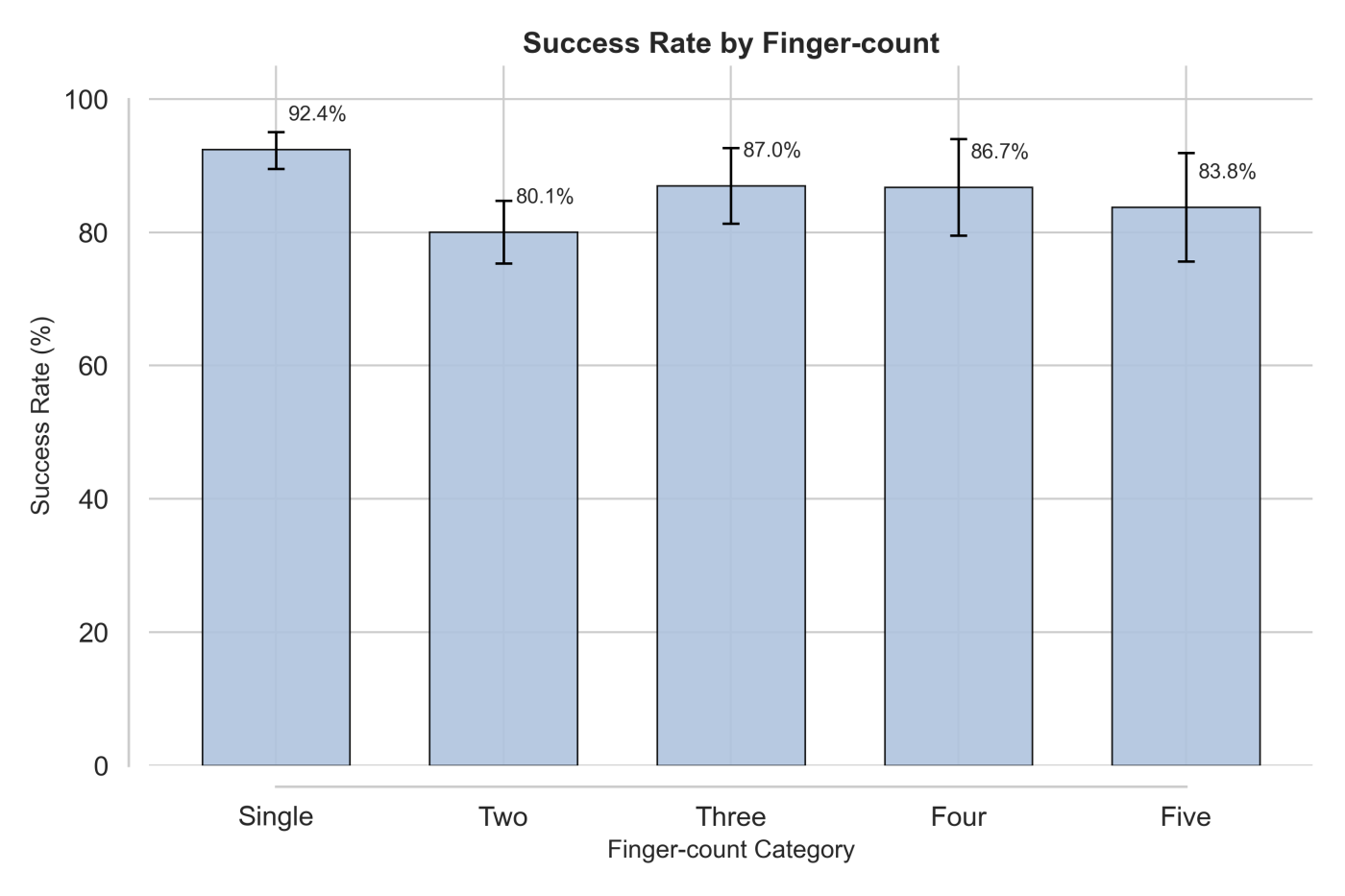}
    \caption{Success rate across finger-count categories.}
    \label{fig:sr_only}
  \end{subfigure}
  \hfill
  % 右图: MCT
  \begin{subfigure}[t]{0.42\textwidth}
    \centering
    \includegraphics[width=\textwidth]{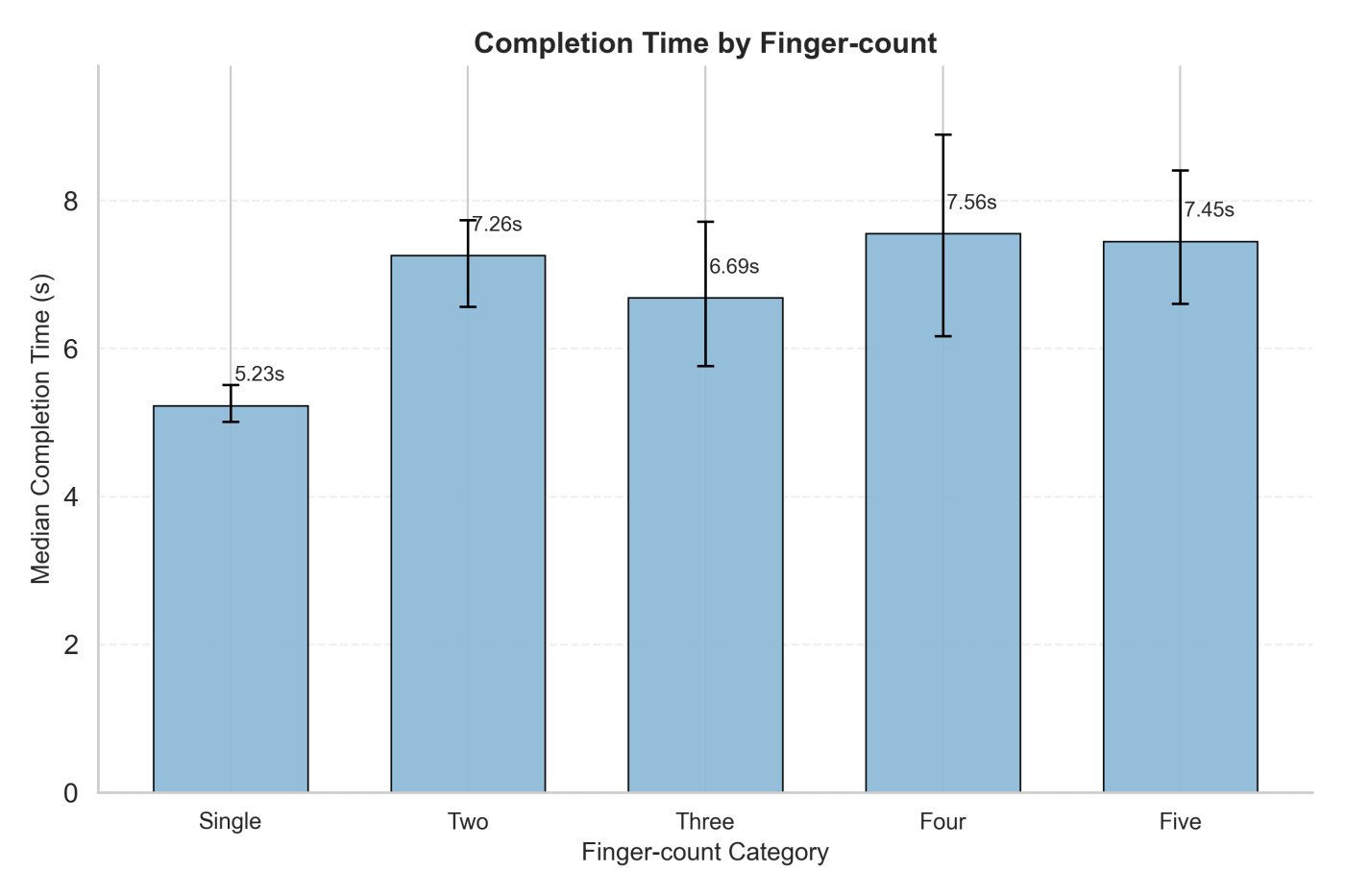}
    \caption{Median completion time across finger-count categories.}
    \label{fig:mct_only}
  \end{subfigure}

  \caption{\textbf{Performance summary across all subjects.}
  (a) Success rate (SR) with 95\% CI. 
  (b) Median completion time (MCT) with 95\% CI, computed on successful trials only.}
\Description{
The figure contains two subplots summarizing performance in the multi-finger pressure control study.
The left subplot shows success rates for single-, two-, three-, four-, and five-finger conditions, 
with single-finger input achieving the highest success rate and multi-finger conditions slightly lower. 
The right subplot shows median completion time for the same finger-count categories, with single-finger 
trials being fastest and completion time increasing as more fingers are involved. Both plots display 
95\% confidence intervals.
}
  \label{fig:sr_mct_split}
\end{figure}

We recorded the trial outcome (success/failure) and the completion time (CT, s). CT was defined as the time from trial onset until the participant successfully held the pressure within the target range for 3\,$\mathrm{s}$. CT was computed only for successful trials. For reporting we summarize success rate (SR, \%), as well as the median and interquartile range (IQR) of CT. Figures~\ref{fig:sr_mct_split} visualize the median CT and success rate by finger-count. Table~\ref{tab:cond_summary_all} summarizes outcomes across finger-count categories.

\subsubsection{Results}
Overall, participants were consistently able to reach and maintain the target pressure, achieving an average success rate of 86.7\% with median completion times around 6\,s. This indicates that the system reliably supports stable pressure control under varied interaction settings.

While single-finger input yielded the highest performance (SR$ = 92.4\%$, median CT$ = 5.23$\,s), multi-finger input also reached high success rates in the 80--87\% range, albeit with longer completion times (median CT up to 7.56\,s) and greater variability. These trends reflect the natural increase in coordination demands as more fingers are involved, yet the overall performance remained robust.

\noindent\textit{Takeaway.} The study demonstrates that the proposed system enables accurate and efficient pressure targeting across different finger combinations. Despite the expected increase in difficulty for multi-finger coordination, participants maintained high success rates and reasonable completion times, underscoring the technique's suitability for integration into pressure-sensitive interactive applications.

% Requires: \usepackage{booktabs}
\begin{table}[t]
  \centering
  \caption{Performance summary aggregated by finger-count with total $N{=}1800$ distributed by combination sampling ratios (Single/Two/Three/Four/Five $=$ 5/4/2/1/1 out of 13).
  SR = success rate with 95\% CI (Wilson); CT = completion time (successful trials only).}
  \label{tab:cond_summary_all}

  \small
  \setlength{\tabcolsep}{3.5pt}
  \renewcommand{\arraystretch}{1.15}

  \begin{tabular}{lrrrrr}
    \toprule
    Finger &
    \makecell{Trials\\($N$)} &
    \makecell{Success\\($N_{\text{succ}}$)} &
    \makecell{SR\\(\%, 95\% CI)} &
    \makecell{Median\\CT (s)} &
    \makecell{IQR\\CT (s)} \\
    \midrule
    Single & 692 & 639 & 92.4 & 5.23 & 2.99 \\
    Two    & 554 & 444 & 80.1 & 7.26 & 4.42 \\
    Three  & 277 & 241 & 87.0 & 6.69 & 4.24 \\
    Four   & 138 & 120 & 86.7 & 7.56 & 5.24 \\
    Five   & 139 & 116 & 83.8 & 7.45 & 4.07 \\
    \midrule
    \textbf{Overall} & 1800 & 1560 & 86.7 & 6.10 & 4.28 \\
    \bottomrule
  \end{tabular}

  \vspace{2pt}
  \footnotesize
  \textit{Notes:} CT statistics are reported on successful trials only.
\end{table}

\subsection{USER STUDY 3: Virtual Pressure-Sensitive Touchpad}
\label{sec:STUDY 3}
\subsubsection{Design and Procedure}
The third study evaluated the pressure-sensitive touchpad functionality of \wristp{} in a composite challenging task that couples drag control with pressure. In this study, participants treated the desktop as a virtual pressure-sensitive touchpad. Prior to the trials, participants performed a five-finger contact on the desktop to simultaneously signal session start and define the interaction plane. Using the 3D positions of all five fingertips, we fitted a plane in wrist camera coordinate frame via an SVD-based least-squares procedure: the plane origin was set to the fingertip centroid, and the unit normal $\mathbf{n}$ was taken as the right-singular vector of the centered fingertip coordinates. We then constructed a local orthonormal frame $\{\mathbf{u},\mathbf{v},\mathbf{n}\}$ on the plane by projecting the negative camera $x$-axes onto the plane and normalizing to obtain $\mathbf{u}$ (aligned to the right in the GUI), and setting $\mathbf{v}=\mathbf{n}\times\mathbf{u}$ (pointing downward). Cursor mapping used the fingertip position projected onto this plane and expressed in $(\mathbf{u},\mathbf{v})$ coordinates.

\begin{figure}[t]
    \centering
    \includegraphics[width=0.4\textwidth]{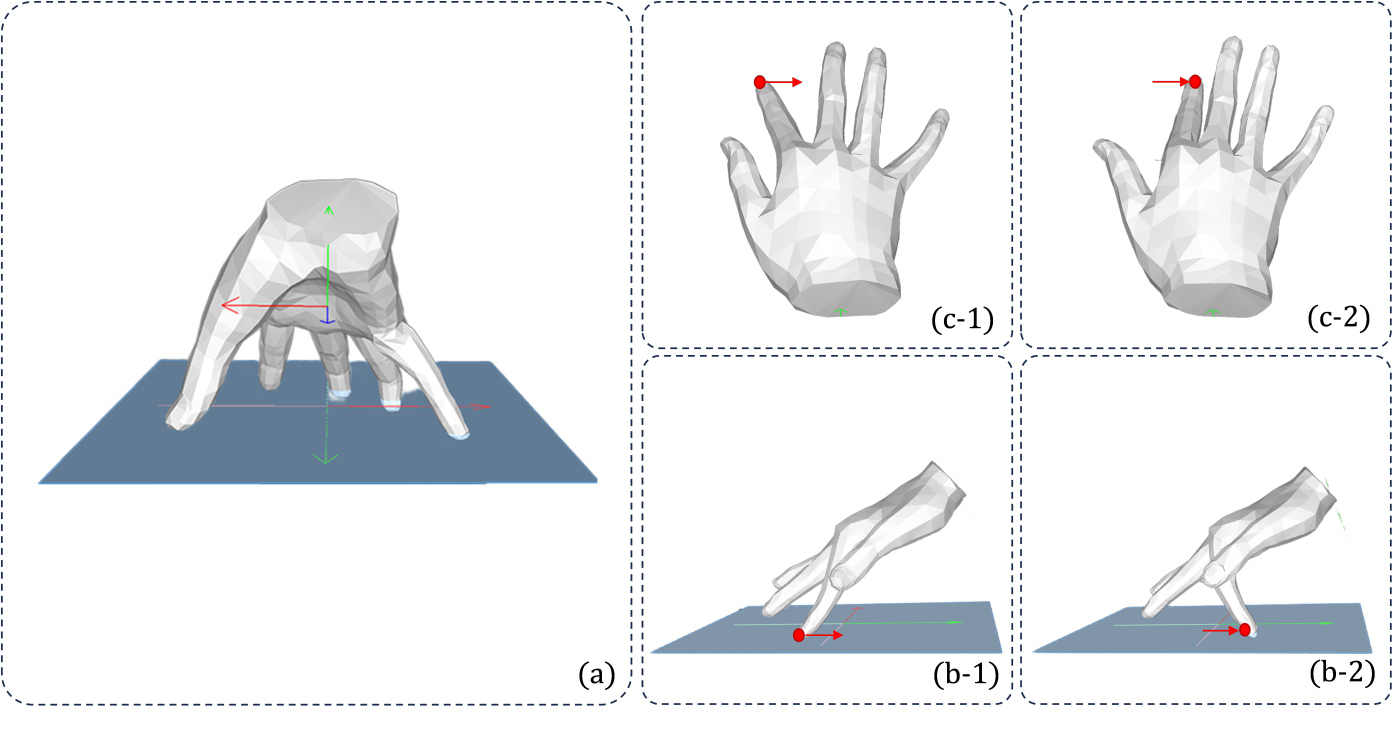}
    \caption{\textbf{Virtual pressure-sensitive touchpad:} (a) definition of the virtual interactive plane; (b–c) scrolling on the plane using the index fingertip.}
  \Description{
The figure shows the virtual pressure-sensitive touchpad setup. 
Panel (a) illustrates how a virtual interaction plane is defined from the positions of five fingertips. 
Panels (b) and (c) depict the index finger performing scrolling gestures on the plane, with fingertip 
contact and movement mapped to on-screen actions.
}
    \label{fig:hardware_prototype_touchpad}
\end{figure}

In each trial, two on-screen balls were shown: a controllable starting ball with a fixed diameter of 10\,mm, and a target ball with a radius uniformly sampled from 30\,mm to 90\,mm. The center-to-center distance between the balls was uniformly sampled from 40\,mm to 400\,mm, which remained the same as the settings in Section~\ref{sec:STUDY_1}.
To ensure ecological validity while avoiding rarely used gestures, the controlling finger was randomly selected on each trial from \emph{index}, \emph{middle}, and \emph{ring}. 
These three fingers were chosen because they are the most frequently used for touchpad interaction in daily computer use. Participants were asked to drag the starting ball into the 
target ball's area using the corresponding fingertip (entry when the starting-ball center lay within the target radius). Upon entry, a pressure target range was displayed (same sampling 
as in Study~2: center uniformly from $[500,1500]\,\mathrm{g}$ and width from $[500,1500]\,\mathrm{g}$). Participants adjusted their pressure so the measured value fell within the range and maintained it for $3$\,$\mathrm{s}$; 
the trial was marked \emph{successful} once the 3\,s hold completed. Trials that did not reach and stabilize within the range by $15$\,s were marked \emph{failures}. Each participant completed 60 trials with randomized finger, target size, and distance.

\subsubsection{Measures and Analysis}

\begin{table}[t]
  \centering
  \caption{Performance summary of drag-and-press study with decomposition into total completion time (CT), pressure stabilization time (PST), and movement time (MT).
  SR = success rate (\%); all time statistics are reported on successful trials only.}
  \label{tab:us2_decompose}

  \small
  \setlength{\tabcolsep}{3.2pt}
  \renewcommand{\arraystretch}{1.15}

  \begin{tabular}{lrrrrrr}
    \toprule
    Group &
    \makecell{Trials\\($N$)} &
    \makecell{Success\\($N_{\text{succ}}$)} &
    \makecell{SR\\(\%)} &
    \makecell{Median\\CT (s)} &
    \makecell{Median\\PST (s)} &
    \makecell{Median\\MT (s)} \\
    \midrule
    Index  & 339 & 326 & 96.2 & 9.75 & 5.64 & 3.36 \\
    Middle & 367 & 363 & 99.0 & 9.59 & 5.52 & 3.45 \\
    Ring   & 374 & 369 & 98.7 & 9.54 & 5.45 & 3.88 \\
    \midrule
    \textbf{Overall} & 1080 & 1058 & 98.0 & 9.59 & 5.52 & 3.50 \\
    \bottomrule
  \end{tabular}
\end{table}

\begin{figure}[t]
  \centering

  % 左图: SR
  \begin{subfigure}[t]{0.42\textwidth}
    \centering
    \includegraphics[width=\textwidth]{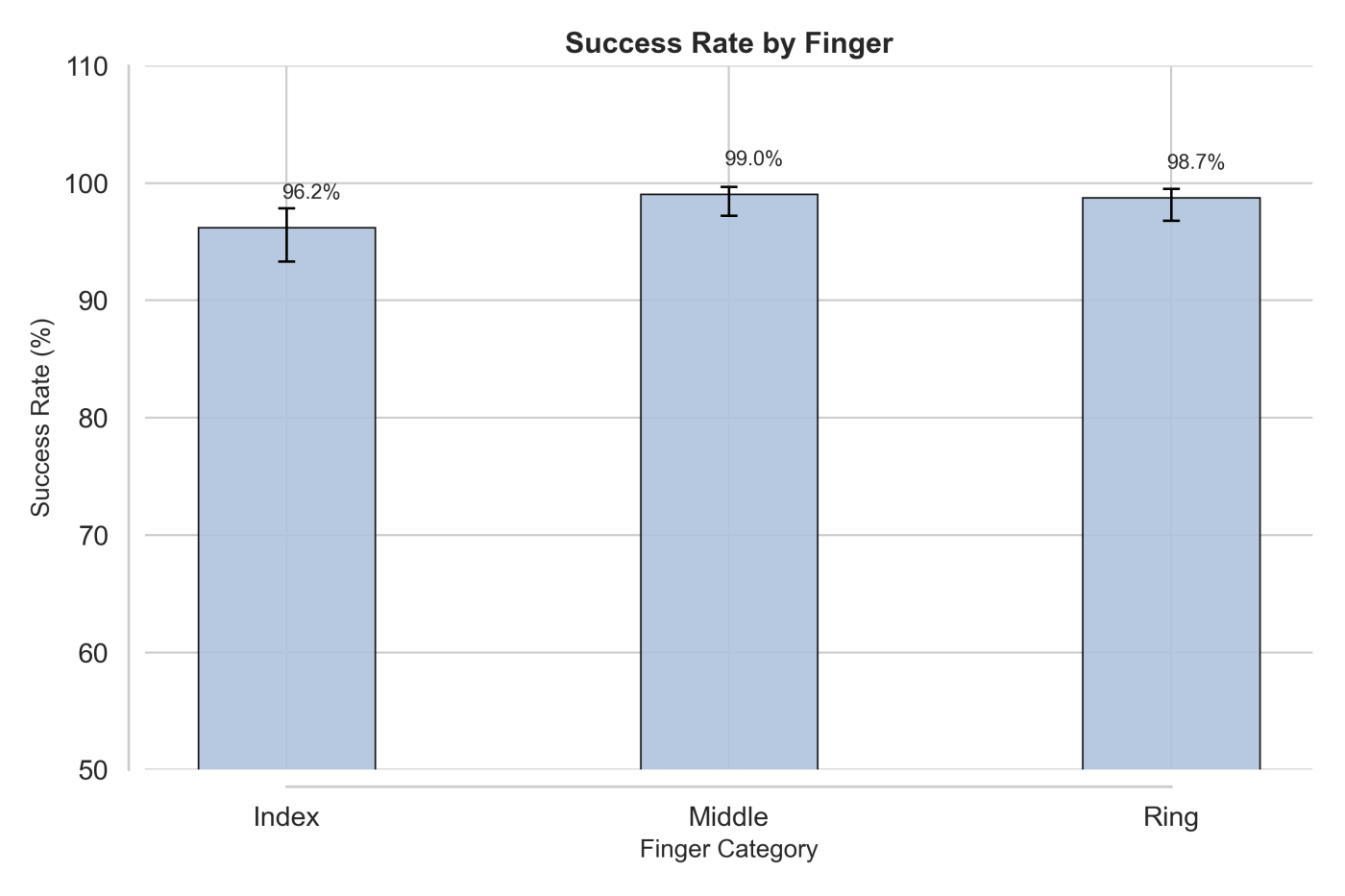}
    \caption{Success rate across three finger groups.}
    \label{fig:us2_sr}
  \end{subfigure}
  \hfill
  % 右图: PST/MT
  \begin{subfigure}[t]{0.42\textwidth}
    \centering
    \includegraphics[width=\textwidth]{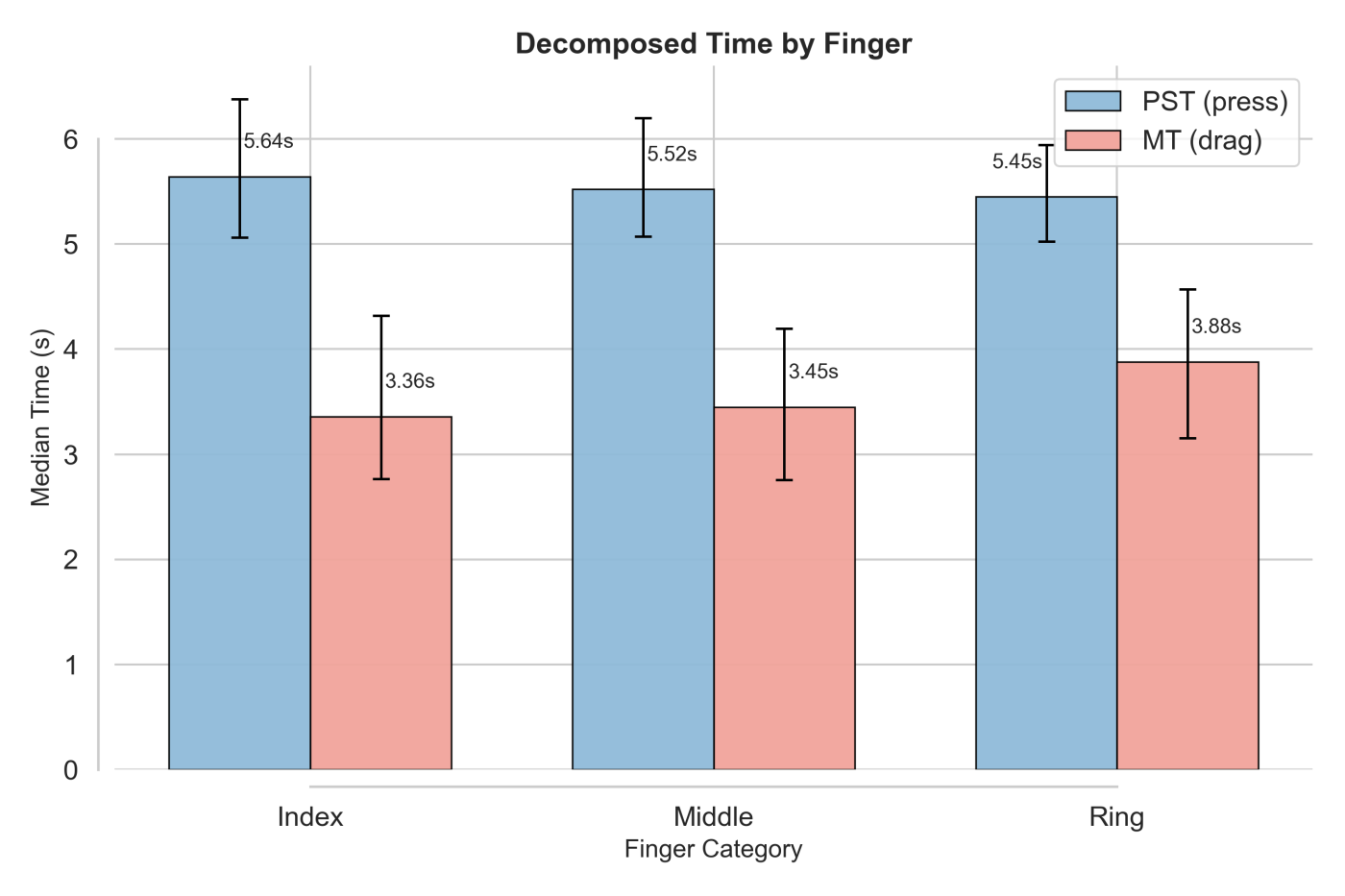}
    \caption{Decomposed completion time across three finger groups.}
    \label{fig:us2_time}
  \end{subfigure}

  \caption{\textbf{Performance summary across all subjects.} 
  (a) Success rate (SR) with 95\% confidence intervals. 
  (b) Median completion time (CT) decomposed into pressure stabilization time (PST) and movement time (MT), each with 95\% confidence intervals, computed on successful trials only.}
\Description{
Two subfigures summarizing performance in the drag-and-press touchpad task.
Panel (a) shows success rates and 95\% confidence intervals for index, middle, 
and ring fingers. Panel (b) shows median completion time decomposed into movement 
time and pressure stabilization time, each with 95\% confidence intervals, for 
the same three finger groups.
}
  \label{fig:sr_mct_split_touchpad}
\end{figure}

We recorded the trial outcome (success/failure), the movement time (MT, s) and the press stabilization time (PST, s). MT was defined as the time from trial onset until the starting ball entered the target ball. 
PST was defined as the time from the moment the starting ball entered the target until the pressure was successfully maintained within the target range for 3\,$\mathrm{s}$. CT was the sum of MT and PST. Figure~\ref{fig:sr_mct_split_touchpad} 
visualizes the median MT, PST, CT and success rate by finger categories. Table~\ref{tab:us2_decompose} summarizes outcomes across finger categories.

\subsubsection{Results}
Overall, participants achieved a high success rate of 98.0\% across all trials, with a median total completion time ($CT_{\text{total}}$) of 9.59\,s.
Notably, this success rate is even higher than in Study~1, despite the drag-and-press task being more challenging; this improvement can be attributed to the longer timeout threshold (15\,s vs.\ 10\,s), which allowed participants more time to stabilize their input. 
The pressure stabilization time (PST) contributed about 5.5\,s and the movement time (MT) about 3.5\,s. 
Performance was consistent across index, middle, and ring fingers, with only minor differences.

In terms of success rate, the middle finger (99.0\%) and the ring finger (98.7\%) performed similarly, while the index finger was slightly lower (96.2\%). 
For pressure stabilization, all three fingers were nearly identical (PST $\approx$ 5.5\,s), 
indicating that the pressure control functionality of \wristp{} works reliably and uniformly across fingers. 
For movement, however, the ring finger required the longest MT (3.88\,s), while the index finger was fastest (3.36\,s), consistent with their relative flexibility. 
The middle finger balanced both high success rate and short MT (3.45\,s), achieving the best overall performance. 
This advantage may stem from its position in the fitted interaction plane, where its movements are most naturally aligned with the plane's horizontal and vertical axes. 

\noindent\textit{Takeaway.} 
All three fingers enabled stable drag-and-press performance with \wristp{}.
While pressure control was equivalent across fingers, drag performance showed finger-specific variation: 
the index benefited from flexibility, the ring was slower due to reduced dexterity, and the middle finger provided the best balance. 
These results suggest that, for pressure-sensitive touchpad applications, the middle finger is a reasonable default choice, balancing speed and accuracy.

This study demonstrates the practical value of \wristp{} as a \emph{virtual pressure-sensitive touchpad}, 
allowing users to perform cursor manipulation and pressure-based input without relying on physical touchpads or mice, 
thereby enabling more flexible and portable interaction scenarios.  

\subsection{Subject Feedback}
\label{subjective_feedback}
After completing the three lab studies, participants were asked to fill out a subjective questionnaire designed to compare their experiences using \wristp{} with those of traditional mouse and touchpad devices. They were also asked to briefly discuss their overall experience with \wristp{}. The questionnaire was based on a 7-point Likert scale (1 = "Strongly Disagree", 7 = "Strongly Agree") and focused on evaluating the interaction experience across the following 
five dimensions: 1) comfort (whether prolonged use of the system led to fatigue); 2) ease of use (whether participants were able to quickly master the interaction method); 3) intuitiveness (whether it aligned with human instinctive responses or natural thinking processes); 4) self-efficacy (whether participants felt confident in completing their intended tasks using this interaction method); 5) overall evaluation (participants' holistic assessment of the interaction experience).

\begin{figure}[t]
    \centering
    \includegraphics[width=0.4\textwidth]{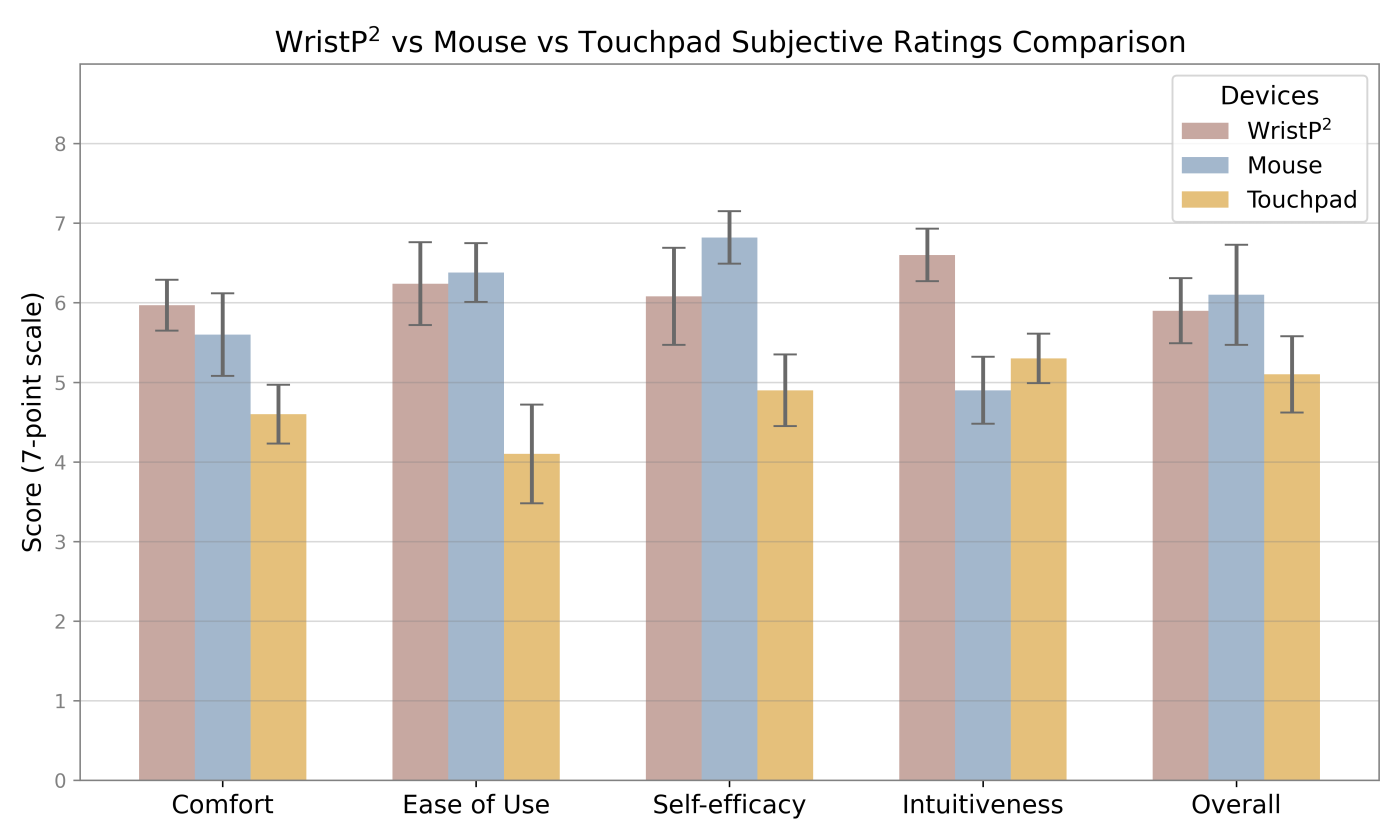}
    \caption{The subjective ratings for different input methods.}
\Description{
Bar chart showing subjective ratings for three input methods across five 
dimensions: comfort, ease of use, intuitiveness, self-efficacy, and overall 
evaluation. WristP achieves higher scores than the laptop touchpad on all 
dimensions, and matches or exceeds the mouse on most metrics.
}
    \label{fig:Feedback}
\end{figure}

The results of the questionnaire are shown in Fig.~\ref{fig:Feedback}. We conducted statistical analysis of the questionnaire data using the Wilcoxon signed-rank test. The results indicate that \wristp{} outperformed the laptop touchpad on all dimensions, exceeded the mouse in the intuitiveness dimension $(Z = -3.45, \, p < 0.01)$, and showed no significant difference compared to the mouse in the other dimensions. 

In follow-up interviews, participants frequently highlighted the intuitiveness and naturalness of interacting with \wristp{}. Six participants felt that mid-air cursor control with their fingers better matched proprioception than using a touchpad or mouse; seven appreciated the continuous pressure control, especially for tasks such as adjusting sliders; and five believed that integrating \wristp{} into AR/VR could significantly enhance immersion. One participant also noted that resting the wrist while moving a single finger felt easier than using a laptop touchpad. Across tasks, participants’ speeds improved by 5.7\%, 10.2\%, and 14.8\% from the first to the second half of the session, suggesting rapid learning.
\subsection{Discussion}
Across the three lab studies we evaluated \wristp{} as (i) a mid-air cursor device (Study~1), (ii) a multi-finger pressure controller (Study~2), and (iii) a virtual pressure-sensitive touchpad (Study~3). Study~1 showed that while a conventional mouse remained fastest, the virtual air mouse achieved touchpad-level throughput, indicating that \wristp{} can support cursor movement with efficiency close to everyday laptop interaction, albeit with a higher intercept due to 3D hand motion and click confirmation. Study~2 and Study~3 demonstrated that participants could reliably reach and maintain target pressures across single- and multi-finger combinations (overall SR $=86.7\%$ in Study~2 and $98.0\%$ in Study~3).

These findings indicate that \wristp{} affords pose and pressure estimation that is sufficient for both efficient pointing and stable pressure targeting without a physical touch surface. The touchpad-level throughput suggests that \wristp{} can act as a practical alternative when a mouse or touchpad is unavailable (e.g., mobile or standing work, large displays, or XR with table proxies), while the robustness of single- and multi-finger pressure control supports pressure-sensitive shortcuts, sliders, and confirmations on uninstrumented surfaces. Finger-specific trade-offs observed in Study~3 also point to simple defaults, with the middle finger offering a good balance between speed and stability, and the index and ring fingers better suited to fast motion and more relaxed targets, respectively.

Taken together with the large-display application study (Sec.~\ref{sec:STUDY 4}), these results suggest that \wristp{} is a portable and precise input modality for both pointing and pressure control.

\section{Application Scenarios}
\label{sec:application}
\begin{figure}[htbp]
  \centering
  \includegraphics[width=0.45\textwidth]{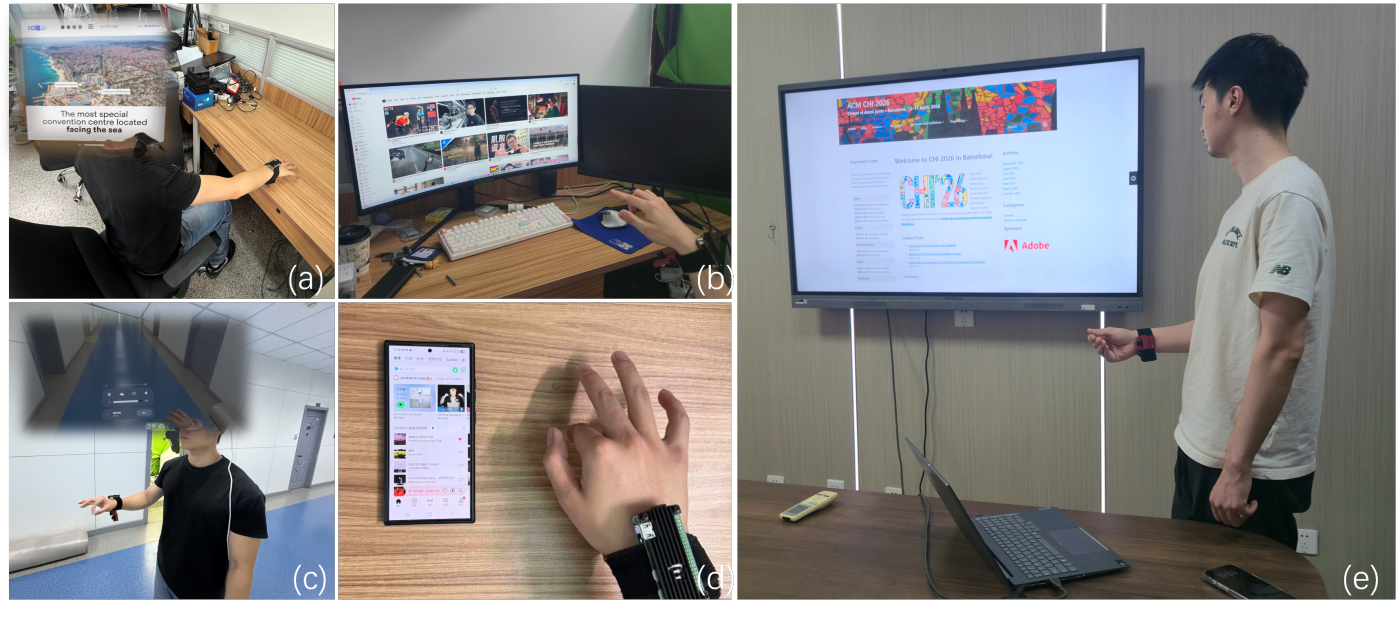}
  \caption{\textbf{Example application scenarios for \wristp{}.} 
  (a) Planar virtual touchpad input in extended reality (XR). 
  (b) Air pointing for desktop/PC. 
  (c) Mid-air gesture input in XR. 
  (d) Virtual touchpad input for mobile device control. 
  (e) Controlling slide presentations on a large screen.}
  \label{fig:application-scenarios}
  \Description{(a) XR planar virtual touchpad input; (b) air pointing for desktop; 
  (c) XR mid-air gesture input; (d) planar touchpad input for mobile control; 
  (e) controlling slide presentations on a large display.}
\end{figure}

Beyond our controlled studies (Section~\ref{sec:user-study}), we illustrate how \wristp{} can be embedded into everyday workflows and evaluate one such scenario in more depth (Study~4). As shown in Fig.~\ref{fig:application-scenarios}, \wristp{} supports planar virtual touchpad input and mid-air gestures in XR, air pointing and virtual touchpad control on desktop and mobile devices, and large-display input. These examples show how \wristp{} can span XR, desktop, mobile, and large-display scenarios.

\subsection{Application Study: Real-World Large-Display Input}
\label{sec:STUDY 4}
To move beyond lab-style tasks and assess one of the large-display scenarios in Fig.~\ref{fig:application-scenarios}(e), we conducted a fourth study in a living-room-style setup, comparing \wristp{} against head-mounted, camera-based hand tracking in a Whac-A-Mole game.
\subsubsection{Participants}
This experiment recruited 12 new participants (2 females, aged 20-26, $M = 23.7$), none of whom had prior experience with the \wristp{} system.
\subsubsection{Design and Procedure}

\begin{figure}[t]
  \centering
  \includegraphics[width=\linewidth]{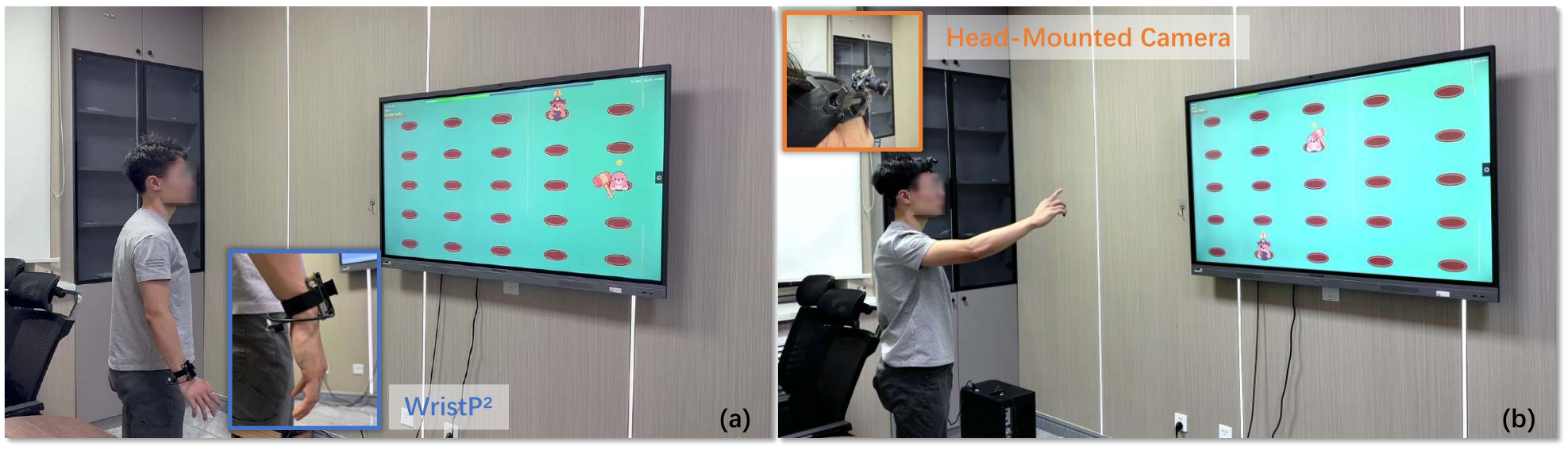}
  \caption{
    \textbf{Whac-A-Mole task setup.} 
  (a) Participant playing the Whac-A-Mole game using \wristp{}. 
  (b) Participant playing the Whac-A-Mole game using a custom head-mounted camera rig.
  }

\Description{
Illustration of the large-display Whac-A-Mole study setup: a participant stands 
in front of a 65-inch screen at a 2–2.5\,m distance, using hand-pose tracking to 
control an on-screen cursor in a living-room-style environment.
}
  \label{fig:mole-setup}
\end{figure}

As shown in Fig.~\ref{fig:mole-setup}, participants stood approximately 2--2.5\,m from a 65-inch display and controlled an on-screen cursor via hand-pose tracking. We compared \wristp{} with two representative monocular vision-based hand-tracking methods—MediaPipe~\cite{lugaresi2019mediapipe} and WiLoR~\cite{WiLoR_2025_CVPR}—both configured with a head-mounted RGB camera to emulate typical AR/VR usage. Detailed descriptions of the baselines are provided in the Appendix~\ref{sec:appendix_eval}.

All three methods rely on monocular RGB input and were integrated into a unified cursor-mapping pipeline, in which the on-screen cursor was controlled using the index-fingertip position under identical control logic. This design ensures that performance differences can be attributed primarily to hand-tracking quality rather than inconsistencies in input-mapping strategies.

The game board consisted of a 5$\times$5 grid of target locations. On each trial, either a mole or a bomb appeared at a randomly chosen cell. Participants were instructed to hit moles as quickly as possible while avoiding bombs; hitting a mole awarded $+1$ point, whereas hitting a bomb incurred a $-5$ point penalty.

Each input method was tested in a single 5-minute game session. The order of methods was counterbalanced across participants, and a 5-minute rest interval was provided between sessions to mitigate fatigue. Game difficulty (e.g., target timing and speed) and session duration were kept constant across conditions to enable a fair comparison between the three input methods.

We aim to answer the following research questions:

\textbf{RQ1 (Performance):} In a real-world large-display task, can \wristp{} achieve better performance (e.g., higher success, shorter time, lower error rate) than head-view approaches?

\textbf{RQ2 (Effort \& Fatigue):} Over prolonged use, does \wristp{} reduce perceived workload and fatigue compared to head-view approaches?

\subsubsection{Measures and Analysis}

In the Whac-A-Mole task, we evaluated task performance using four quantitative metrics: hit rate (\textbf{HR}), the proportion of successful mole hits; error rate (\textbf{ER}), the proportion of bomb targets mistakenly hit; reaction time (\textbf{RT}), the time elapsed between a mole's appearance and its hit; and the mean game score (\textbf{Score}). 

After completing all three experimental conditions, participants completed a short subjective questionnaire and were asked to briefly discuss their experiences. Beyond the subjective measures described in Section 7.6, we additionally incorporated a social acceptability measure, evaluating participants' willingness to use the device in public settings.
\subsubsection{Results}

For \textbf{RQ1 (Performance)}, Table~\ref{tab:performance} shows that \wristp{} outperformed both head-view baselines on all metrics: it achieved the highest hit rate and game score (HR $= 94.47\%$, Score $= 78.92$), the lowest error rate (ER $= 2.07\%$), and the shortest reaction time (RT $= 1.43$\,s), with smaller standard deviations. Relative to the two head-mounted methods, this corresponds to an HR gain of about 8--11 percentage points, roughly one-third to one-half lower ER, and 35--40\% faster RT, yielding a 30--40\% score increase. These gains mainly reflect differences in tracking quality: head-mounted cameras observe the hand from 0.5\,m away, where the hand occupies relatively few pixels and is more affected by self-occlusion, whereas \wristp{} maintains a close, wrist-anchored view that produces smoother fingertip trajectories and fewer tracking dropouts. This directly improves rapid target acquisition in the Whac-A-Mole task and answers RQ1 in favor of \wristp{}.

\begin{figure}[t]
    \centering

    % Left: Reaction time violin plot
    \begin{subfigure}[t]{0.38\textwidth}
        \centering
        \includegraphics[width=\textwidth]{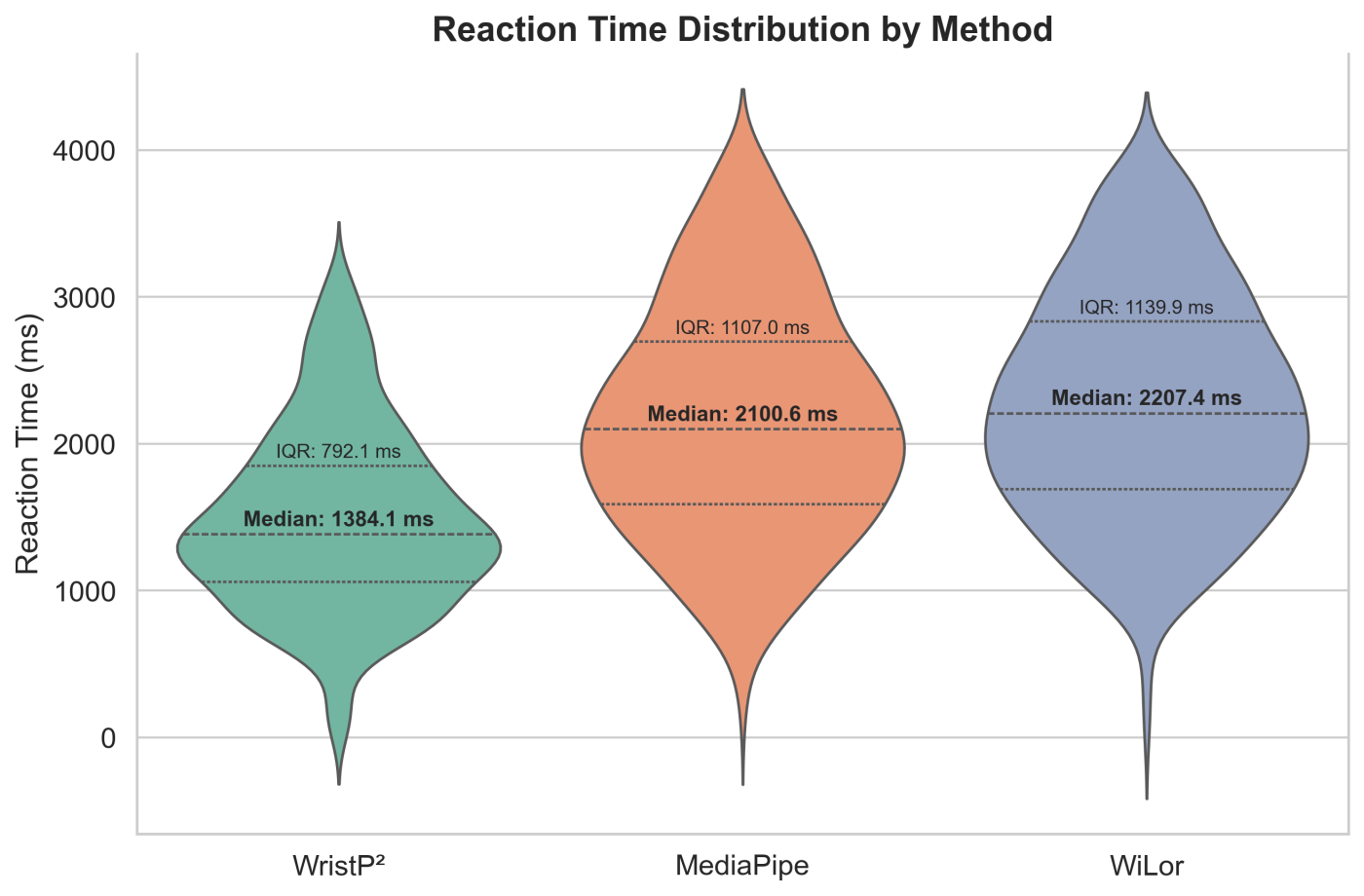}
        \caption{Reaction time distributions across input methods.}
        \Description{
Violin plots showing the distribution of reaction times for three input methods, 
with WRISTP$^2$ exhibiting the lowest median and smallest spread.
}
        \label{fig:reaction_time_violin}
    \end{subfigure}
    \hfill
    % Right: Subjective ratings
    \begin{subfigure}[t]{0.45\textwidth}
        \centering
        \includegraphics[width=\textwidth]{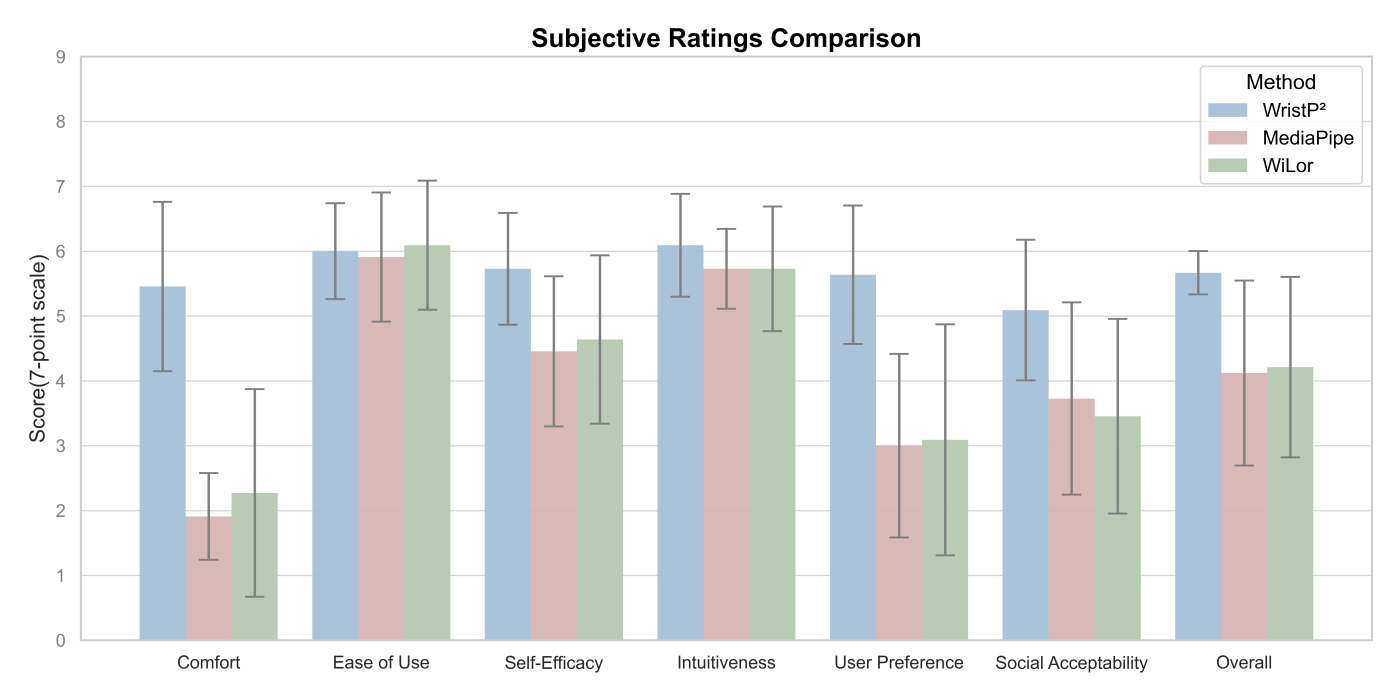}
        \caption{Subjective ratings for different input methods.}
        \Description{
Bar chart comparing subjective questionnaire ratings across three input methods, 
showing comfort, ease of use, and other factors, with WRISTP$^2$ rated highest overall.
}
        \label{fig:Feedback-b}
    \end{subfigure}

    \caption{\textbf{Evaluation results for the Whac-A-Mole study comparing WRISTP$^2$ with head-mounted baselines (MediaPipe, WiLoR).} (a) WRISTP$^2$ achieves \textbf{significantly lower reaction time} (mean 1.43s) and lower variance. (b) Subjective ratings indicate that WRISTP$^2$ induces \textbf{lower fatigue (better Comfort)} and is perceived as easier to use compared to head-view approaches. Error bars denote 95\% CI.}
    \label{fig:reaction_subjective_combined}
\end{figure}

For \textbf{RQ2 (Effort \& Fatigue)}, the configurations differ in ergonomics. Head-mounted tracking requires users to keep the hand within a camera frustum tied to head pose, often inducing larger arm movements to stay in view, while \wristp{} decouples hand input from head orientation and allows small wrist and finger motions in comfortable hand-down postures. Subjective ratings (Fig.~\ref{fig:Feedback-b}) are consistent with this: participants reported lower fatigue and workload and a clear preference for \wristp{} over the head-view conditions. Together, the objective gains and subjective feedback indicate that \wristp{} can reduce perceived effort and fatigue during multi-minute large-display use.

Furthermore, in follow-up interviews, all participants emphasized the smaller movement range and more discreet interaction style afforded by \wristp{}, and expressed greater willingness to use it in public settings—consistent with the subjective social acceptability ratings.
\begin{table}[t]
\centering
\caption{Task performance comparison (mean $\pm$ SD)}
\label{tab:performance}

\small
\setlength{\tabcolsep}{3pt}
\renewcommand{\arraystretch}{1.12}

\begin{tabular}{lcccc}
\toprule
Method & HR$\uparrow$\% & ER$\downarrow$\% & RT$\downarrow$s & Score$\uparrow$ \\
\midrule
MediaPipe \cite{lugaresi2019mediapipe} & $86.51 \pm 9.79$ & $3.04 \pm 2.72$ & $2.19 \pm 0.21$ & $61.17 \pm 12.66$ \\
WiLoR \cite{WiLoR_2025_CVPR}           & $83.70 \pm 8.43$ & $4.57 \pm 3.83$ & $2.29 \pm 0.20$ & $55.58 \pm 12.21$ \\
\textbf{\wristp{} (ours)}             & $\mathbf{94.47 \pm 3.64}$ & $\mathbf{2.07 \pm 0.92}$ & $\mathbf{1.43 \pm 0.10}$ & $\mathbf{78.92 \pm 8.30}$ \\
\bottomrule
\end{tabular}
\end{table}

\section{Limitations and Future Work}
Although \wristp{} demonstrates strong performance in 3D hand pose and pressure estimation, it still has certain limitations and areas that warrant further investigation.

\textit{Hand-Object Interaction.}
At present, our work focuses on estimating contact and pressure between the hand and rigid planar or quasi-planar surfaces.
However, everyday interaction frequently involves grasping, squeezing, or stabilizing objects of diverse shapes and compliances, where pressure is distributed across curved contact patches and between fingers.
Extending \wristp{} to such hand-object scenarios would enable richer applications such as pressure-aware manipulation, adaptive grip control, and safety-critical feedback (e.g., avoiding excessive force on fragile or deformable items).
In future work, we plan to expand our dataset by incorporating more object-centric and within-hand interactions, and to estimate pressure both between fingers (e.g., during pinch or clasping) and between the hand and arbitrary objects.

\textit{Visual Obstruction.}
When performing extreme wrist rotations or holding bulky objects, some fingers may be partially or fully occluded from the wrist-view camera.
A natural next step is to fuse the wrist view with complementary perspectives such as head-mounted or environment-mounted cameras, or with non-visual signals (e.g., EMG, acoustics), so that the system can remain robust when visual coverage is incomplete.

\textit{Bimanual Occlusion.}
Many real-world tasks involve coordinated use of both hands, which often leads to mutual occlusion.
While some recent methods~\cite{MeMaHand,meng20223d} can predict bimanual hand poses from external viewpoints in such cases, there is still limited work in HCI that addresses bimanual interaction from ego-centric perspectives.
Our current dataset and model focus on a single hand; in future work, we plan to collect bimanual interaction data and extend the model to jointly reason about two hands, enabling pressure-aware collaboration such as two-handed manipulation and cooperative gestures.

\textit{Form Factor and On-Device Deployment.}
Although our prototype achieves wireless deployment and relies on compact off-the-shelf components, the current form factor and power profile still lag behind those of commercial wearable devices that integrate low-power camera modules~\cite{BrilliantLabs2025,tonsen2020high,TapXR1}.
The under-wrist camera and electronics can still interfere with certain tabletop postures, and real-world long-term use will require slimmer packaging and tighter integration into form factors such as smart wristbands or watches.
Achieving this will likely require custom hardware (e.g., higher-integration SoCs, flexible PCBs, folded optics) and more stringent power management.
In addition, the current model runs at about 22\,fps on an NVIDIA RTX4060 GPU, which may limit interaction smoothness and makes direct deployment on low-power wearables challenging.
Future work could explore model compression, architecture simplification, and adaptive sampling strategies to support real-time, on-device inference under mobile power and compute budgets.

\section{Conclusion}

We presented \wristp{}, a wrist-worn, camera-based system that reconstructs a full 3D hand mesh and per-vertex pressure from a single RGB frame. Trained on a large wrist-view dataset with high-fidelity pose and pressure annotations, our model achieves low-millimeter pose error and reliable contact and pressure estimation across diverse lighting and surfaces. Four studies show that \wristp{} delivers touchpad-level pointing efficiency, robust multi-finger pressure control, and, in a large-display Whac-A-Mole task, higher task success with lower arm fatigue than head-mounted hand tracking, positioning \wristp{} as a practical wristband platform for pressure-aware interaction on uninstrumented surfaces and in mid-air.

\bibliographystyle{ACM-Reference-Format}
\bibliography{sample-base}
\appendix

\section{Extended Details About Dataset}
\label{sec:extended_dataset}
In this section, we provide a complete list and detailed definitions of gestures in the \wristp{} dataset in Section~\ref{sec:dataset}.

\subsection{On-plane Gestures}

The on-plane gestures we designed include 48 actions, consisting of common actions using pressure-sensitive touchpads, as well as additional complex actions to increase the diversity of touch data. The descriptions of these gestures are shown in Table~\ref{tab:pressure_series_gestures} below:
\begin{table*}[htbp]
\centering
\caption{On-plane Series Gestures}
\resizebox{\textwidth}{!}{
\begin{tabular}{|c|l|c|l|}
\hline
\textbf{No.} & \textbf{Action Description} & \textbf{No.} & \textbf{Action Description} \\
\hline
1 & Thumb pulp press & 2 & Thumb tip press \\
\hline
3 & Middle pulp press & 4 & Middle tip press \\
\hline
5 & Index pulp press & 6 & Index tip press \\
\hline
7 & Ring pulp press & 8 & Ring tip press \\
\hline
9 & Pinky pulp press & 10 & Pinky tip press \\
\hline
11 & Two-finger pulp press & 12 & Two-finger tip press \\
\hline
13 & Three-finger pulp press & 14 & Three-finger tip press \\
\hline
15 & Four-finger pulp press & 16 & Four-finger tip press \\
\hline
17 & Five-finger pulp press & 18 & Five-finger tip press \\
\hline
19 & Thumb press, move horizontally and vertically & 20 & Index press, move horizontally and vertically \\
\hline
21 & Middle press, move horizontally and vertically & 22 & Ring press, move horizontally and vertically \\
\hline
23 & Pinky press, move horizontally and vertically & 24 & Two-finger press, move horizontally and vertically \\
\hline
25 & Three-finger press, move horizontally and vertically & 26 & Four-finger press, move horizontally and vertically \\
\hline
27 & Five-finger press, move horizontally and vertically & 28 & Thumb press, circular motion \\
\hline
29 & Index press, circular motion & 30 & Middle press, circular motion \\
\hline
31 & Ring press, circular motion & 32 & Pinky press, circular motion \\
\hline
33 & Two-finger light touch, pinch & 34 & Two-finger press, pinch \\
\hline
35 & Five-finger light touch, clench & 36 & Five-finger press, clench \\
\hline
37 & Thumb rotation, continuous press & 38 & Index rotation, continuous press \\
\hline
39 & Middle rotation, continuous press & 40 & Ring rotation, continuous press \\
\hline
41 & Pinky rotation, continuous press & 42 & Thumb roll press \\
\hline
43 & Middle roll press & 44 & Index roll press \\
\hline
45 & Ring roll press & 46 & Pinky roll press \\
\hline
47 & Palm side press & 48 & Palm center press \\
\hline
\end{tabular}
}
\label{tab:pressure_series_gestures}
\end{table*}

%Table S1: Names and standardized definitions of the 48 planar-interaction actions (including start/stop conditions, suggested duration, and failure criteria).
\subsection{Mid-air Gestures}
Mid-air gestures encompass static standard sign language, including 10 American Sign Language (ASL) alphabet letters and 18 common daily interaction gestures. The names and illustrations of these gestures are shown in Fig.~\ref{fig:Mid-air Series Gestures}  below:
%Table S2: Names and standardized definitions of the 28 in-air gestures (including 10 ASL letters and 18 common interaction gestures).
\begin{figure*}[htbp]
    \centering
    \includegraphics[width=\textwidth]{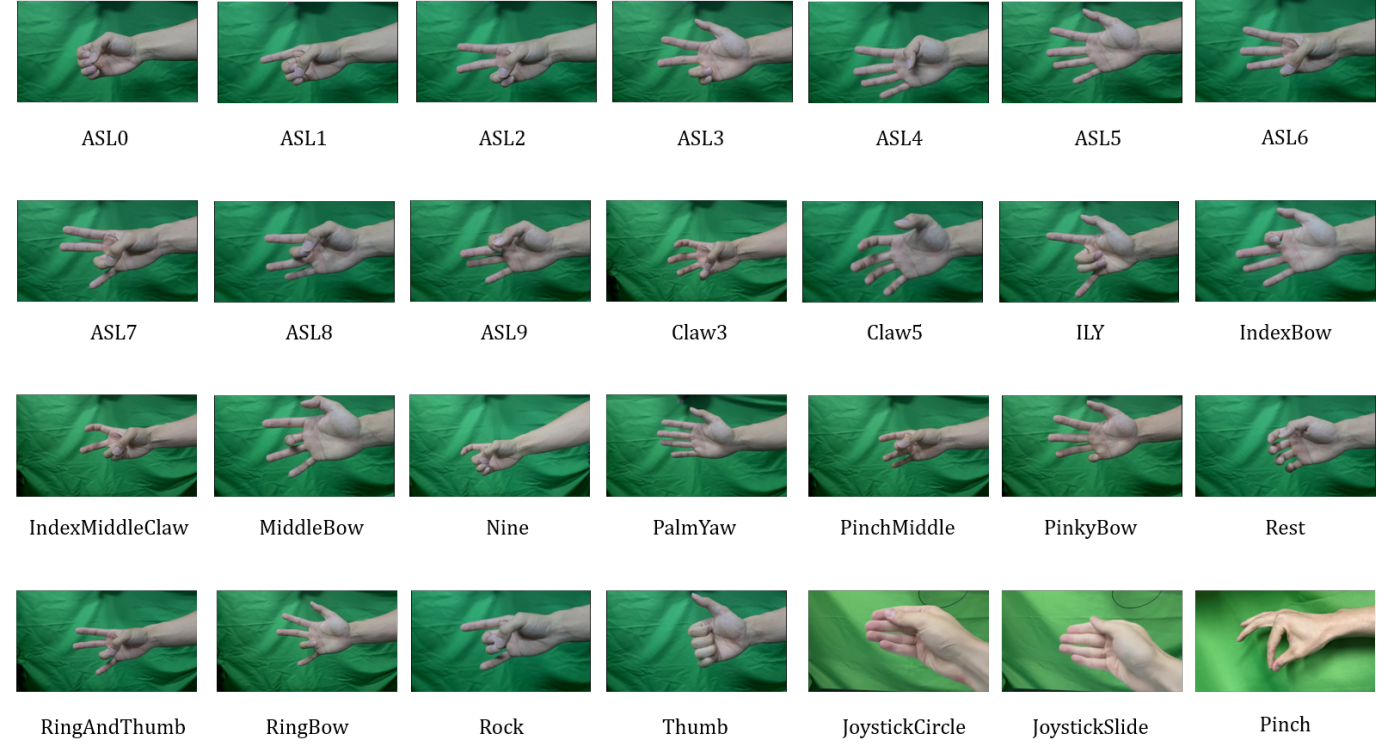}
    \caption{Mid-air Series Gestures}
    \Description{
A grid of images illustrating the set of mid-air gestures used in the study. 
Each gesture shows a hand performing a distinct posture or motion in free space, 
including finger extensions, pinches, spreads, rotations, and multi-finger configurations. 
The gestures represent the full mid-air gesture vocabulary collected for WRISTP$^2$.
}
    \label{fig:Mid-air Series Gestures}
\end{figure*}
\section{Detailed Algorithmic Design}
\subsection{Canonical Hand-Local Coordinate Frame}
As described in Section~\ref{subsec:hand-local-frame}, to provide a consistent reference frame across participants and sessions, we introduce a hand-local canonical coordinate system, which serves as the basis for both annotation optimization and network prediction. Unlike externally measured frames (e.g., world or camera coordinates), this local frame is defined algorithmically from anatomical landmarks on the hand mesh. Specifically, the origin is placed at the wrist joint, the $x$-axis is aligned with the vector from the wrist to the index MCP joint, and the palm normal---computed as the cross product between the wrist--index vector and the wrist--pinky vector---defines the $z$-axis. The $y$-axis is then derived to complete a right-handed orthogonal system. We additionally enforce $\det(R)=1$ and flip the palm normal if necessary to ensure consistent orientation across subjects and sessions.

We denote this canonical hand-local frame as $\{H\}$, the world frame as $\{W\}$, and the camera frame as $\{C\}$. A point $\mathbf{x}_W$ expressed in the world frame is mapped into $\{H\}$ by
\[
\mathbf{x}_H = \mathbf{T}^{H}_{W}\,\mathbf{x}_W, \quad 
\mathbf{T}^{H}_{W} = 
\begin{bmatrix}
R_{H \leftarrow W} & t_{H \leftarrow W} \\
\mathbf{0}^\top & 1
\end{bmatrix},
\]
where $R_{H \leftarrow W}$ and $t_{H \leftarrow W}$ denote the rotation and translation from the world to the hand-local canonical frame. Similarly, the wrist-mounted camera extrinsics are expressed with respect to $\{H\}$ as $\mathbf{T}^{C}_{H} = [R_{C \leftarrow H}\mid t_{C \leftarrow H}]$, enabling standard projection into $\{C\}$.

\emph{Rationale.} Performing optimization and prediction in this canonical hand-local frame removes global translation and rotation degrees of freedom, thereby improving numerical stability and convergence. It also normalizes all subjects into a unified representation, facilitating cross-participant alignment and enhancing generalization. Finally, canonicalizing meshes in $\{H\}$ ensures that our Mesh-VQ-VAE quantization and multitask predictor operate on geometry that is invariant to camera placement or donning/doffing variations, yielding more robust supervision and transferability.

\subsection{Shape Parameter Calibration (MANO \texorpdfstring{$\beta$}{beta})}
In this section, we describe in detail how to apply the approach of \cite{karunratanakul2023harp} to calibrate the MANO parameter $\beta$ for participants' hands before the annotation optimization step. We first calibrate $\beta$ by optimizing them using the following composite objective. This calibration ensures that the hand mesh is closely aligned with the ground-truth markers in the world frame, which is crucial for reducing the optimization difficulty in later stages. The resulting calibrated shape parameters $\beta^*$ are fixed for subsequent optimization of the pose parameters $\theta$ and the rigid transformations.
\begin{equation}
\label{eq:shape-loss}
\begin{aligned}
\mathcal{L}_{\text{shape}}(\beta,\theta)
&= \omega_{\!3\text{D}}\,\mathcal{L}^{3\text{D}}_{\text{marker}}
 + \omega_{\!2\text{D}}\,\mathcal{L}^{2\text{D}}_{\text{marker}}
 + \omega_{\text{mask}}\,\mathcal{L}_{\text{mask}}
\\
&\quad+ \omega_{\text{anat}}\,\mathcal{L}_{\text{anat}}(\beta,\theta)
 + \omega_{\text{reg}}\,\mathcal{L}_{\text{reg}}(\beta,\theta).
\end{aligned}
\end{equation}

Here, $\mathcal{L}^{3\text{D}}_{\text{marker}}$ enforces 3D consistency between mocap markers in the world coordinate system and the corresponding mesh vertices in the hand-local frame, while $\mathcal{L}^{2\text{D}}_{\text{marker}}$ provides projection-consistency supervision in the image plane. The marker losses primarily constrain \emph{bone lengths} and skeletal proportions, whereas the mask term $\mathcal{L}_{\text{mask}}$ constrains the \emph{global hand shape/fatness} via silhouette alignment.

Let $J^{\text{gt},W}_{m}\in\mathbb{R}^{N_J\times 3}$ be the 3D ground-truth marker coordinates in the world frame, and let $[R_{\text{kine}}\mid t_{\text{kine}}]$ and $K$ denote the Kinect extrinsics and intrinsics, respectively. Denote by $\mathrm{MANO}(\beta,\theta)$ the differentiable MANO layer producing the hand mesh in the hand-local frame and by $J^{\text{pred},C}_{v}$ the corresponding 3D mesh joint locations. We define the conversion between world and local coordinate systems as follows:
\begin{align}
J^{\text{gt},C}_{m} &= R_{\text{kine}}\,J^{\text{gt},W}_{m} + t_{\text{kine}},\\[2pt]
\mathcal{L}^{3\text{D}}_{\text{marker}}(\beta,\theta) &= \frac{1}{N_J}\sum_{j=1}^{N_J}\big\|\,J^{\text{pred},C}_{v,j} - J^{\text{gt},C}_{m,j}\,\big\|_2^2,\\[2pt]
\mathcal{L}^{2\text{D}}_{\text{marker}}(\beta,\theta) &= \frac{1}{N_J}\sum_{j=1}^{N_J}\big\|\,\Pi_{K}\!\left(J^{\text{pred},C}_{v,j}\right) - \Pi_{K}\!\left(J^{\text{gt},C}_{m,j}\right)\,\big\|_2^2,
\end{align}
where $\Pi_{K}(\cdot)$ denotes projection with intrinsics $K$.

Note that the 3D marker loss involves the ground-truth markers in the world frame and the corresponding mesh vertices in the local frame. To obtain the world-to-local transformation matrix, we use the coordinate system conversion function to map the ground-truth markers into the hand-local frame. However, this transformation matrix is not directly applicable for converting the predicted mesh from the hand-local frame to the world frame. Instead, we use this computed matrix as an initial guess for optimization, and then maintain a delta transformation that includes a delta rotation matrix and a delta translation vector. This optimization process refines the transformation matrix, which is then used to project the predicted mesh into the world frame for final supervision against the 3D marker ground truth.

The binary hand masks are obtained by a two-stage pipeline (YOLO detection followed by SAM2 segmentation). Let $M_{\text{gt}}$ be the resulting ground-truth mask and $\tilde{M}(\beta,\theta)$ the predicted silhouette rendered by projecting the mesh produced by $\mathrm{MANO}(\beta,\theta)$. We use a BCE+Dice hybrid loss:
\begin{equation}
\mathcal{L}_{\text{mask}}(\beta,\theta)
= \lambda_{\text{bce}}\;\mathrm{BCE}\!\left(\tilde{M},\, M_{\text{gt}}\right)
+ \lambda_{\text{dice}}\;\Big(1 - \mathrm{Dice}\!\left(\tilde{M},\, M_{\text{gt}}\right)\Big).
\end{equation}
The anatomical prior $\mathcal{L}_{\text{anat}}(\beta,\theta)$ follows MANO's joint-limit and plausibility constraints, and $\mathcal{L}_{\text{reg}}(\beta,\theta)$ regularizes parameter magnitudes. After convergence, the calibrated shape $\beta^{*}$ is fixed for the subsequent stages of our pipeline.

We use WiLoR \cite{WiLoR_2025_CVPR} to process the calibration sequence of the hand image to provide the initial values $\beta_{\text{init}}$ and  $\theta_{\text{init}}$. The calibrated shape parameters $\beta^{*}$ will be used as fixed values for the subsequent optimization of this participant.

\subsection{Pre-training Dataset Construction and Statistics (for GNN initialization and Hand-VQ-VAE)}
\begin{figure}[htbp]
    \centering
    \includegraphics[width=0.35\textwidth]{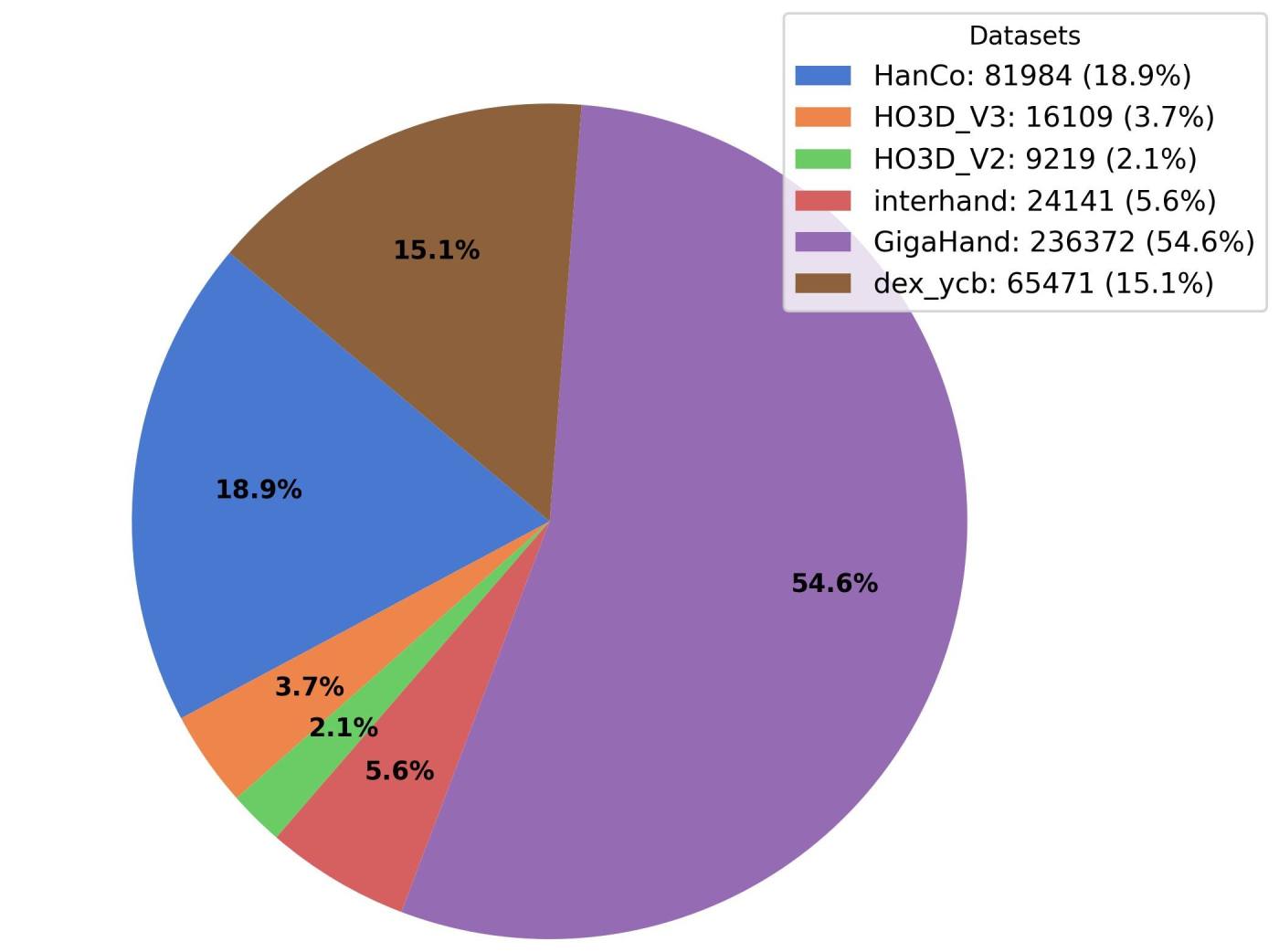}
    \caption{Distribution of various datasets used for GNN initialization and Hand-VQ-VAE prior construction.}
    \Description{
A bar chart showing the size distribution of multiple hand-related datasets. 
Each bar corresponds to a different dataset used for GNN initialization and for training 
the Hand-VQ-VAE prior, illustrating the variation in dataset scale.
}
    \label{fig:data_set_distribution}
\end{figure}

The pie chart in Fig.~\ref{fig:data_set_distribution} represents the distribution of various datasets used for GNN initialization and Hand-VQ-VAE prior construction. The total number of samples across all datasets is 433,296. The largest portion of the data comes from the GigaHand dataset, contributing 54.6\% (236,372 samples). Other significant datasets include HanCo (18.9\%, 81,984 samples) and DexYCB(15.1\%, 65,471 samples). Smaller contributions come from Interhand (5.6\%, 24,141 samples), HO3D V2 (2.1\%, 9,219 samples), and HO3D V3 (3.7\%, 16,109 samples).
\subsection{Fisheye Background Replacement Data Augmentation Details}

To introduce diversity into the background and improve the model's generalization across different environments, we design a background replacement strategy. The algorithm first randomly selects a background image \( I_{\text{pano}} \) from a pre-downloaded 360° panoramic image database \( \mathcal{D}_{\text{pano}} \), which contains over 1550 high-quality indoor panorama images. To simulate different viewpoints, a random 3D rotation matrix \( R_{\text{view}} \) is generated. This step ensures that the background is observed from various angles, enriching the training data diversity.

The selected panoramic image \( I_{\text{pano}} \) is then transformed into a fisheye perspective based on the calibrated fisheye camera model \( \mathcal{C}_{\text{fish}} \) and the virtual camera pose \( R_{\text{view}} \). The projection transformation process is as follows:

\subsubsection{(Ray Generation for Fisheye View:}  
For each pixel \( (u_{\text{fish}}, v_{\text{fish}}) \) of the target fisheye image (with dimensions \( H_{\text{out}} \times W_{\text{out}} \)), the inverse projection function \( \mathcal{P}^{-1} \) of the fisheye camera model is used to compute the unit observation ray \( \vec{d}_{\text{cam}}(u_{\text{fish}}, v_{\text{fish}}) \) in the camera's 3D coordinate system:
\[
\vec{d}_{\text{cam}}(u_{\text{fish}}, v_{\text{fish}}) = \mathcal{P}^{-1}(u_{\text{fish}}, v_{\text{fish}}, 1)
\]
The collection of all rays forms the ray bundle \( \mathbf{D}_{\text{cam}} \in \mathbb{R}^{H_{\text{out}} \times W_{\text{out}} \times 3} \).

\subsubsection{Ray Bundle Transformation:}  
The ray bundle \( \mathbf{D}_{\text{cam}} \) in the camera's standard coordinate system is then transformed using the randomly generated rotation matrix \( R_{\text{view}} \) to obtain the ray bundle \( \mathbf{D}_{\text{world}} \) in the panoramic world coordinate system. The transformation for each pixel's corresponding ray is:
\[
\vec{d}_{\text{world}}(u_{\text{fish}}, v_{\text{fish}}) = R_{\text{view}} \vec{d}_{\text{cam}}(u_{\text{fish}}, v_{\text{fish}})
\]

\subsubsection{World Coordinates to Spherical Coordinates Conversion:}  
For each ray \( \vec{d}_{\text{world}} = (x_w, y_w, z_w)^T \), it is converted into spherical coordinates \( (\lambda, \phi) \), where:
\[
\lambda = \operatorname{atan2}(x_w, z_w)
\]
\[
\phi = \operatorname{asin}\left(\frac{y_w}{\|\vec{d}_{\text{world}}\|_2}\right)
\]
Here, \( \lambda \) is the longitude (azimuth) and \( \phi \) is the latitude (elevation).

\subsubsection{Spherical Coordinates to Panoramic Image Mapping:}  
The spherical coordinates \( (\lambda, \phi) \) are mapped to the 2D pixel coordinates \( (u_{\text{pano}}, v_{\text{pano}}) \) in the selected panoramic image \( I_{\text{pano}} \) using equirectangular projection:
\[
u_{\text{pano}} = \left(\frac{\lambda}{2\pi} + \frac{1}{2}\right) \cdot (W_{\text{pano}} - 1)
\]
\[
v_{\text{pano}} = \left(\frac{1}{2} - \frac{\phi}{\pi}\right) \cdot (H_{\text{pano}} - 1)
\]

\subsubsection{Background Pixel Resampling and Composition:}  
The corresponding pixel colors from the panoramic image \( I_{\text{pano}} \) are extracted using bilinear interpolation at the coordinates \( (u_{\text{pano}}, v_{\text{pano}}) \). The final fisheye background image \( I_{\text{bg\_fish}} \) is obtained by combining the resampled pixels with the hand mask \( M_{\text{hand}} \) and the original hand foreground \( I_{\text{hand}} \):
\[
I_{\text{final}} = I_{\text{bg\_fish}} \odot M_{\text{hand}} + I_{\text{hand}} \odot (1 - M_{\text{hand}})
\]
where \( \odot \) denotes element-wise multiplication.

This projection transformation process generates a background-replaced image that is visually consistent with fisheye lens distortion, improving the model's performance in complex real-world scenarios.

\section{Comparison with Baseline Methods}
\label{sec:supp-baseline}

In this section, we present additional comparisons between \wristp{} and
existing methods for 3D hand pose reconstruction and
contact/pressure estimation.
All numbers are reported on the held-out test split consisting of
sequences from five subjects for whom we collected synchronized
head-mounted and environment-mounted video, along with full 3D hand pose
and pressure annotations.
All methods are evaluated on the same test split as \wristp{} without
any additional training or fine-tuning.
Our goal is to systematically compare wrist-mounted sensing against
state-of-the-art head-mounted and environment-mounted approaches, and to
analyze their robustness under different illumination.
We acknowledge that fine-tuning these baselines on our dataset would yield a fairer comparison, and we leave this investigation to future work.
\subsection{Experimental Setup for Baseline Methods}
\label{sec:appendix_eval}
\paragraph{Baselines.}
We consider three representative baselines:
\emph{MediaPipe} \cite{lugaresi2019mediapipe}, \emph{WiLoR} \cite{WiLoR_2025_CVPR}, and \emph{PressureVision++} \cite{pressurevision++}.
\emph{MediaPipe} is a widely used lightweight hand pose estimator in HCI
research and is often employed to obtain proxy ground-truth 2D/3D hand
joint annotations from RGB images.
However, as our experiments show, its 3D accuracy is limited in our
setting, which calls into question the common practice of using its
predictions as training labels for downstream models.
\emph{WiLoR} is a model-based state-of-the-art approach that directly
regresses 3D MANO hand parameters from monocular RGB frames, and
constitutes a strong environment-/head-mounted baseline for pose
reconstruction.

For pressure estimation, although \emph{EgoPressure} \cite{Egopressure} reports strong
single-view hand pressure prediction performance, neither its code nor
its dataset is publicly available.
We therefore adopt its predecessor, \emph{PressureVision++}, as our
baseline.
PressureVision++ takes an RGB image of the hand as input and regresses a
per-pixel pressure map in kPa.
Because our dataset does not provide pixel-wise pressure annotations for
the head-mounted camera (and obtaining such annotations would be
non-trivial), we evaluate PressureVision++ only on binary contact
prediction for both environment and head views, and do not report dense
pressure errors for this method.
For all baselines, we report metrics for both head-mounted and
environment-mounted camera views whenever applicable and further break
down the results by illumination condition.

\paragraph{Evaluation protocol.}
We evaluate 3D hand pose reconstruction in the hand-local frame using
the same metrics as in the main paper:
MPJPE, PA-MPJPE, PVE, PA-PVE, and MJAE.
To study robustness to lighting, we report results under three
illumination conditions (High/Medium/Low, denoted H/M/L) as well as the
overall performance across all conditions.

For contact/pressure estimation we focus on binary contact prediction.
A mesh vertex is considered in contact if the maximum of its per-vertex pressure
exceeds 10\,g; a sample is classified as contact if at least one vertex
is in contact.
Given this ground-truth labeling, we evaluate each method using
Accuracy (Acc), Precision (Prec), Recall (Rec), and F1-score (F1).
Unless otherwise stated, all metrics are reported on the held-out test
set.

\subsection{Quantitative Results}
\label{sec:supp-quantitative}

\paragraph{Pose reconstruction across illumination conditions.}
Table~\ref{tab:supp-pose-env-illum} summarizes pose reconstruction
accuracy under different illumination conditions for the environment
view baselines and our wrist-mounted model.
Across all settings, \wristp{} achieves the lowest errors on MPJPE,
PA-MPJPE, PVE, PA-PVE, and MJAE.

Aggregated over all illumination conditions (\emph{All} row),
\wristp{} reduces MPJPE from 46.8\,mm (MediaPipe) and 15.0\,mm (WiLoR)
to 3.0\,mm, corresponding to roughly $15\times$ and $5\times$ lower
error, respectively.
PA-MPJPE decreases from 23.8\,mm and 9.2\,mm to 2.3\,mm, and MJAE drops
from 36.8$^\circ$ and 13.8$^\circ$ to 3.4$^\circ$.
Similar trends hold for the vertex-based metrics PVE and PA-PVE.

The benefits of wrist-mounted sensing become even more apparent under
Low illumination.
In this regime, \wristp{} attains 3.2\,mm MPJPE and 3.5$^\circ$ MJAE,
while MediaPipe and WiLoR incur 45.2\,mm / 36.2$^\circ$ and
15.4\,mm / 13.8$^\circ$ errors, respectively.

\begin{table*}[t]
  \centering
  \caption{\textbf{Pose accuracy across illumination conditions (env view).}
  Position errors are reported in mm; MJAE in degrees ($^\circ$).
  Lower is better for all metrics. PVE and PA-PVE are not available
  for MediaPipe and are marked as ``--''.}
  \label{tab:supp-pose-env-illum}
  \renewcommand{\arraystretch}{1.2}
  \begin{tabular}{llccccc}
    \toprule
    Illumination & Method & MPJPE & PA-MPJPE & PVE & PA-PVE & MJAE [$^\circ$] \\
    \midrule
    High (H) & MediaPipe & 47.8 & 23.5 & --   & --   & 36.5 \\
             & WiLoR     & 14.6 &  9.2 & 14.1 &  9.0 & 13.6 \\
             & \wristp{} (Ours)
                      & \textbf{2.8} & \textbf{2.1} & \textbf{2.8} & \textbf{2.2} & \textbf{3.2} \\
    \midrule
    Medium (M) & MediaPipe & 47.0 & 24.3 & --   & --   & 37.7 \\
               & WiLoR     & 15.2 &  9.3 & 14.6 &  9.1 & 14.1 \\
               & \wristp{} (Ours)
                        & \textbf{3.1} & \textbf{2.3} & \textbf{3.0} & \textbf{2.4} & \textbf{3.5} \\
    \midrule
    Low (L) & MediaPipe & 45.2 & 23.4 & --   & --   & 36.2 \\
            & WiLoR     & 15.4 &  9.1 & 14.9 &  8.8 & 13.8 \\
            & \wristp{} (Ours)
                     & \textbf{3.2} & \textbf{2.3} & \textbf{3.1} & \textbf{2.4} & \textbf{3.5} \\
    \midrule
    All & MediaPipe & 46.8 & 23.8 & --   & --   & 36.8 \\
        & WiLoR     & 15.0 &  9.2 & 14.5 &  9.0 & 13.8 \\
        & \wristp{} (Ours)
                 & \textbf{3.0} & \textbf{2.3} & \textbf{2.9} & \textbf{2.3} & \textbf{3.4} \\
    \bottomrule
  \end{tabular}
\end{table*}

\paragraph{Effect of camera view.}
Table~\ref{tab:supp-pose-view} compares pose accuracy across different
camera configurations when aggregating over all illumination
conditions.
For MediaPipe and WiLoR, we evaluate both environment and head-mounted
views, while \wristp{} uses only the wrist-mounted camera.

The performance of envirionment- and head-view remains comparable.
For instance, WiLoR's error shifts from 15.0\,mm MPJPE (env) to 15.3\,mm (head), and MJAE increases from 13.8$^\circ$ to 14.8$^\circ$.
This indicates that the distinction between these two external perspectives is minimal.
In contrast, \wristp{} achieves 3.0\,mm MPJPE and 3.4$^\circ$ MJAE from
the wrist-mounted view, outperforming both baseline configurations.
We attribute this performance gap to the inherent advantages of the wrist-mounted perspective: its proximal placement enables high-fidelity observation of fine-grained finger movements, while the viewing angle reduces self-occlusion compared to external views.

\begin{table}[t]
  \centering
  \caption{\textbf{Effect of camera view on pose accuracy (All illumination).}
  Position errors are reported in mm; MJAE in degrees ($^\circ$).
  Lower is better.}
  \label{tab:supp-pose-view}

  \small
  \setlength{\tabcolsep}{3pt}
  \renewcommand{\arraystretch}{1.1}

  \begin{tabular}{llccccc}
    \toprule
    Method & View & MPJPE & PA\textendash MPJPE & PVE & PA\textendash PVE & MJAE ($^\circ$) \\
    \midrule
    MediaPipe        & env   & 46.80 & 23.80 & --   & --   & 36.80 \\
                     & head  & 39.10 & 19.40 & --   & --   & 31.20 \\
    WiLoR            & env   & 15.00 &  9.20 & 14.50 &  9.00 & 13.80 \\
                     & head  & 15.30 &  9.30 & 14.70 &  9.20 & 14.80 \\
    \wristp{} (Ours) & wrist & \textbf{3.00} & \textbf{2.30} & \textbf{2.90} & \textbf{2.30} & \textbf{3.40} \\
    \bottomrule
  \end{tabular}
\end{table}

\paragraph{Contact and pressure estimation.}
Table~\ref{tab:supp-press-baseline} reports binary contact prediction
metrics for PressureVision++ and \wristp{}, split by illumination and
view.
Under the environment view, PressureVision++ achieves reasonably high
precision (92.4\% on average) but very low recall (20.9\%), resulting in
an F1-score of only 34.1\%.
This pattern indicates that PressureVision++ behaves as a highly
conservative classifier that predicts contact only in very confident
cases and misses the majority of true contact events.

In contrast, \wristp{} attains both high precision (92.6\%) and very
high recall (97.9\%) on average, yielding a 95.2\% F1-score.
The difference is most striking under Low illumination: for the
environment view, PressureVision++ reaches 31.5\% accuracy but only
0.66\% recall and 1.32\% F1-score, essentially failing to detect
contacts, whereas \wristp{} maintains 97.3\% accuracy, 97.9\% recall,
and 94.2\% F1-score.
For the head-mounted view, PressureVision++ degrades further to an
average recall of 7.3\% and F1-score of 13.5\%.
These results highlight that reliable contact prediction from distant
views is highly challenging, especially under poor lighting, and that
our wrist-mounted configuration is crucial for capturing fine-grained
contact events.

\begin{table*}[t]
  \centering
  \caption{\textbf{Contact prediction performance under different lighting conditions and camera views.}
  We report Accuracy (Acc), Precision (Prec), Recall (Rec), and F1-score (\%) for
  PressureVision++ and \wristp{} (ours). Bold numbers indicate the best results
  under each lighting condition for the env view.}
  \label{tab:supp-press-baseline}

  \setlength{\tabcolsep}{8pt}
  \renewcommand{\arraystretch}{1.25}

  \begin{tabular}{lllcccc}
    \toprule
    View & Method & Lighting & Acc & Prec & Rec & F1 \\
    \midrule
    \multirow{4}{*}{env} & \multirow{4}{*}{PressureVision++}
      & High (H)   & 43.3 & 87.9 & 27.6 & 42.0 \\
    & & Medium (M) & 56.0 & 98.0 & 31.3 & 47.4 \\
    & & Low (L)    & 31.5 & 100.0 & 0.7 & 1.3 \\
    & & Avg.       & 44.1 & 92.4 & 20.9 & 34.1 \\
    \midrule
    \multirow{4}{*}{head} & \multirow{4}{*}{PressureVision++}
      & High (H)   & 34.4 & 96.3 & 12.4 & 21.9 \\
    & & Medium (M) & 40.1 & 80.0 & 7.5  & 13.7 \\
    & & Low (L)    & 31.1 & 0.0  & 0.0  & 0.0  \\
    & & Avg.       & 35.3 & 90.5 & 7.3  & 13.5 \\
    \midrule
    \multirow{4}{*}{wrist} & \multirow{4}{*}{\wristp{} (Ours)}
      & High (H)   & \textbf{98.7} & \textbf{96.1} & \textbf{98.5} & \textbf{97.3} \\
    & & Medium (M) & \textbf{97.0} & \textbf{90.1} & \textbf{97.3} & \textbf{93.6} \\
    & & Low (L)    & \textbf{97.3} & \textbf{90.7} & \textbf{97.9} & \textbf{94.2} \\
    & & Avg.       & \textbf{97.7} & \textbf{92.6} & \textbf{98.0} & \textbf{95.2} \\
    \bottomrule
  \end{tabular}
\end{table*}

\subsection{Qualitative Results}
\label{sec:supp-qualitative}

\begin{figure*}[t]
  \centering
  \includegraphics[width=0.95\linewidth]{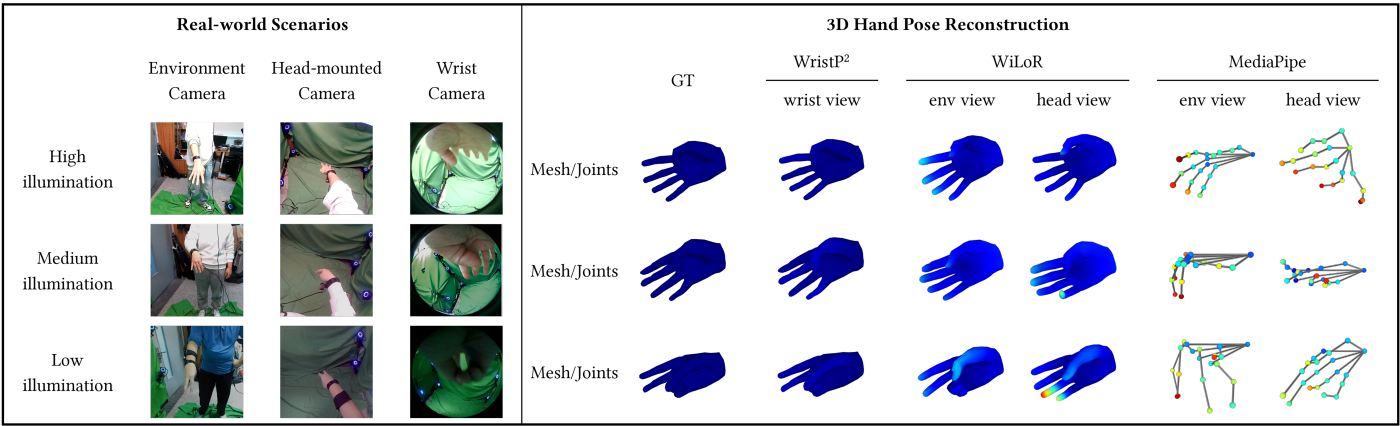}
  \caption{\textbf{Qualitative comparison on 3D hand pose reconstruction.}
  From left to right: input images from the environment camera, head-mounted camera, and wrist camera, 
  ground-truth mesh,
  and pose predicted by \wristp{} (ours, reconstructed mesh),
  WiLoR (reconstructed mesh),
  and MediaPipe (3D joints, no mesh support).
  Per-vertex or per-joint position errors are visualized as color-coded overlays 
  on the predicted meshes or 3D joints, 
  utilizing a consistent jet colormap with a minimum value (vmin) of 0 mm 
  and a maximum value (vmax) of 100 mm.
  Both MediaPipe and WiLoR suffer from inaccurate joint configurations, 
  while MediaPipe additionally exhibits finger entanglement 
  that results in physically implausible 3D hand pose reconstructions.
  Conversely, \wristp{} better preserves fingertip articulation and reduces depth
  ambiguities.}
  \Description{
A row of nine images showing a qualitative comparison of 3D hand pose reconstruction. 
The sequence includes: the input images from three distinct views, the ground-truth hand mesh, 
and the predicted meshes/joints from MediaPipe, WiLoR, and WRISTP2. 
WRISTP2's reconstruction shows more accurate fingertip articulation and fewer depth errors, 
especially in challenging lighting or wrist-angle conditions.
}
  \label{fig:supp-pose-qual}
\end{figure*}

\paragraph{Pose reconstruction.}
Figure~\ref{fig:supp-pose-qual} provides qualitative comparisons for 3D
hand pose reconstruction under challenging illumination and wrist
orientations.
Each example shows input camera images from distinct views, the ground-truth mesh,
and predictions from MediaPipe, WiLoR, and \wristp{}.
The baselines frequently exhibit depth ambiguities, finger interpenetration,
and noisy fingertip articulation, especially under low illumination.
In contrast, \wristp{} produces meshes that closely match the ground
truth in both global pose and fine-grained articulation, consistent with
the quantitative advantages reported in
Tables~\ref{tab:supp-pose-env-illum} and~\ref{tab:supp-pose-view}.

\begin{figure*}[t]
  \centering
  \includegraphics[width=\linewidth]{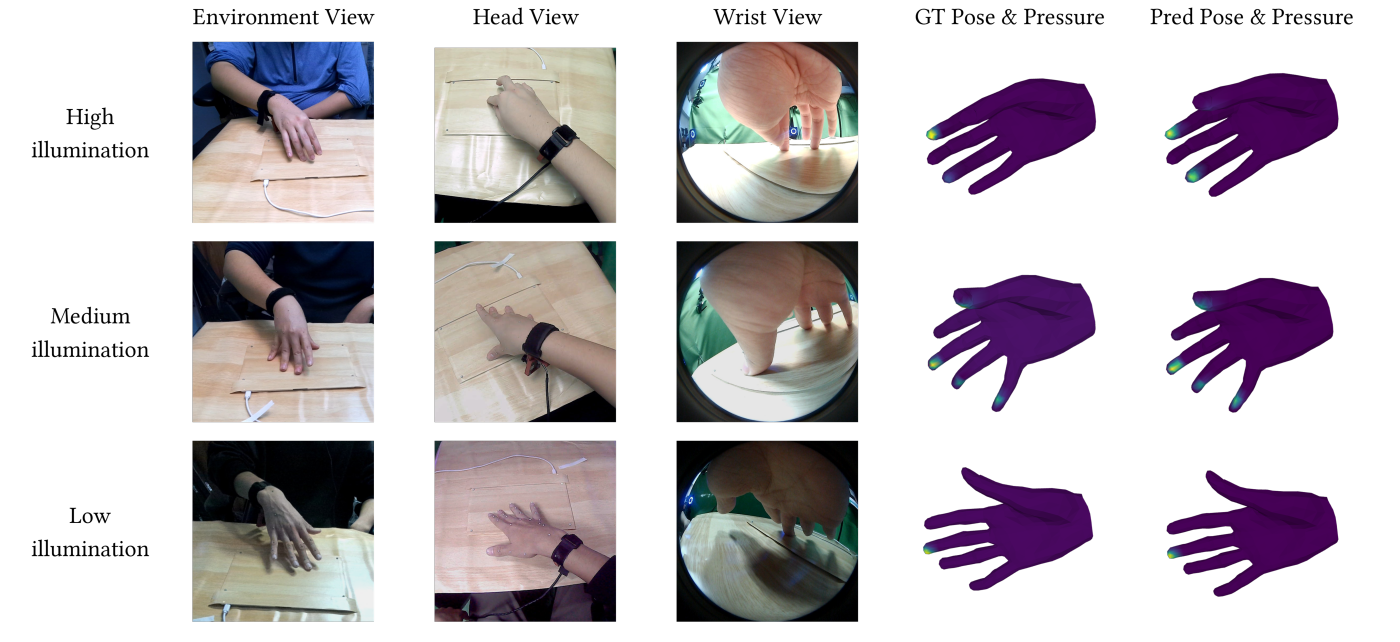}
  \caption{\textbf{Qualitative comparison on contact and pressure
  estimation.}
  From left to right: input images from distinct views,
  ground-truth hand meshes overlaid with pressure visualization, and prediction of \wristp{}.
  \wristp{} accurately localizes contact regions and reproduces
  plausible pressure magnitudes. 
  Meanwhile, PressureVision++ predicts all-zero pressure values 
  across the 2D pixel space for both the environment view and head view in all three cases, and is not included in the figure.}
  \Description{
A qualitative comparison of contact and pressure estimation across five columns:
the input images from three distinct views, the ground-truth pressure-overlaid hand mesh,
and the prediction from WRISTP2. 
PressureVision++ predicts zero pressure across all these scenes and thus cannot be included in the figure. In contrast, WRISTP2 shows accurate contact localization and realistic pressure magnitudes, especially around fingertip regions. 
}
  \label{fig:supp-press-qual}
\end{figure*}

\paragraph{Contact and pressure estimation.}
Figure~\ref{fig:supp-press-qual} shows planar interaction examples with
contact and pressure maps.
For each scene, we visualize ground-truth pressure, followed by
the predictions of \wristp{}. 
Notably, PressureVision++ exhibits overly conservative predictions, 
predicting zero pressure across all these scenes and thus cannot be included in the figure.
Consistent with Table~\ref{tab:supp-press-baseline},
\wristp{} produces contact regions that closely align with the ground
truth and captures coherent pressure patterns across the fingers and
palm.
PressureVision++ often misses light contacts at the fingertips, predicts
overly sparse contact regions, or fails completely under low
illumination, leading to very low recall and F1-scores.

\begin{figure*}[t]
  \centering
  \includegraphics[width=0.45\linewidth]{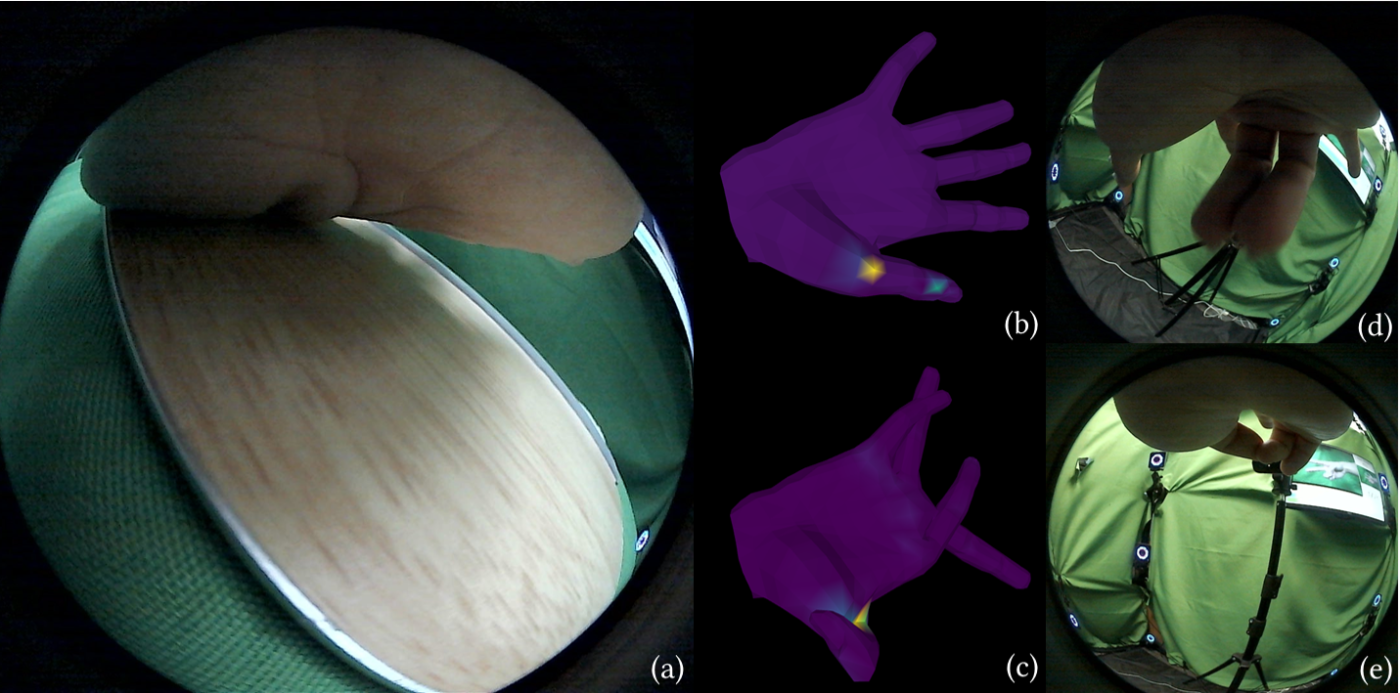}
  \caption{\textbf{Representative failure cases of \wristp{}.}
  Under challenging visual conditions, both pose and pressure estimation can degrade. 
  (a) Complete occlusion of multiple finger segments leads to a MPJPE of 31.7 mm 
      and notable pressure misestimation, 
      with ground-truth and predicted pose/pressure reconstructions shown in (b) and (c), respectively. 
      Specifically, the occlusion of the thumb pad in (c) stems from the 2D rendering angle, 
      and the model predicts no pressure here. 
  (d) Partial occlusion of the index finger coupled with 
      motion blur on the middle and ring fingers results in a MPJPE of 18.1 mm. 
  (e) Occlusion at the base of the middle finger and 
      full occlusion of the pinky yield a MPJPE of 19.4 mm. 
  We include these examples for transparency and to motivate future 
  research on handling such challenging scenarios.}
  \Description{
Representative failure cases of WRISTP2 shown across several examples.
Images illustrate scenarios with severe self-occlusion, object-induced
occlusion, or motion blur. In these cases, both the
reconstructed 3D hand pose and the estimated contact/pressure map exhibit
noticeable errors, such as mislocalized fingertips, incomplete contacts,
or incorrect pressure magnitudes. These examples highlight challenging
conditions where the system’s performance degrades.
}
  \label{fig:supp-failure}
\end{figure*}

\paragraph{Failure cases.}
Finally, Figure~\ref{fig:supp-failure} highlights typical failure cases
for \wristp{}.
Severe occlusions, motion blur, and hand
poses far outside the training distribution can still lead to noticeable
errors in pose or pressure estimation.
These examples complement our quantitative results and suggest promising
directions for future work, such as combining wrist-mounted cameras with
additional sensing modalities or incorporating explicit reasoning about
object appearance and contact dynamics.

\section{Additional Qualitative Results}
\label{app:qua}
\begin{figure*}[t]
    \centering
    \includegraphics[width=0.95\textwidth]{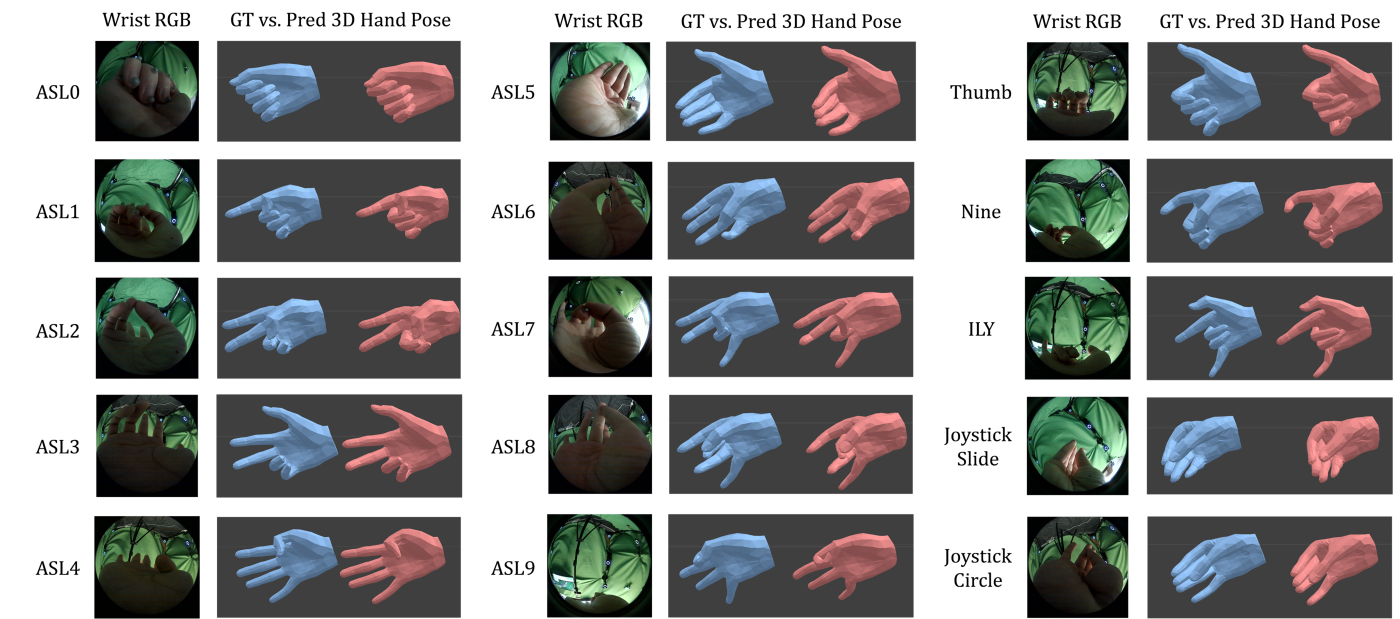}
    \caption{\textbf{Additional qualitative results of mid-air hand poses.} Images of wrist camera views for various mid-air gestures, with a comparison between the ground truth 3D hand poses (in blue) and the predicted 3D hand poses (in red).}
    \Description{
Examples of mid-air hand pose reconstruction. For each gesture, the
figure shows the wrist-camera RGB input alongside a 3D visualization of
the ground-truth hand mesh (blue) and the predicted hand mesh (red).
Gestures include single-finger articulation, multi-finger extension,
pinching, and dynamic mid-air poses. Differences between the blue and red
meshes highlight reconstruction accuracy across diverse poses and view
conditions.
}
    \label{fig:pose results}
\end{figure*}
As a supplement to Section~\ref{sec:offline-eval}, the model's predictions for 15 different mid-air interaction gestures at varying wrist pitch angles are presented in Fig.~\ref{fig:pose results}, demonstrating the model's high accuracy in predicting complex 3D hand postures.

\section{Specification of Multi-Finger Combinations in Pressure Control}
\begin{figure*}[t]
    \centering
    \includegraphics[width=\textwidth]{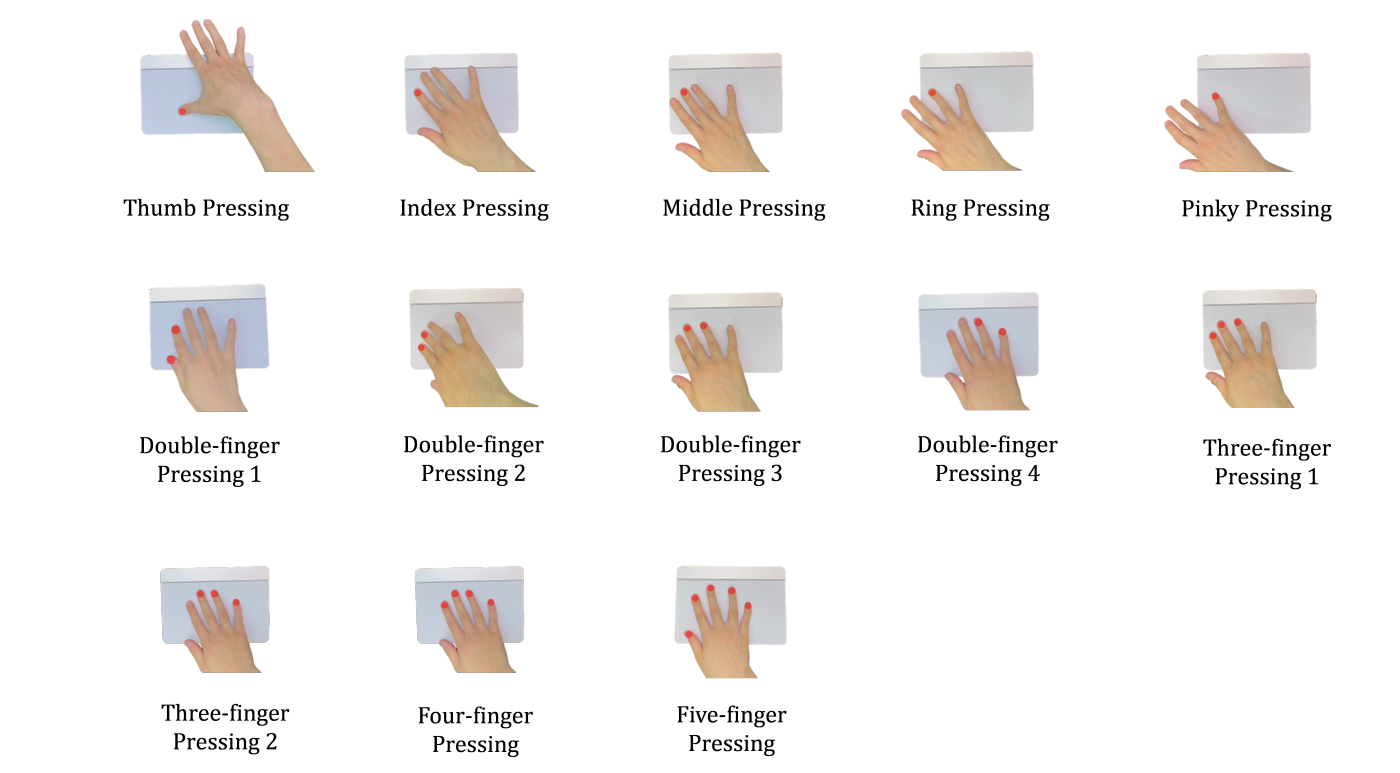}
    \caption{\textbf{The 13 finger combinations in the Multi-finger Pressure Control user study.} The combinations of single, double, triple, quadruple, and quintuple finger configurations represent natural and commonly used finger postures in planar hand interactions.}
    \Description{
Visualization of the 13 finger combinations used in the Multi-finger
Pressure Control study. The set spans single-, two-, three-, four-, and
five-finger configurations that commonly occur in planar hand
interactions. Each image illustrates the specific fingers involved in that
trial condition.
}
    \label{fig:Multi-Finger Combinations}
\end{figure*}
In Section~\ref{sec:STUDY 2}, thirteen different finger combinations were randomly selected, encompassing single-finger, two-finger, three-finger, four-finger, and five-finger configurations. These combinations represent common and natural finger postures in planar hand interactions, with the specific combinations illustrated in Fig.~\ref{fig:Multi-Finger Combinations}.

\makeatletter
\let\balance\relax
\makeatother
\end{document}